\documentclass[11pt,a4paper]{article}

\usepackage{fullpage}
\usepackage{epsfig}
\usepackage{gensymb}
\usepackage[english]{babel}
\usepackage{graphicx}
\usepackage{amssymb}
\usepackage{multirow}
\usepackage{times}

\usepackage{amsmath,amscd,amssymb,german}
\usepackage{mathrsfs}
\usepackage{graphicx}
\usepackage[scaled]{helvet}
\usepackage{dsfont} 
\usepackage[T1]{fontenc} 
\usepackage[ansinew]{inputenc} 

\newcommand{\remove}[1]{}

\newtheorem{problem}{Open problem}

\newtheorem{conjecture}{Conjecture}

\begin{document}

\title{A Survey on Small-Area Planar Graph Drawing\\
\vspace{3mm}
\large{Giuseppe Di Battista$^1$ and Fabrizio Frati$^{2}$\\}
\vspace{2mm}
\small{$1$ Dipartimento di Informatica e Automazione - Roma Tre University, Italy\\
$2$ School of Information Technoologies - The University of Sydney, Australia\\
\vspace{2mm}
\{gdb,frati\}@dia.uniroma3.it
\vspace{-10mm}
}}

\date{}
\maketitle

\begin{abstract}
We survey algorithms and bounds for constructing planar drawings of graphs in small area.
\end{abstract}

\section{Introduction}\label{se:intro}

It is typical in Computer Science to classify problems according to the amount of resources that are needed to solve them. Hence, problems are usually classified according to the amount of time or to the amount of memory that a specific model of computation requires for their solution.

This epistemological need of classifying problems finds, in the Graph Drawing field, a very original interpretation. A Graph Drawing problem can be broadly described as follows: Given a graph of a certain family and a drawing convention (e.g.\ all edges should be straight-line segments), draw the graph optimizing some specific features. Among those features a fundamental one is the amount of geometric space that the drawing spans and a natural question is: Which is the amount of space that is required for drawing a planar graph, or a tree, or a bipartite graph? Hence, besides classifying problems according to the above classical coordinates, Graph Drawing classifies problems according to the amount of geometric space that a drawing that solves that problem requires.

Of course, such a space requirement can be influenced by the class of graphs (one can expect that the area required to draw an $n$-vertex tree is less than the one required to draw an $n$-vertex general planar graph) and by the drawing convention (straight-line drawings look more constrained than drawings where edges can be polygonal lines).

The attempt of classifying graph drawing problems with respect to the space required spurred, over the last fifty years, a large body of research. On one hand, techniques have been devised to compute geometric lower bounds that are completely original and do not find counterparts in the techniques adopted in Computer Science to find time or memory lower bounds. On the other hand, the uninterrupted upper bound hunting has produced several elegant algorithmic techniques.

In this paper we survey the state of the art on such algorithmic and lower bound techniques for several families of planar graphs. Indeed, drawing planar graphs without crossings is probably the most classical Graph Drawing topic and many researches gave fundamental contributions on planar drawings of trees, outerplanar graphs, series-parallel graphs, etc.

We survey the state of the art focusing on the impact of the most popular drawing conventions on the geometric space requirements. In Section~\ref{se:straight-line} we discuss straight-line drawings. In Section~\ref{se:poly-line} we analyze drawings where edges can be polygonal lines. In Section~\ref{se:upward} we describe upward drawings, i.e.\ drawings of directed acyclic graphs where edges follow a common vertical direction. In Section~\ref{se:convex} we describe convex drawings, where the faces of a planar drawing are constrained to be convex polygons. Proximity drawings, where vertices and edges should enforce some proximity constraints, are discussed in Section~\ref{se:proximity}. Section~\ref{se:clustered} is devoted to drawings of clustered graphs.

We devote special attention to put in evidence those that we consider the main open problems of the~field.

\section{Preliminaries}\label{se:preliminaries}

In this section we present preliminaries and definitions. For more about graph drawing, see~\cite{dett-gd-99,kw-dgmm-01}.

\subsection*{Planar Drawings, Planar Embeddings, and Planar Graphs} \label{se:graphs-planarembeddings}

All the graphs that we consider are \emph{simple}, i.e., they contain no multiple edges and loops. A \emph{drawing} of a graph $G(V,E)$ is a mapping of each vertex of $V$ to a point in the plane and of each edge of $E$ to a simple curve connecting its endpoints. A drawing is \emph{planar} if no two edges intersect except, possibly, at common endpoints. A \emph{planar graph} is a graph admitting a planar drawing.

A planar drawing of a graph determines a circular ordering of the edges incident to each vertex. Two drawings of the same graph are \emph{equivalent} if they determine the same circular ordering around each vertex and a \emph{planar embedding} (sometimes also called {\em combinatorial embedding}) is an equivalence class of planar drawings. A graph is \emph{embedded} when an embedding of it has been decided. A planar drawing partitions the plane into topologically connected regions, called \emph{faces}. The unbounded face is the \emph{outer face}, while the bounded faces are the \emph{internal faces}. The outer face of a graph $G$ is denoted by $f(G)$. A graph together with a planar embedding and a choice for its outer face is a \emph{plane graph}. In a plane graph, \emph{external} and \emph{internal} vertices are defined as the vertices incident and not incident to the outer face, respectively. Sometimes, the distinction is made between \emph{planar embedding} and \emph{plane embedding}, where the former is an equivalence class of planar drawings and the latter is a planar embedding together with a choice for the outer face. The \emph{dual graph} of an embedded planar graph $G$ has a vertex for each face of $G$ and has an edge $(f,g)$ for each two faces $f$ and $g$ of $G$ sharing an edge.

\subsection*{Maximality and Connectivity} \label{se:graphs-connectivity}

A plane graph is \emph{maximal} (or equivalently is a \emph{triangulation}) when all its faces are delimited by \emph{$3$-cycles}, that is, by cycles of three vertices. A planar graph is \emph{maximal} when it can be embedded as a triangulation. Algorithms for drawing planar graphs usually assume to deal with maximal planar graphs. In fact, any planar graph can be augmented to a maximal planar graph by adding some ``dummy'' edges to the graph. Then the algorithm can draw the maximal planar graph and finally the inserted dummy edges can be removed obtaining a drawing of the input graph.

A graph is \emph{connected} if every pair of vertices is connected by a path. A graph with at least $k+1$ vertices is \emph{$k$-connected} if removing any (at most) $k-1$ vertices leaves the graph connected; $3$-connected, $2$-connected, and $1$-connected graphs are also called \emph{triconnected}, \emph{biconnected}, and \emph{connected} graphs, respectively. A \emph{separating cycle} is a cycle whose removal disconnects the graph.

\subsection*{Classes of Planar Graphs} \label{se:graphs-classes}
A \emph{tree} is a connected acyclic graph. A \emph{leaf} in a tree is a node of degree one. A \emph{caterpillar} $C$ is a tree such that the removal from $C$ of all the leaves and of their incident edges turns $C$ into a path, called the \emph{backbone} of the caterpillar.

A \emph{rooted tree} is a tree with one distinguished node called \emph{root}. In a rooted tree each node $v$ at distance (i.e., length of the shortest path) $d$ from the root is the \emph{child} of the only node at distance $d-1$ from the root $v$ is connected to. A \emph{binary tree} (a \emph{ternary tree}) is a rooted tree such that each node has at most two children (resp. three children). Binary and ternary trees can be supposed to be rooted at any node of degree at most two and three, respectively. The \emph{height} of a rooted tree is the maximum number of nodes in any path from the root to a leaf. Removing a non-leaf node $u$ from a tree disconnects the tree into connected components. Those containing children of $u$ are the \emph{subtrees} of $u$.

A \emph{complete tree} is a rooted tree such that each non-leaf node has the same number of children and such that each leaf has the same distance from the root. Complete trees of degree three and four are also called \emph{complete binary trees} and \emph{complete ternary trees}, respectively.

A rooted tree is \emph{ordered} if a clockwise order of the neighbors of each node (i.e., a planar embedding) is specified. In an ordered binary tree and in an ordered ternary tree, fixing a linear ordering of the children of the root yields to define the \emph{left} and \emph{right child} of a node, and the \emph{left}, \emph{middle}, and \emph{right child} of a node, respectively. If the tree is ordered and binary (ternary), the subtrees rooted at the left and right child (at the left, middle, and right child) of a node $u$ are the \emph{left} and the \emph{right subtree} of $u$ (the \emph{left}, the \emph{middle}, and the \emph{right subtree} of $u$), respectively. Removing a path $P$ from a tree disconnects the tree into connected components. The ones containing children of nodes in $P$ are the \emph{subtrees} of $P$. If the tree is ordered and binary (ternary), then each component is a \emph{left} or \emph{right subtree} (a \emph{left}, \emph{middle}, or \emph{right subtree}) of $\mathcal P$, depending on whether the root of such subtree is a left or right child (is a left, middle, or right child) of a node in $\mathcal P$, respectively.

An \emph{outerplane graph} is a plane graph such that all the vertices are incident to the outer face. An \emph{outerplanar embedding} is a planar embedding such that all the vertices are incident to the same face. An \emph{outerplanar graph} is a graph that admits an outerplanar embedding. A \emph{maximal outerplane graph} is an outerplane graph such that all its internal faces are delimited by cycles of three vertices. A \emph{maximal outerplanar embedding} is an outerplanar embedding such that all its faces, except for the one to which all the vertices are incident, are delimited by cycles of three vertices. A \emph{maximal outerplanar graph} is a graph that admits a maximal outerplanar embedding. Every outerplanar graph can be augmented to maximal by adding dummy edges to it.

If we do not consider the vertex corresponding to the outer face of $G$ and its incident edges then the dual graph of an outerplane graph $G$ is a tree.  Hence, when dealing with outerplanar graphs, we talk about the \emph{dual tree} of an outerplanar graph (meaning the dual graph of an outerplane embedding of the outerplanar graph). The nodes of the dual tree of a maximal outerplane graph $G$ have degree at most three. Hence the dual tree of $G$ can be rooted to be a binary tree.

\emph{Series-parallel graphs} are the graphs that can be inductively constructed as follows. An edge $(u,v)$ is a series-parallel graph with \emph{poles} $u$ and $v$. Denote by $u_i$ and $v_i$ the poles of a series-parallel graph $G_i$. Then, a \emph{series composition} of a sequence $G_1,G_2,\dots,G_k$ of series-parallel graphs, with $k\geq 2$, constructs a series-parallel graph that has poles $u=u_1$ and $v=v_k$, that contains graphs $G_i$ as subgraphs, and such that vertices $v_i$ and $u_{i+1}$ have been identified to be the same vertex, for each $i=1,2,\dots,k-1$. A \emph{parallel composition} of a set $G_1,G_2,\dots,G_k$ of series-parallel graphs, with $k\geq 2$, constructs a series-parallel graph that has poles $u=u_1=u_2=\cdots=u_k$ and $v=v_1=v_2=\cdots=v_k$, that contains graphs $G_i$ as subgraphs, and such that vertices $u_1,u_2,\cdots,u_k$ (vertices $v_1,v_2,\cdots,v_k$) have been identified to be the same vertex. A \emph{maximal series-parallel graph} is such that all its series compositions construct a graph out of exactly two smaller series-parallel graphs $G_1$ and $G_2$, and such that all its parallel compositions have a component which is the edge between the two poles. Every series-parallel graph can be augmented to maximal by adding dummy edges to it. The \emph{fan-out} of a series-parallel graph is the maximum number of components in a parallel composition.

A graph $G$ is \emph{bipartite} if its vertex set $V$ can be partitioned into two subsets $V_1$ and $V_2$ so that every edge of $G$ is incident to a vertex of $V_1$ and to a vertex of $V_2$. A \emph{bipartite planar graph} is both bipartite and planar. A \emph{maximal bipartite planar graph} admits a planar embedding in which all its faces have exactly four incident vertices. Every bipartite planar graph with at least four vertices can be augmented to maximal by adding dummy edges to it.

\subsection*{Drawing Standards}

A {\em straight-line drawing} is a drawing such that each edge is represented by a straight-line segment. A {\em poly-line drawing} is a drawing such that each edge is represented by a sequence of consecutive segments. The points in which two consecutive segments of the same edge touch are called \emph{bends}. A {\em grid drawing} is a drawing such that vertices and bends have integer coordinates. An {\em orthogonal drawing} is a poly-line drawing such that each edge is represented by a sequence of horizontal and vertical segments. A {\em convex drawing} (resp. {\em strictly-convex drawing}) is a planar drawing such that each face is delimited by a convex polygon (resp. strictly-convex polygon), that is, every interior angle of the drawing is at most $180^{\circ}$ (resp. less than $180^{\circ}$) and every exterior angle is at least $180^{\circ}$ (resp. more than $180^{\circ}$). An \emph{order-preserving drawing} is a drawing such that the order of the edges incident to each vertex respects an order fixed in advance. An \emph{upward drawing} (resp. \emph{strictly-upward drawing}) of a rooted tree is a drawing such that each edge is represented by a non-decreasing curve (resp. increasing curve). A \emph{visibility representation} is a drawing such that each vertex is represented by a horizontal segment $\sigma(u)$, each edge $(u,v)$ is represented by a vertical segment connecting a point of $\sigma(u)$ with a point of $\sigma(v)$, and no two segments cross, except if they represent a vertex and one of its incident edges.

\subsection*{Area of a Drawing}

The \emph{bounding box} of a drawing is the smallest rectangle with sides parallel to the axes that contains the drawing completely. The \emph{height} and \emph{width} of a drawing are the height and width of its bounding box. The \emph{area} of a drawing is the area of its bounding box. The \emph{aspect ratio} of a drawing is the ratio between the maximum and the minimum of the height and width of the drawing. Observe that the concept of area of a drawing only makes sense once a \emph{resolution rule} is fixed, i.e., a rule that does not allow vertices to be arbitrarily close (\emph{vertex resolution rule}), or edges to be arbitrarily short (\emph{edge resolution rule}). Without any of such rules, one could just construct drawings with arbitrarily small area. It is usually assumed in the literature that graph drawings in small area have to be constructed on a grid. In fact all the algorithms we will present in Sects.~\ref{se:straight-line},~\ref{se:poly-line},~\ref{se:upward},~\ref{se:convex}, and~\ref{se:clustered} assign integer coordinates to vertices. The assumption of constructing drawings on the grid is usually relaxed in the context of proximity drawings (hence in Sect.~\ref{se:proximity}), where in fact it is assumed that no two vertices have distance less than one unit.

\subsection*{Directed Graphs and Planar Upward Drawings}

A \emph{directed acyclic graph} (\emph{DAG} for short) is a graph whose edges are oriented and containing no cycle $(v_1,\dots,v_n)$ such that edge $(v_i,v_{i+1})$ is directed from $v_i$ to $v_{i+1}$, for $i=1,\dots,n-1$, and edge $(v_n,v_1)$ is directed from $v_n$ to $v_1$. The \emph{underlying graph} of a DAG $G$ is the undirected graph obtained from $G$ by removing the directions on its edges. An \emph{upward drawing} of a DAG is such that each edge is represented by an increasing curve. An \emph{upward planar drawing} is a drawing which is both upward and planar. An \emph{upward planar DAG} is a DAG that admits an upward planar drawing. In a directed graph, the \emph{outdegree} of a vertex is the number of edges leaving the vertex and the \emph{indegree} of a vertex is the number of edges entering the vertex. A \emph{source} (resp. \emph{sink}) is a vertex with indegree zero (resp. with outdegree zero). An \emph{st-planar DAG} is a DAG with exactly one source $s$ and one sink $t$ that admits an upward planar embedding in which $s$ and $t$ are on the outer face. \emph{Bipartite DAGs} and \emph{directed trees} are DAGs whose underlying graphs are bipartite graphs and trees, respectively. A \emph{series-parallel DAG} is a DAG that can be inductively constructed as follows. An edge $(u,v)$ directed from $u$ to $v$ is a series-parallel DAG with \emph{starting pole} $u$ and \emph{ending pole} $v$. Denote by $u_i$ and $v_i$ the starting and ending poles of a series-parallel DAG $G_i$, respectively. Then, a \emph{series composition} of a sequence $G_1,G_2,\dots,G_k$ of series-parallel DAGs, with $k\geq 2$, constructs a series-parallel DAG that has starting pole $u=u_1$, that has ending pole $v=v_k$, that contains DAGs $G_i$ as subgraphs, and such that vertices $v_i$ and $u_{i+1}$ have been identified to be the same vertex, for each $i=1,2,\dots,k-1$. A \emph{parallel composition} of a set $G_1,G_2,\dots,G_k$ of series-parallel DAGs, with $k\geq 2$, constructs a series-parallel DAG that has starting pole $u=u_1=u_2=\cdots=u_k$, that has ending pole $v=v_1=v_2=\cdots=v_k$, that contains DAGs $G_i$ as subgraphs, and such that vertices $u_1,u_2,\cdots,u_k$ (vertices $v_1,v_2,\cdots,v_k$) have been identified to be the same vertex. We remark that series-parallel DAGs are a subclass of the upward planar DAGs whose underlying graph is a series-parallel graph.

\subsection*{Proximity Drawings}

A \emph{Delaunay drawing} of a graph $G$ is a straight-line drawing such that no three vertices are on the same line, no four vertices are on the same circle, and three vertices $u$, $v$, and $z$ form a $3$-cycle $(u,v,z)$ in $G$ if and only if the circle passing through $u$, $v$, and $z$ in the drawing contains no vertex other than $u$, $v$, and $z$. A \emph{Delaunay triangulation} is a graph that admits a Delaunay drawing.

The \emph{Gabriel region} of two vertices $x$ and $y$ is the disk having segment $\overline{xy}$ as diameter. A \emph{Gabriel drawing} of a graph $G$ is a straight-line drawing of $G$ having the property that two vertices $x$ and $y$ of the drawing are connected by an edge if and only if the Gabriel region of $x$ and $y$ does not contain any other vertex. A \emph{Gabriel graph} is a graph admitting a Gabriel drawing.

A \emph{relative neighborhood drawing} of a graph $G$ is a straight-line drawing such that two vertices $x$ and $y$ are adjacent if and only if there is no vertex whose distance to both $x$ and $y$ is less than the distance between $x$ and $y$. A \emph{relative neighborhood graph} is a graph admitting a relative neighborhood drawing.

A \emph{nearest neighbor drawing} of a graph $G$ is a straight-line drawing of $G$ such that each vertex has a unique closest vertex and such that two vertices $x$ and $y$ of the drawing are connected by an edge if and only if $x$ is the vertex of $G$ closest to $y$ or viceversa. A \emph{nearest neighbor graph} is a graph admitting a nearest neighbor drawing.

A \emph{$\beta$-drawing} is a straight-line drawing of $G$ having the property that two vertices $x$ and $y$ of the drawing are connected by an edge if and only if the $\beta$-region of $x$ and $y$ does not contain any other vertex. The \emph{$\beta$-region} of $x$ and $y$ is the line segment $\overline{xy}$ if $\beta=0$, it is the intersection of the two closed disks of radius $d(x,y)/(2\beta)$ passing through both $x$ and $y$ if $0<\beta <1$, it is the intersection of the two closed disks of radius $d(x,y)/(2\beta)$ that are centered on the line through $x$ and $y$ and that respectively pass through $x$ and through $y$ if $1\leq \beta <\infty$, and it is the closed infinite strip perpendicular to the line segment $\overline{xy}$ if $\beta =\infty$.

\emph{Weak proximity drawings} are such that there is no geometric requirement on the pairs of vertices not connected by an edge. For example, a \emph{weak Gabriel drawing} of a graph $G$  is a straight-line drawing of $G$ having the property that if two vertices $x$ and $y$ of the drawing are connected by an edge then the Gabriel region of $x$ and $y$ does not contain any other vertex, while there might exist two vertices whose Gabriel region is empty and that are not connected by an edge.

A \emph{Euclidean minimum spanning tree} $T$ of a set $P$ of points is a tree spanning the points in $P$ (that is, the nodes of $T$ coincide with the points of $P$ and no ``Steiner points'' are allowed) and having minimum total edge length.

A \emph{greedy drawing} of a graph $G$ is a straight-line drawing of $G$ such that, for every pair of nodes $u$ and $v$, there exists a \emph{distance-decreasing path}, where a path $(v_0,v_1,\ldots,v_m)$ is distance-decreasing if $d(v_{i},v_{m})<d(v_{i-1},v_{m})$, for $i=1,\ldots,m$, where $d(p,q)$ denotes the Euclidean distance between two points $p$ and $q$.

For more about proximity drawings, see Chapter 7 in~\cite{gd-handbook}.

\subsection*{Clustered Graphs and $c$-Planar Drawings}

A \emph{clustered graph} is a pair $C(G,T)$, where $G$ is a graph, called \emph{underlying graph}, and $T$ is a rooted tree, called \emph{inclusion tree}, such that the leaves of $T$ are the vertices of $G$. Each internal node $\nu$ of $T$ corresponds to the subset of vertices of $G$, called \emph{cluster}, that are the leaves of the subtree of $T$ rooted at $\nu$. A clustered graph $C(G,T)$ is \emph{$c$-connected} if each cluster induces a connected subgraph of $G$, it is \emph{non-$c$-connected} otherwise.

A \emph{drawing} $\Gamma$ of a clustered graph $C(G,T)$ consists of a drawing of $G$ (each vertex is a point in the plane and each edge is as Jordan curve between its endvertices) and of a representation of each node $\mu$ of $T$ as a simple closed region containing all and only the vertices that belong to $\mu$. A drawing is \emph{$c$-planar} if it has no edge crossings (i.e., the drawing of the underlying graph is planar), no edge-region crossings (i.e., an edge intersects the boundary of a cluster at most once), and no region-region crossings (i.e., no two cluster boundaries cross).

A \emph{$c$-planar embedding} is an equivalence class of $c$-planar drawings of $C$, where two $c$-planar drawings are equivalent if they have the same order of the edges incident to each vertex and the same order of the edges incident to each cluster.

\section{Straight-line Drawings}\label{se:straight-line}

In this section, we discuss algorithms and bounds for constructing small-area planar straight-line drawings %
of planar graphs and their subclasses.
In Sect.~\ref{se:straight-planar} we deal with general planar graphs,
in Sect.~\ref{se:straight-bipartite} we deal with $4$-connected and bipartite graphs,
in Sect.~\ref{se:straight-series-parallel} we deal with series-parallel graphs,
in Sect.~\ref{se:straight-outerplanar} we deal with outerplanar graphs,
and in Sect.~\ref{se:straight-trees} we deal with trees.
Table~\ref{ta:straight-line} summarizes the best known area bounds for straight-line planar drawings of planar graphs and their subclasses. Observe that the lower bounds of the table that refer to general planar graphs, $4$-connected planar graphs, and bipartite planar graphs hold true for {\em plane} graphs.

\begin{table}[!htb]\footnotesize
\centering
  \linespread{1.2}
  \selectfont
  \begin{tabular}{|c|c|c|c|c|}
    \cline{2-5}
    \multicolumn{1}{c|}{} & \emph{Upper Bound} & \emph{Refs.} & \emph{Lower Bound} & \emph{Refs.} \\
    \hline
    {\em General Planar Graphs} & $\frac{8n^2}{9}+O(n)$ & \cite{fpp-hdpgg-90,s-epgg-90,b-dpg89a-08} & $\frac{4n^2}{9}-O(n)$ & \cite{Val81,fpp-hdpgg-90,FratiP07,mnra-madp3t-10}\\
    \hline
    {\em $4$-Connected Planar Graphs} & $\lfloor\frac{n}{2}\rfloor \times (\lceil\frac{n}{2}\rceil-1)$ & \cite{mnn-gd4pg-01} & $\lfloor\frac{n}{2}\rfloor \times (\lceil\frac{n}{2}\rceil-1)$ & \cite{mnn-gd4pg-01}\\
    \hline
    {\em Bipartite Planar Graphs} & $\lfloor\frac{n}{2}\rfloor \times (\lceil\frac{n}{2}\rceil-1)$ & \cite{bb-dpbgsa-05} & $\lfloor\frac{n}{2}\rfloor \times (\lceil\frac{n}{2}\rceil-1)$ & \cite{bb-dpbgsa-05}\\
    \hline
    {\em Series-Parallel Graphs} & $O(n^2)$ & \cite{fpp-hdpgg-90,s-epgg-90,zhn-sgdpgbb-10} & $\Omega(n 2^{\sqrt{\log n}})$ & \cite{f-lbarspg-j10}\\
    \hline
    {\em Outerplanar Graphs} & $O(n^{1.48})$ & \cite{df-sadog-j09} & $\Omega(n)$ & \emph{trivial}\\
    \hline
    {\em Trees} & $O(n \log n)$ & \cite{cdp-noad-92} & $\Omega(n)$ & \emph{trivial}\\
    \hline
 \end{tabular}
 \vspace{2mm}
 \caption{\small A table summarizing the area requirements for straight-line planar drawings of several classes of planar graphs. Notice that $4$-connected planar graphs have been studied only with the additional constraint of having at least four vertices on the outer face.}
  \label{ta:straight-line}
\end{table}

\subsection{General Planar Graphs} \label{se:straight-planar}

In this section, we discuss algorithms and bounds for constructing small-area planar straight-line drawings of general planar graphs. Observe that, in order to derive bounds on the area requirements of general planar graphs, it suffices to restrict the attention to maximal planar graphs, as every planar graph can be augmented to maximal by the insertion of ``dummy'' edges. Moreover, such an augmentation can be performed in linear time~\cite{r-nmdg-87}.

We start by proving that every plane graph admits a planar straight-line drawing~\cite{w-bzv-36,s-cm-51}. The simplest and most elegant proof of such a statement is, in our opinion, the one presented by F\'ary in 1948~\cite{f-srpg-48}.

\begin{figure}[htb]
  \centering
  \begin{tabular}{c c c}
	\mbox{\epsfig{figure=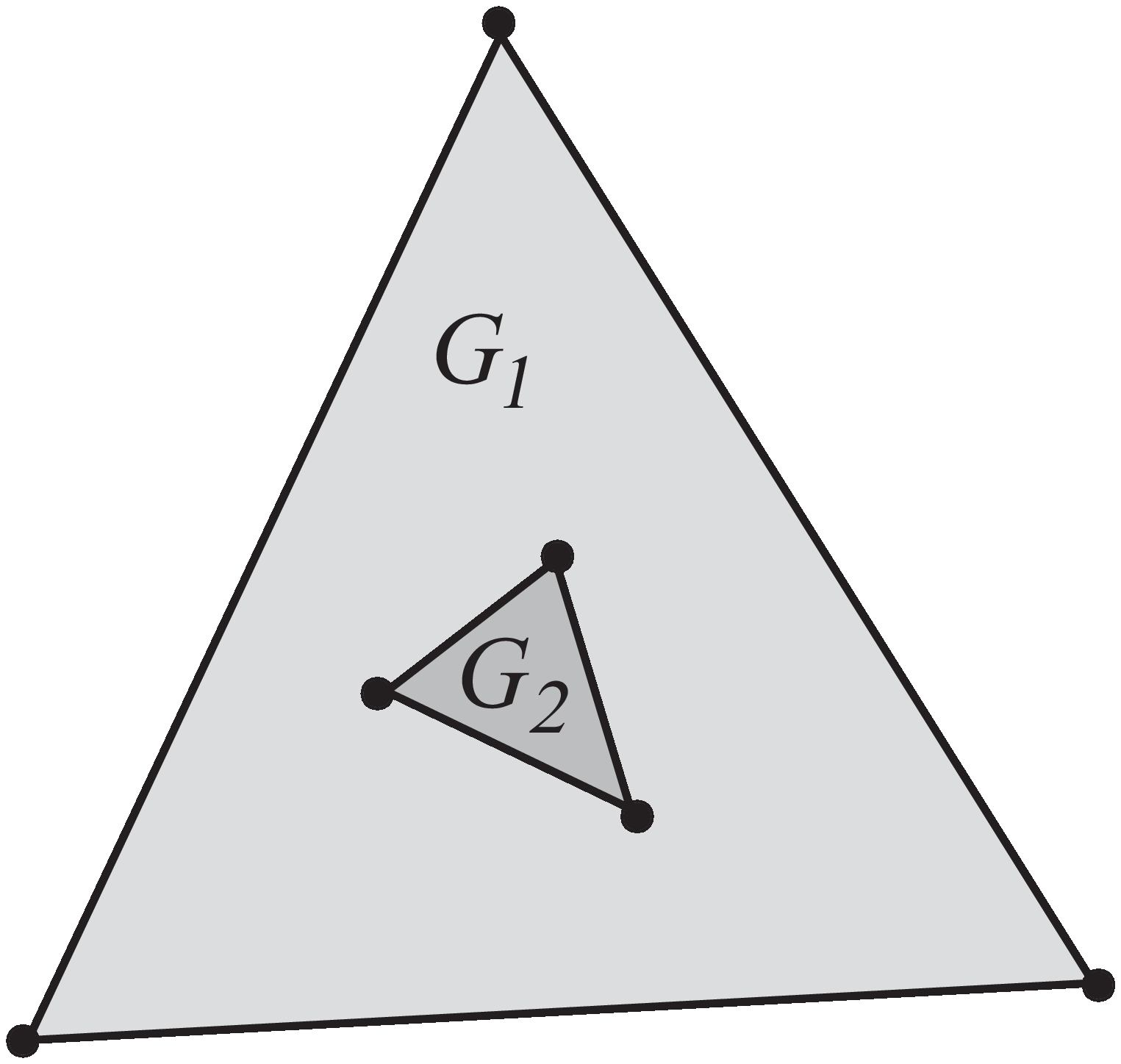,scale=0.2,clip=}} \hspace{5mm} &
	\mbox{\epsfig{figure=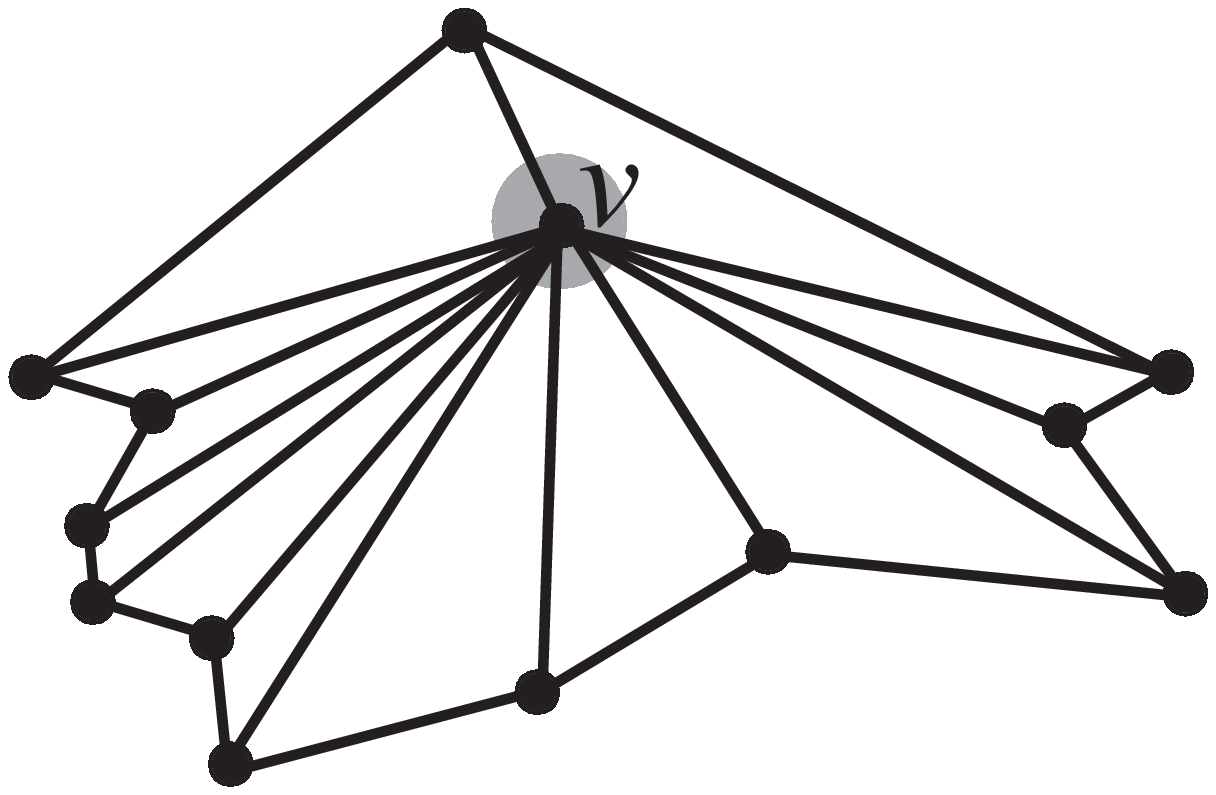,scale=0.27,clip=}} \hspace{5mm} &
	\mbox{\epsfig{figure=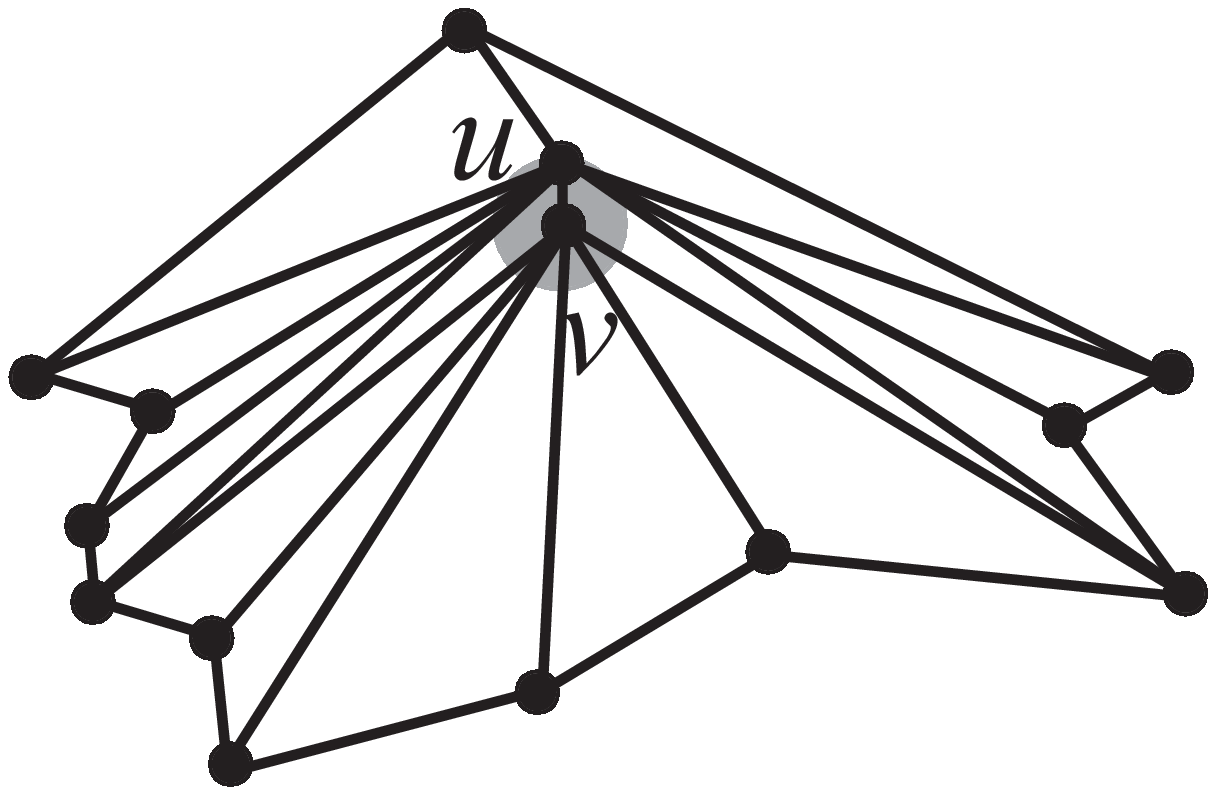,scale=0.27,clip=}} \\
        (a) \hspace{5mm} & (b) \hspace{5mm} & (c)\\
  \end{tabular}
  \caption{(a) Induction in F\'ary's algorithm if $G$ contains a separating $3$-cycle. (b)--(c) Induction in F\'ary's algorithm if $G$ contains no separating $3$-cycle.}
  \label{fig:fary1}
\end{figure}

F\'ary's algorithm works by induction on the number $n$ of vertices of the plane graph $G$; namely, the algorithm inductively assumes that a straight-line planar drawing of $G$ can be constructed with the further constraint that the outer face $f(G)$ is drawn as an arbitrary triangle $\Delta$. The inductive hypothesis is trivially satisfied when $n=3$. If $n>3$, then two cases are possible. In the first case $G$ contains a separating $3$-cycle $c$. Then let $G_1$ (resp. $G_2$) be the graph obtained from $G$ by removing all the vertices internal to $c$ (resp. external to $c$). Both $G_1$ and $G_2$ have less than $n$ vertices, hence the inductive hypothesis applies first to construct a straight-line planar drawing $\Gamma_1$ of $G_1$ in which $f(G_1)$ is drawn as an arbitrary triangle $\Delta$, and second to construct a straight-line planar drawing $\Gamma_2$ of $G_2$ in which $f(G_2)$ is drawn as $\Delta(c)$, where $\Delta(c)$ is the triangle representing $c$ in $\Gamma_1$ (see Fig.~\ref{fig:fary1}(a)). Thus, a straight-line drawing $\Gamma$ of $G$ in which $f(G)$ is represented by $\Delta$ is obtained.
In the second case, $G$ does not contain any separating $3$-cycle, i.e. $G$ is $4$-connected. Then, consider any internal vertex $u$ of $G$ and consider any neighbor $v$ of $u$. Construct an $(n-1)-$vertex plane graph $G'$ by removing $u$ and all its incident edges from $G$, and by inserting ``dummy'' edges between $v$ and all the neighbors of $u$ in $G$, except for the two vertices $v_1$ and $v_2$ forming faces with $u$ and $v$. The graph $G'$ is simple, as $G$ contains no separating $3$-cycle. Hence, the inductive hypothesis applies to construct a straight-line planar drawing $\Gamma'$ of $G'$ in which $f(G')$ is drawn as $\Delta$. Further, dummy edges can be removed and vertex $u$ can be introduced in $\Gamma'$ together with its incident edges, without altering the planarity of $\Gamma'$. In fact, $u$ can be placed at a suitable point in the interior of a small disk centered at $v$, thus obtaining a straight-line drawing $\Gamma$ of $G$ in which $f(G)$ is represented by $\Delta$ (see Figs.~\ref{fig:fary1}(b)--(c)).

The first algorithms for constructing planar straight-line grid drawings of planar graphs in polynomial area were presented (fifty years later than F\'ary's algorithm!) by de Fraysseix, Pach, and Pollack~\cite{fpp-sssfepg-88,fpp-hdpgg-90} and, simultaneously and independently, by Schnyder~\cite{s-epgg-90}. The approaches of the two algorithms, that we sketch below, are still today the base of every known algorithm to construct planar straight-line grid drawings of triangulations.

The algorithm by de Fraysseix \emph{et al.}~\cite{fpp-sssfepg-88,fpp-hdpgg-90} relies on two main ideas.

\begin{figure}[htb]
  \centering
  \begin{tabular}{c c c}
	\mbox{\epsfig{figure=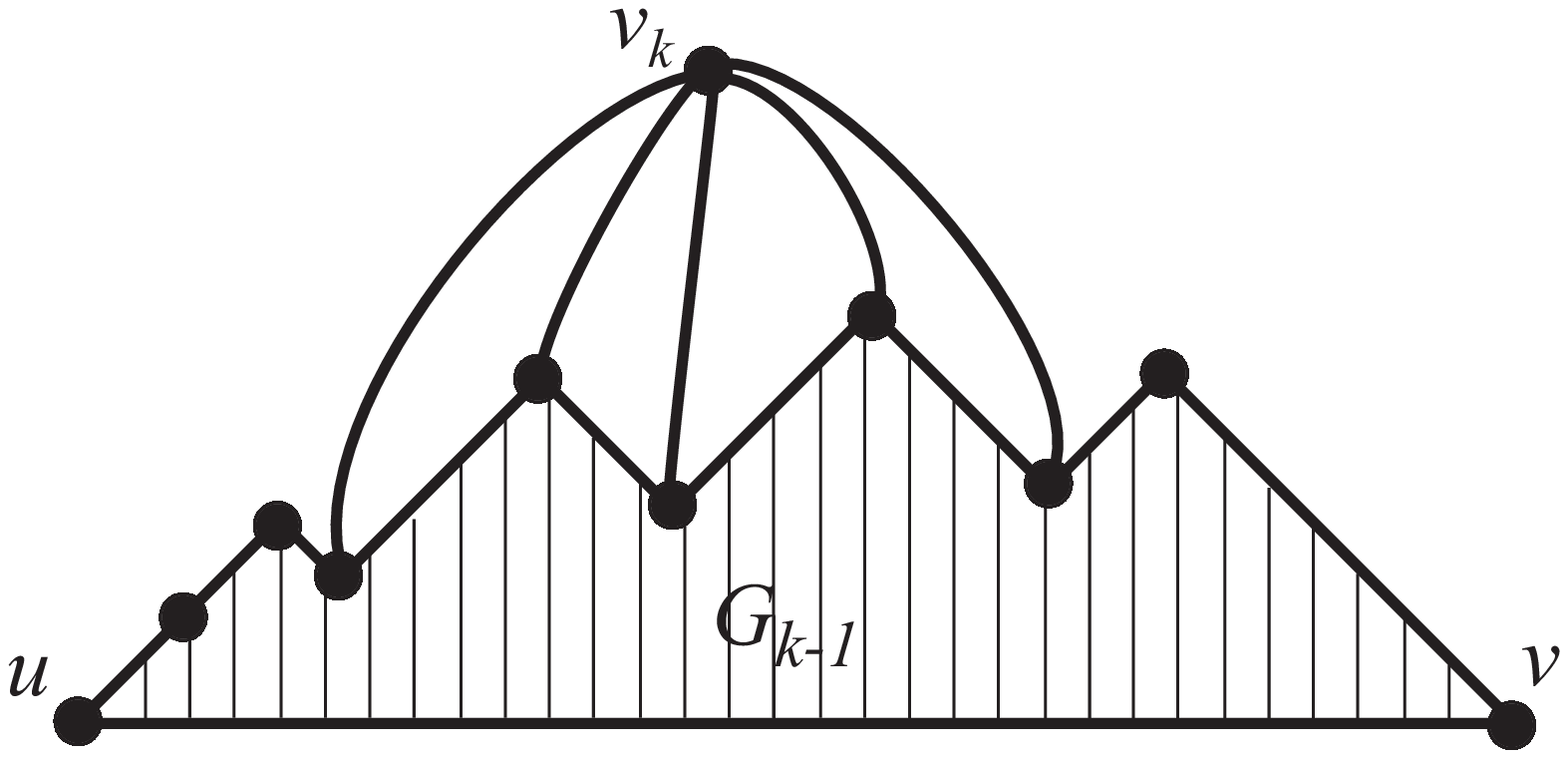,scale=0.28,clip=}} \hspace{3mm} &
	\mbox{\epsfig{figure=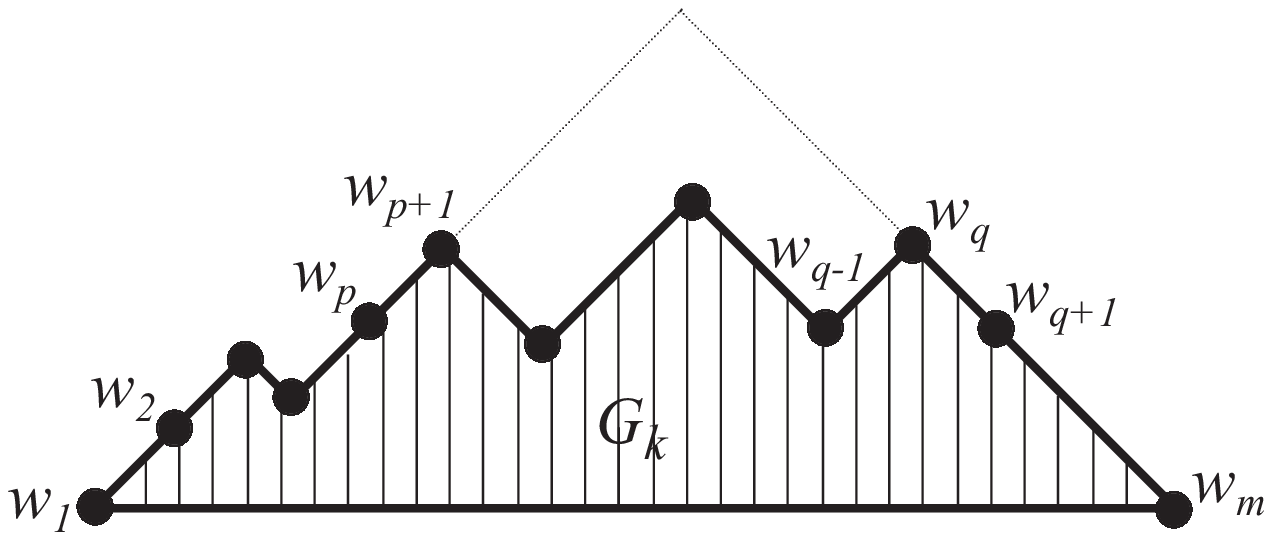,scale=0.35,clip=}} \hspace{3mm} &
	\mbox{\epsfig{figure=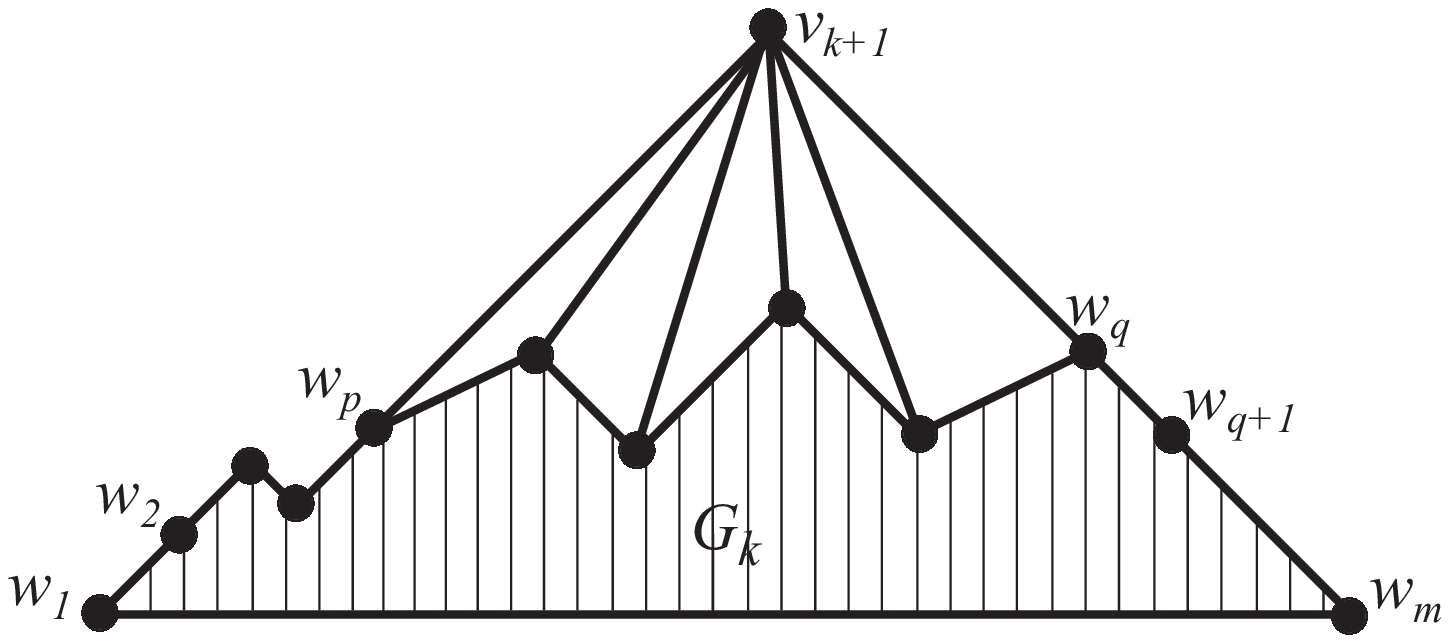,scale=0.35,clip=}} \\
        (a) \hspace{3mm} & (b) \hspace{3mm} & (c)\\
  \end{tabular}
  \caption{(a) A canonical ordering of a maximal plane graph $G$. (b) The drawing of $G_k$ constructed by the algorithm of de Fraysseix~\emph{et~al.} (c) The drawing of $G_{k+1}$ constructed by the algorithm of de Fraysseix~\emph{et~al.}}
  \label{fig:canonical1}
\end{figure}

First, any $n$-vertex maximal plane graph $G$ admits a total ordering $\sigma$ of its vertices, called \emph{canonical ordering}, such that (see Fig.~\ref{fig:canonical1}(a)): (i) the subgraph $G_k$ of $G$ induced by the first $k$ vertices in $\sigma$ is biconnected, for each $k=3,\dots,n$; and (ii) the $k$-th vertex in $\sigma$ lies in the outer face of $G_{k-1}$, for each $k=4,\dots,n$.

Second, a straight-line drawing of an $n$-vertex maximal plane graph $G$ can be constructed starting from a drawing of the $3$-cycle induced by the first three vertices in a canonical ordering $\sigma$ of $G$ and incrementally adding vertices to the partially constructed drawing in the order defined by $\sigma$.
To construct the drawing of $G$ one vertex at a time, the algorithm maintains the invariant that the outer face of $G_k$ is delimited by a polygon composed of a sequence of segments having slopes equal to either $45^{\degree}$ or $-45^{\degree}$. When the next vertex $v_{k+1}$ in $\sigma$ is added to the drawing of $G_k$ to construct a drawing of $G_{k+1}$, a subset of the vertices of $G_k$ undergoes a horizontal shift that allows for $v_{k+1}$ to be introduced in the drawing still maintaining the invariant that the outer face of $G_{k+1}$ is delimited by a polygon composed of a sequence of segments having slopes equal to either $45^{\degree}$ or $-45^{\degree}$ (see Fig.~\ref{fig:canonical1}(b)--(c)).

The area of the constructed drawings is $(2n-4) \times (n-2)$. The described algorithm has been proposed by de Fraysseix~\emph{et~al.} together with an $O(n \log n)$-time implementation. The authors conjectured that its complexity could be improved to $O(n)$. This bound was in fact achieved a few years later by Chrobak and Payne in~\cite{chrobak95lineartime}.

The ideas behind the algorithm by Schnyder~\cite{s-epgg-90} are totally different from the ones of de Fraysseix~\emph{et~al.} In fact, Schnyder's algorithm constructs the drawing by determining the coordinates of all the vertices in one shot. The algorithm relies on results concerning planar graph embeddings that are indeed less intuitive than the canonical ordering of a plane graph used by de Fraysseix~\emph{et~al.}

First, Schnyder introduces the concept of \emph{barycentric representation} of a graph $G$ as an injective function $v\in V(G)\rightarrow (x(v),y(v),z(v))$ such that $x(v)+y(v)+z(v)=1$, for all vertices $v\in V(G)$, and such that, for each edge $(u,v)\in E(G)$ and each vertex $w\notin \{u,v\}$, $x(u)<x(w)$ and $x(v)<x(w)$ hold, or $y(u)<y(w)$ and $y(v)<y(w)$ hold, or $z(u)<z(w)$ and $z(v)<z(w)$ hold. Schnyder proves that, given any graph $G$, given any barycentric representation $v\rightarrow (x(v),y(v),z(v))$ of $G$, and given any three non-collinear points $\alpha$, $\beta$, and $\gamma$ in the three-dimensional space, the mapping $f:v\in V(G)\rightarrow v_1 \alpha +v_2 \beta + v_3 \gamma$ is a straight-line planar embedding of $G$ in the plane spanned by $\alpha$, $\beta$, and $\gamma$.

\begin{figure}[htb]
  \centering
  \begin{tabular}{c c}
	\mbox{\epsfig{figure=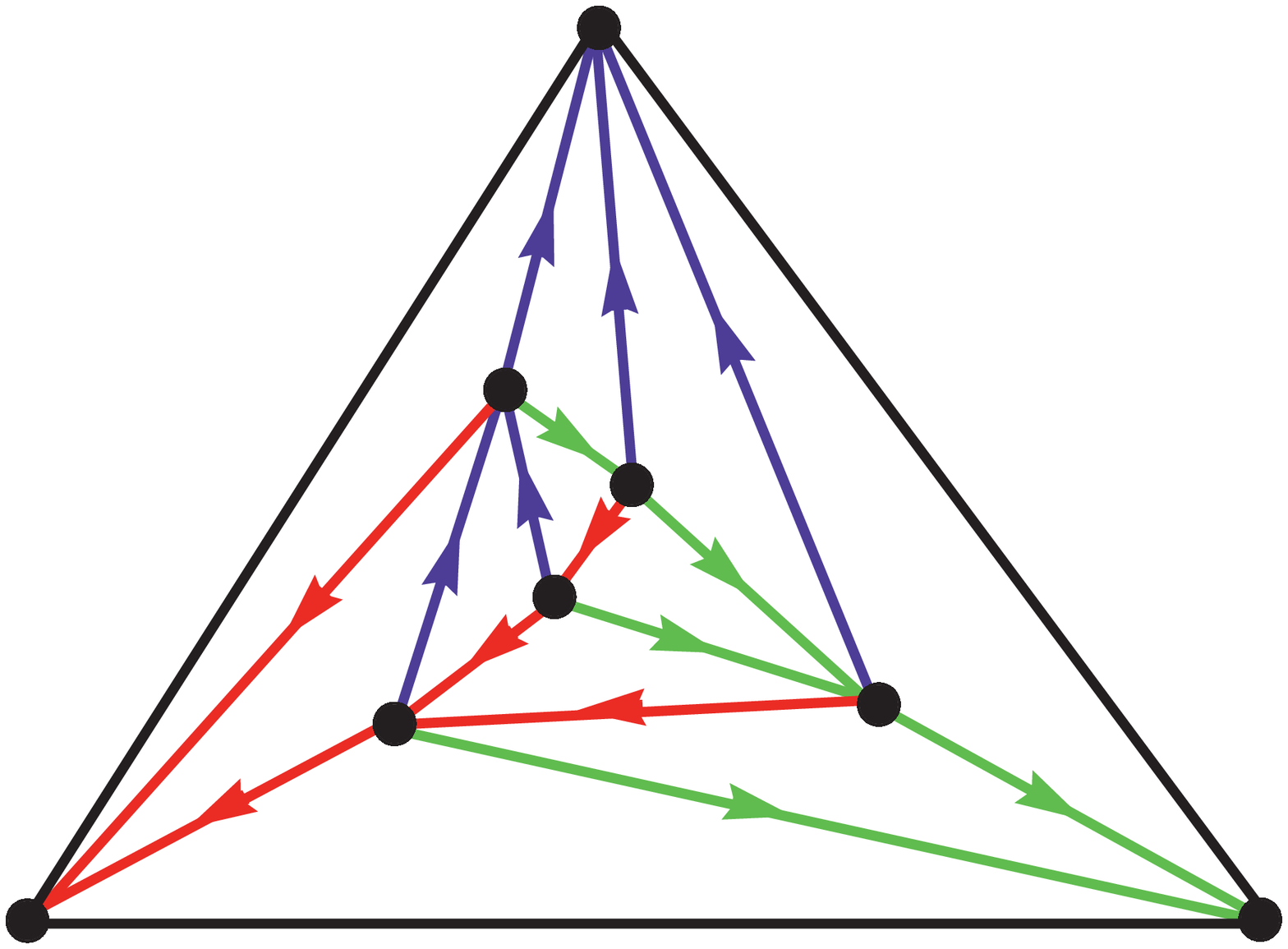,scale=0.25,clip=}} \hspace{5mm} &
	\mbox{\epsfig{figure=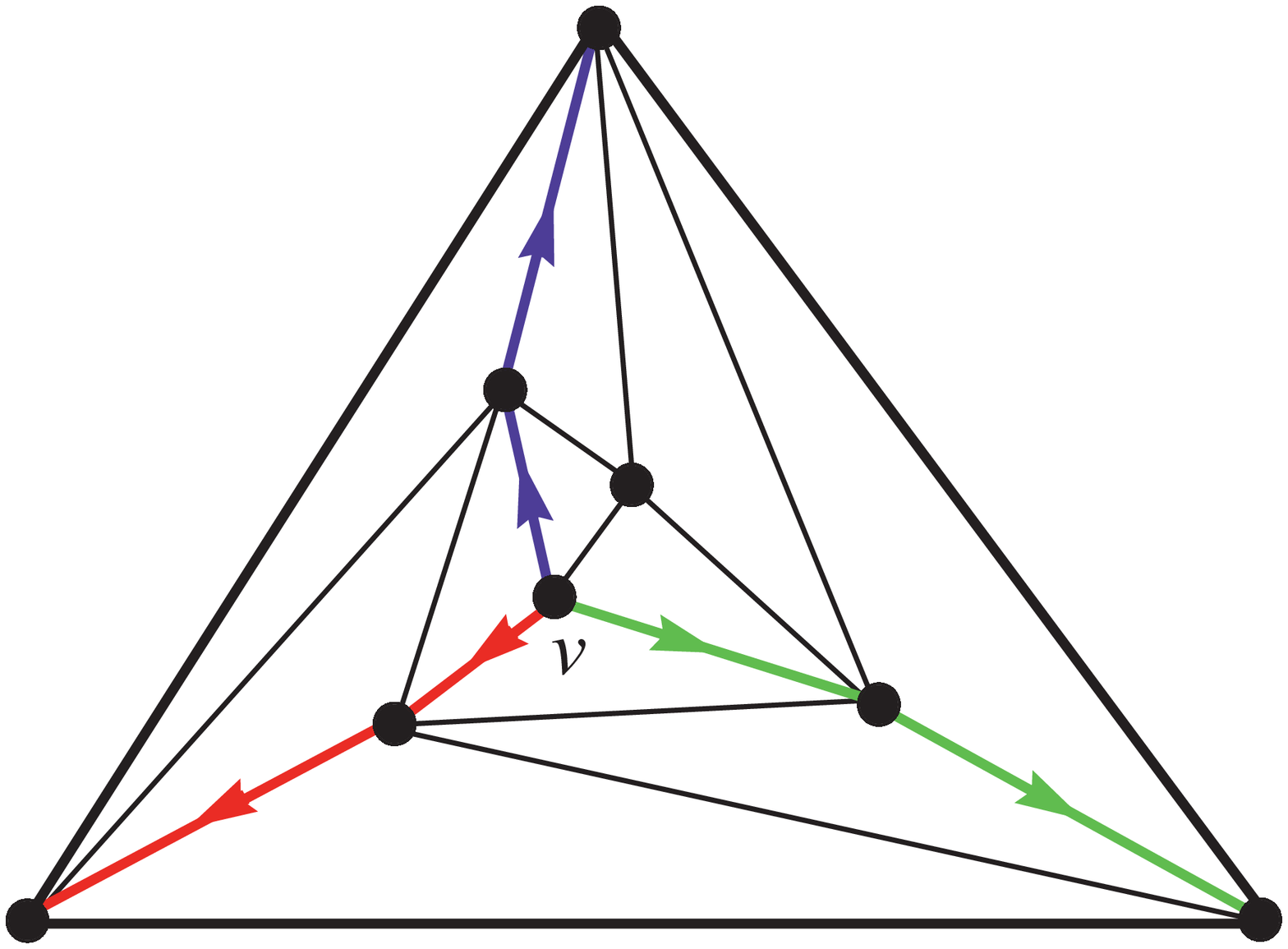,scale=0.25,clip=}} \\
        (a) \hspace{5mm} & (b) \\
  \end{tabular}
  \caption{(a) A realizer for a plane graph $G$. (b) Paths $P_1(v)$, $P_2(v)$, and $P_3(v)$ (represented by green, red, and blue edges, respectively) and regions $R_1(v)$, $R_2(v)$, and $R_3(v)$ (delimited by $P_1(v)$, $P_2(v)$, and $P_3(v)$, and by the edges incident to the outer face of $G$).}
  \label{fig:schnyder}
\end{figure}

Second, Schnyder introduces the concept of a \emph{realizer} of $G$ as an orientation and a partition of the interior edges of a plane graph $G$ into three sets $T_1$, $T_2$, and $T_3$, such that: (i) the set of edges in $T_i$, for each $i=1,2,3$, is a tree spanning all the internal vertices of $G$ and exactly one external vertex; (ii) all the edges of $T_i$ are directed towards this external vertex, which is the root of $T_i$; (iii) the external vertices belonging to $T_1$, to $T_2$, and to $T_3$ are distinct and appear in counter-clockwise order on the border of the outer face of $G$; and (iv) the counter-clockwise order of the edges incident to $v$ is: Leaving $T_1$, entering $T_3$, leaving $T_2$, entering $T_1$, leaving $T_3$, and entering~$T_2$. Fig.~\ref{fig:schnyder}(a) illustrates a realizer for a plane graph~$G$. Trees $T_1$, $T_2$, and $T_3$ are sometimes called \emph{Schnyder woods}.

Third, Schnyder describes how to get a barycentric representation of a plane graph $G$ starting from a realizer of $G$; this is essentially done by looking, for each vertex $v\in V(G)$ at the paths $P_i(v)$, that are the only paths composed entirely of edges of $T_i$ connecting $v$ to the root of $T_i$ (see Fig.~\ref{fig:schnyder}(b)), and counting the number of the faces or the number of the vertices in the regions $R_1(v)$, $R_2(v)$, and $R_3(v)$ that are defined by $P_1(v)$, $P_2(v)$, and $P_3(v)$. The area of the constructed drawings is $(n-2)\times (n-2)$.

Schnyder's upper bound has been unbeaten for almost twenty years. Only recently Brandenburg~\cite{b-dpg89a-08} proposed an algorithm for constructing planar straight-line drawings of triangulations in $\frac{8n^2}{9} + O(n)$ area. Such an algorithm is based on a geometric refinement of the de Fraysseix \emph{et al.}~\cite{fpp-sssfepg-88,fpp-hdpgg-90} algorithm combined with some topological properties of planar triangulations due to Bonichon et al.~\cite{bsm-wtr-02}, that will be discussed in Sect.~\ref{se:poly-line}.

\begin{figure}[htb]
  \centering
  \begin{tabular}{c c c}
	\mbox{\epsfig{figure=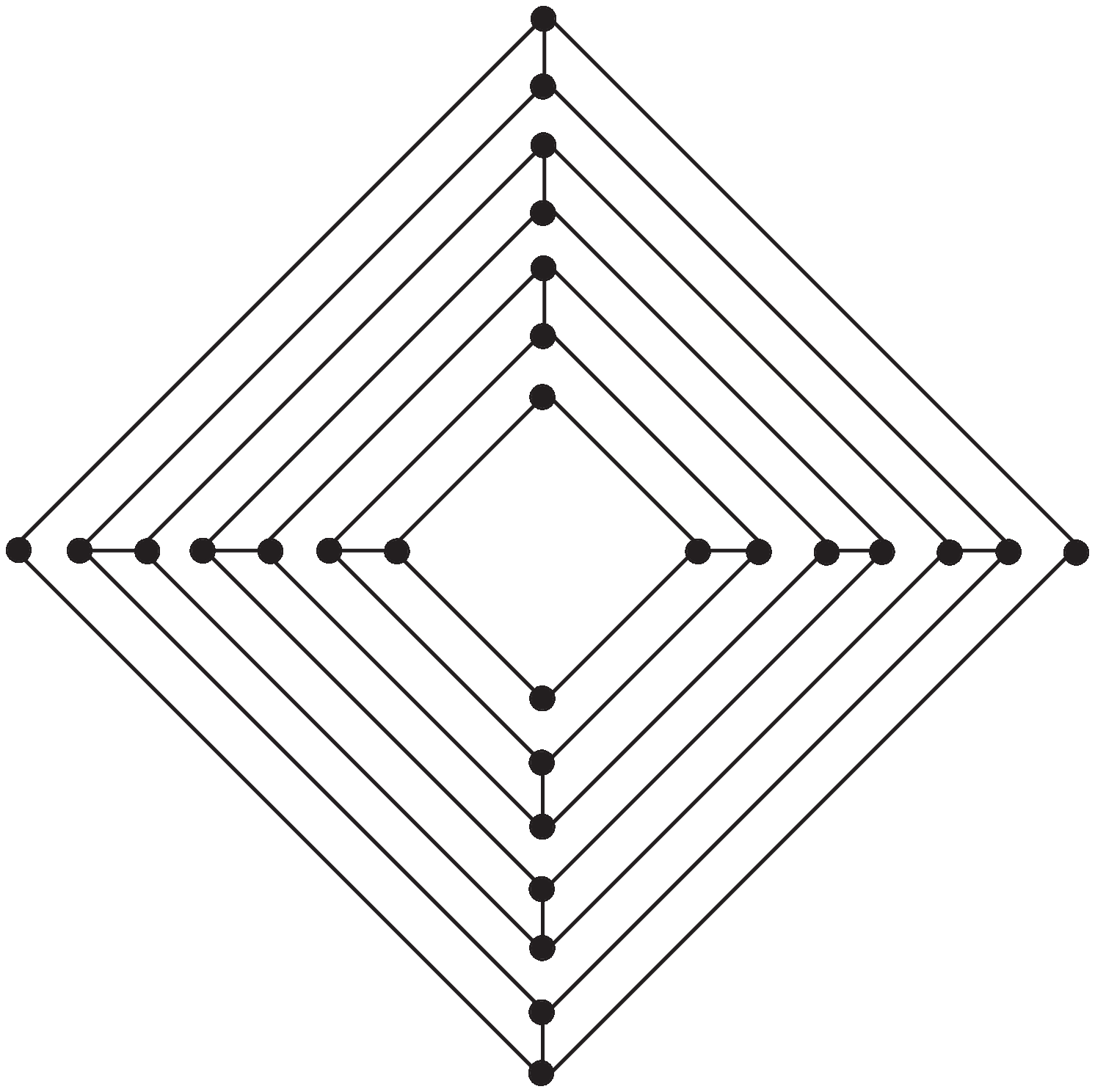,scale=0.3,clip=}} \hspace{5mm} &
	\mbox{\epsfig{figure=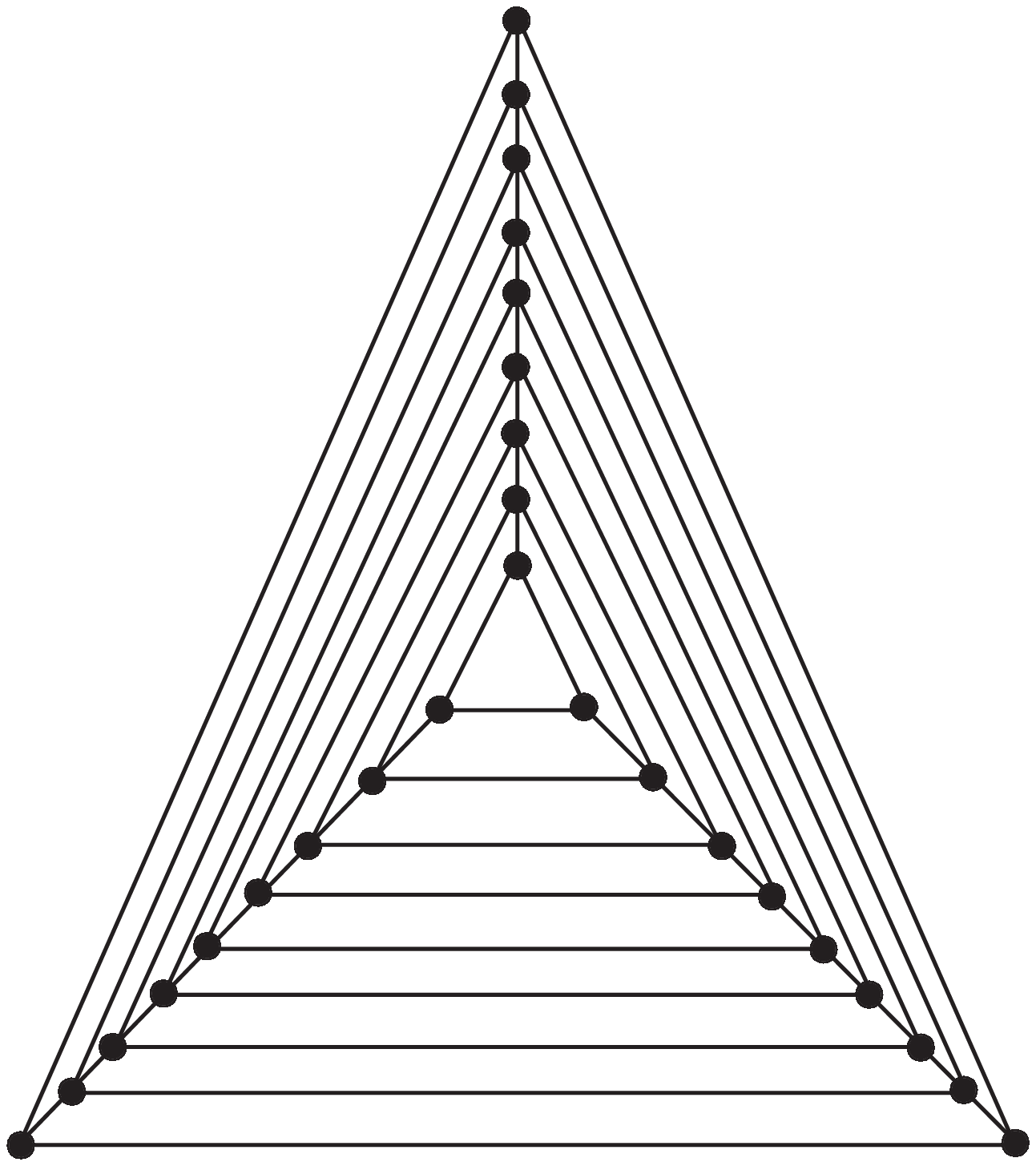,scale=0.3,clip=}} \hspace{5mm} &
    \mbox{\epsfig{figure=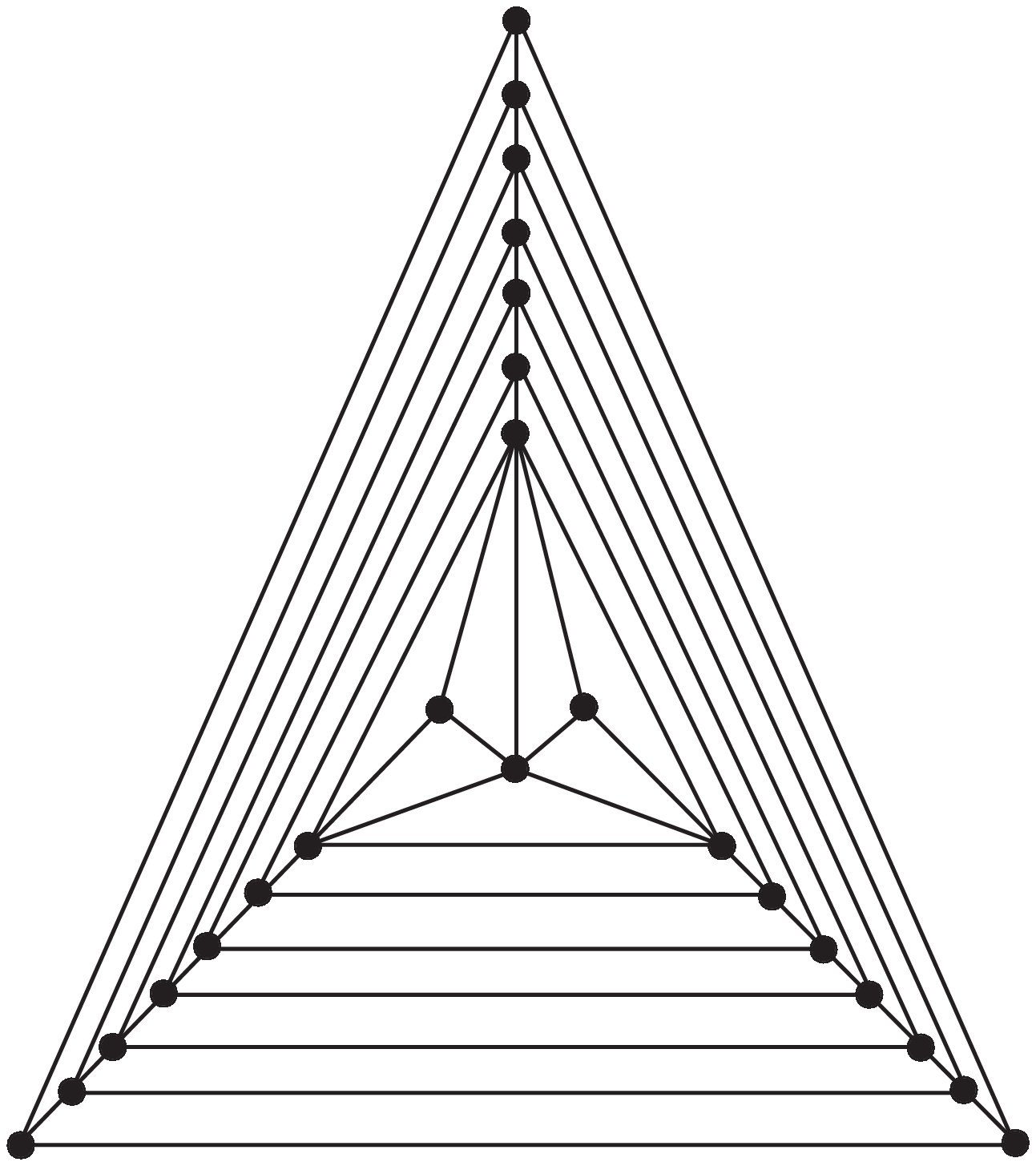,scale=0.3,clip=}}\\
        (a) \hspace{5mm} & (b) \hspace{5mm} & (c) \\
  \end{tabular}
  \caption{(a) A graph~\cite{Val81} requiring quadratic area in any straight-line and poly-line drawing. (b) A graph~\cite{fpp-hdpgg-90} requiring $\left(\frac{2n}{3}-1\right) \times\left(\frac{2n}{3}-1\right)$ area in any straight-line and poly-line drawing. (c) A graph~\cite{FratiP07} requiring $\frac{4n^2}{9}-\frac{2n}{3}$ area in any straight-line drawing.}
  \label{fig:lowerboundnested}
\end{figure}

A quadratic area upper bound for straight-line planar drawings of plane graphs is asymptotically optimal. In fact, almost ten years before the publication of such algorithms, Valiant observed in~\cite{Val81} that there exist $n$-vertex plane graphs (see  Fig.~\ref{fig:lowerboundnested}(a)) requiring $\Omega(n^2)$ area in any straight-line planar drawing (in fact, in every poly-line planar drawing). It was then proved by de Fraysseix \emph{et al.}~in~\cite{fpp-hdpgg-90} that \emph{nested triangles graphs} (see Fig.~\ref{fig:lowerboundnested}(b)) require $\left(\frac{2n}{3}-1\right) \times\left(\frac{2n}{3}-1\right)$ area in any straight-line planar drawing (in fact, in every poly-line planar drawing). Such a lower bound was only recently improved to $\frac{4n^2}{9}-\frac{2n}{3}$ by Frati and Patrignani~\cite{FratiP07}, for all $n$ multiple of $3$ (see Fig.~\ref{fig:lowerboundnested}(c)), and then by Mondal \emph{et al.}~\cite{mnra-madp3t-10} to $\left \lfloor \frac{2n}{3}-1 \right\rfloor  \times \left \lfloor \frac{2n}{3} \right\rfloor $, for all $n\geq 6$.

However, the following remains open:

\begin{problem}
Close the gap between the $\frac{8n^2}{9} + O(n)$ upper bound and the $\frac{4n^2}{9} - O(n)$ lower bound for the area requirements of straight-line drawings of plane graphs.
\end{problem}

\subsection{$4$-Connected and Bipartite Planar Graphs} \label{se:straight-bipartite}

In this section, we discuss algorithms and bounds for constructing planar straight-line drawings of $4$-connected and bipartite planar graphs. Such different families of graphs are discussed in the same section since the best known upper bound for the area requirements of bipartite planar graphs uses a preliminary augmentation to $4$-connected planar graphs.

\begin{figure}[htb]
  \centering
  \begin{tabular}{c c}
	\mbox{\epsfig{figure=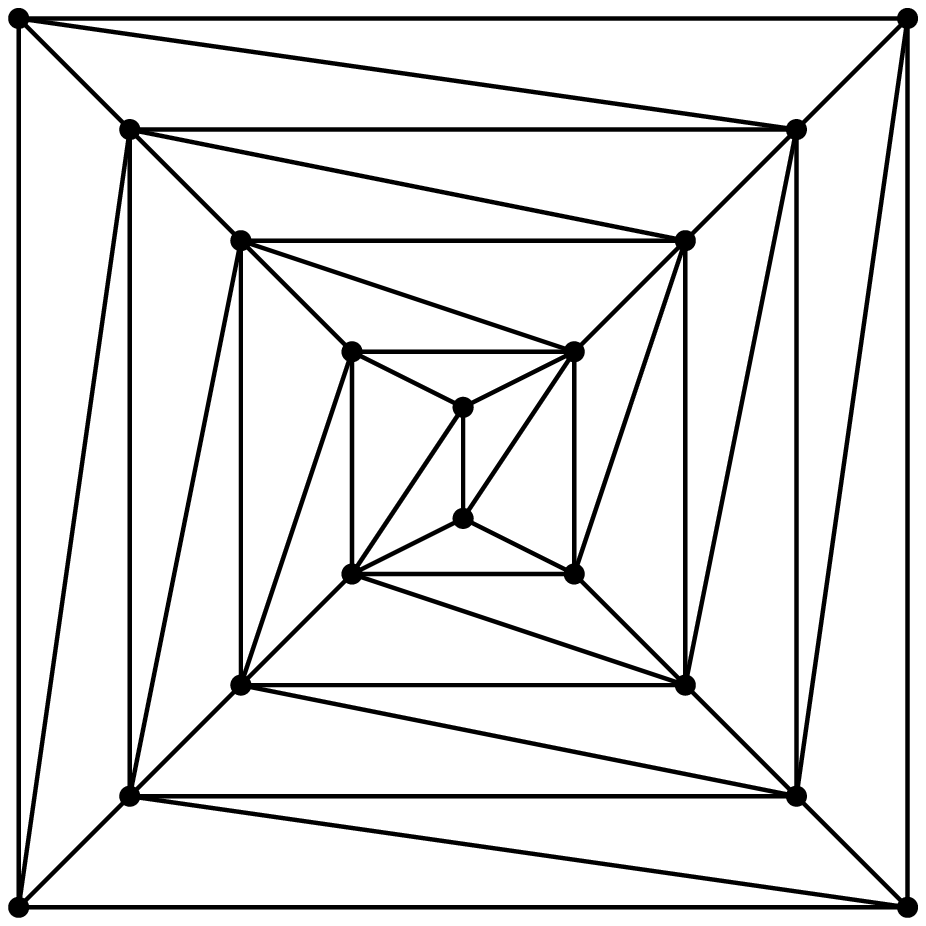,scale=0.4,clip=}} \hspace{5mm} &
	\mbox{\epsfig{figure=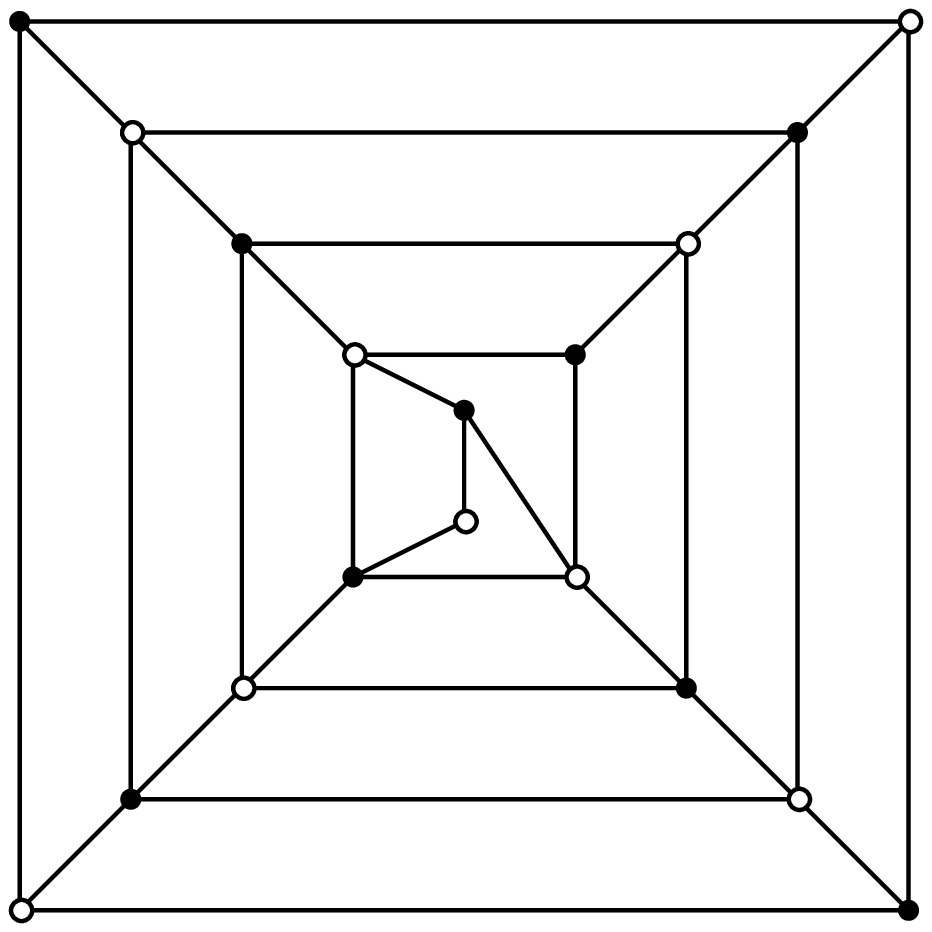,scale=0.4,clip=}} \\
        (a) \hspace{5mm} & (b) \\
  \end{tabular}
  \caption{(a) A $4$-connected plane graph requiring $(\lceil{\frac{n}{2}}\rceil-1) \times (\lfloor{\frac{n}{2}}\rfloor)$ area in any straight-line planar drawing. (b) A bipartite plane graph requiring $(\lceil{\frac{n}{2}}\rceil-1) \times (\lfloor{\frac{n}{2}}\rfloor)$ area in any straight-line planar drawing.}
  \label{fig:straight-line-fourconnected}
\end{figure}

Concerning $4$-connected plane graphs, tight bounds are known for the area requirements of planar straight-line drawings if the graph has at least four vertices incident to the outer face. Namely,  Miura \emph{et al.}~proved in~\cite{mnn-gd4pg-01} that every such a graph has a planar straight-line drawing in $(\lceil{\frac{n}{2}}\rceil-1) \times (\lfloor{\frac{n}{2}}\rfloor)$ area, improving upon previous results of He~\cite{h-gefcpg-97}. The authors show that this bound is tight, by exhibiting a class of $4$-connected plane graphs with four vertices incident to the outer face requiring $(\lceil{\frac{n}{2}}\rceil-1) \times (\lfloor{\frac{n}{2}}\rfloor)$ area (see Fig.~\ref{fig:straight-line-fourconnected}(a)).

\begin{figure}[htb]
  \centering{
	\mbox{\epsfig{figure=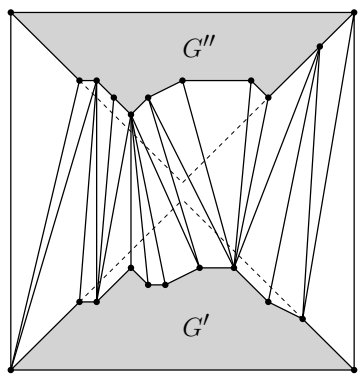,scale=0.4,clip=}}}
  \caption{The algorithm by Miura \emph{et al.}~to construct straight-line drawings of $4$-connected plane graphs~\cite{mnn-gd4pg-01}.}
  \label{fig:Miura}
\end{figure}

The algorithm of Miura \emph{et al.}~divides the input $4$-connected plane graph $G$ into two graphs $G'$ and $G''$ with the same number of vertices. This is done by performing a \emph{4-canonical ordering} of $G$ (see~\cite{kh-rel4cpgiagdp-97}). The graph $G'$ ($G''$, respectively) is then drawn inside an isosceles right triangle $\Delta'$ (resp.\ $\Delta''$) whose width is $\frac{n}{2}-1$ and whose height is half of its width. To construct such drawings of $G'$ and $G''$, Miura \emph{et al.}~design an algorithm that is similar to the algorithm by de Fraysseix \emph{et al.}~\cite{fpp-hdpgg-90}. In the drawings produced by their algorithm the slopes of the edges incident to the outer faces of $G'$ and $G''$ have absolute value which is at most $45\degree$. The drawing of $G''$ is then rotated by $180\degree$ and placed on top of the drawing of $G'$. This allows for drawing the edges connecting $G'$ with $G''$ without creating crossings. Fig.~\ref{fig:Miura} depicts the construction of the Miura \emph{et al.}'s algorithm.

As far as we know, no bound better than the one for general plane graphs is known for $4$-connected plane graphs (possibly having three vertices incident to the outer face), hence the following is open:

\begin{problem}
Close the gap between the $\frac{8n^2}{9} + O(n)$ upper bound and the $\frac{n^2}{4} - O(n)$ lower bound for the area requirements of straight-line drawings of $4$-connected plane graphs.
\end{problem}

Biedl and Brandenburg~\cite{bb-dpbgsa-05} show how to construct planar straight-line drawings of bipartite planar graphs in $(\lceil{\frac{n}{2}}\rceil-1) \times (\lfloor{\frac{n}{2}}\rfloor)$ area. To achieve such a bound, they exploit a result of Biedl \emph{et al.}~\cite{bkk-tpgfcc-98} stating that all planar graphs without separating triangles, except those ``containing a star''  (see~\cite{bb-dpbgsa-05} and observe that in this case a star is not just a vertex plus some incident edges), can be augmented to $4$-connected by the insertion of dummy edges; once such an augmentation is done, Biedl and Brandenburg use the algorithm of Miura \emph{et al.}~\cite{mnn-gd4pg-01} to draw the resulting $4$-connected plane graph. In order to be able to use Miura \emph{et al.}'s algorithm, Biedl and Brandenburg prove that no bipartite plane graph ``contains a star'' and that Miura \emph{et al.}'s algorithm works more in general for plane graphs that become $4$-connected if an edge is added to them. The upper bound of Biedl and Brandenburg is tight as the authors show a bipartite plane graph requiring $(\lceil{\frac{n}{2}}\rceil-1) \times (\lfloor{\frac{n}{2}}\rfloor)$ area in any straight-line planar drawing (see  Fig.~\ref{fig:straight-line-fourconnected}(b)).

\subsection{Series-Parallel Graphs} \label{se:straight-series-parallel}

In this section, we discuss algorithms and bounds for constructing small-area planar straight-line drawings of series-parallel graphs.

No sub-quadratic area upper bound is known for constructing small-area planar straight-line drawings of series-parallel graphs. The best known quadratic upper bound for straight-line drawings is provided in~\cite{zhn-sgdpgbb-10}.

In~\cite{f-lbarspg-j10} Frati proved that there exist series-parallel graphs requiring $\Omega(n 2^{\sqrt{\log n}})$ area in any straight-line or poly-line grid drawing. Such a result is  achieved in two steps. In the first one, an $\Omega(n)$ lower bound for the maximum between the height and the width of any straight-line or poly-line grid drawing of $K_{2,n}$ is proved, thus answering a question of Felsner \emph{et~al.}~\cite{journals/jgaa/FelsnerLW03} and improving upon previous results of Biedl \emph{et~al.}~\cite{journals/ipl/BiedlCL03}. In the second one, an $\Omega(2^{\sqrt{\log n}})$ lower bound for the minimum between the height and the width of any straight-line or poly-line grid drawing of certain series-parallel graphs is proved.

\begin{figure}[htb]
\centering{
	\mbox{\epsfig{figure=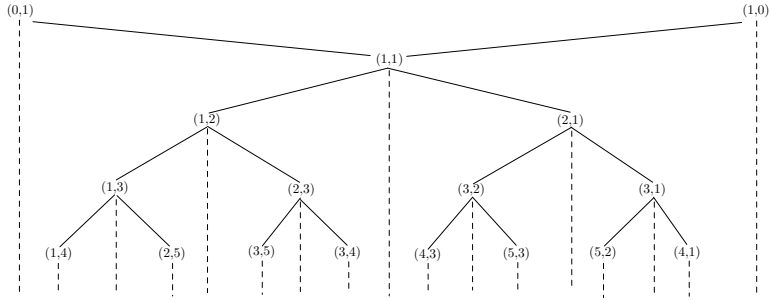,scale =0.5,clip=}}}
	\caption{The Stern-Brocot tree is a tree containing all the pairs of relatively prime numbers.}
  \label{fig:sternbrocot}
\end{figure}

The proof that $K_{2,n}$ requires $\Omega(n)$ height or width in any straight-line or poly-line drawing has several ingredients. First, a simple ``optimal'' drawing algorithm for $K_{2,n}$ is exhibited, that is, an algorithm is presented that computes a drawing of $K_{2,n}$ inside an arbitrary convex polygon if such a drawing exists. Second, the drawings constructed by the mentioned algorithm inside a rectangle are studied. Such a study reveals that the slopes of the segments representing the edges of $K_{2,n}$ have a strong relationship with the relatively prime numbers as ordered in the \emph{Stern-Brocot} tree (see~\cite{s-uzf-58,b-cranm-60} and Fig.~\ref{fig:sternbrocot}). Such a relationship leads to derive some arithmetical properties of the lines passing through infinite grid points in the plane and to achieve the $\Omega(n)$ lower bound.

\begin{figure}[htb]
\centering{
\begin{tabular} {c c c}
    \mbox{\epsfig{figure=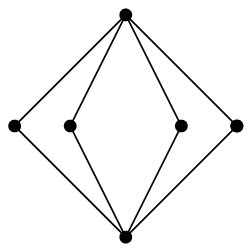,scale =0.5,clip=}} \hspace{3mm} &
    \mbox{\epsfig{figure=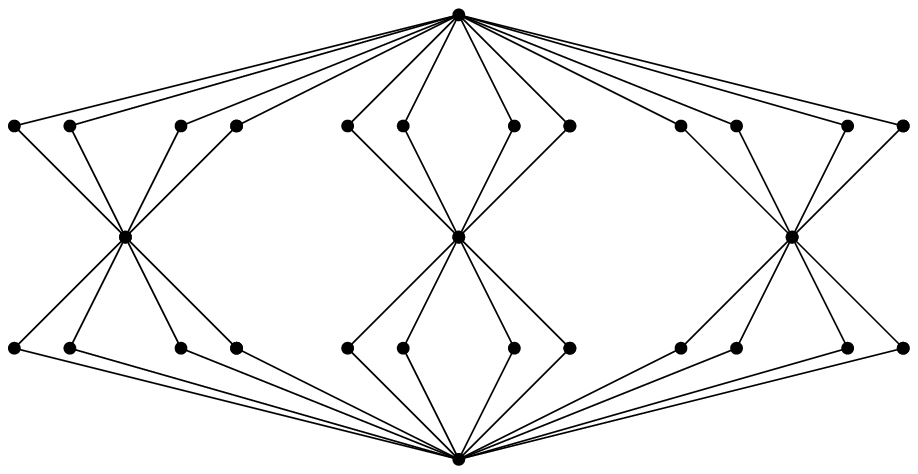,scale =0.5,clip=}} \hspace{3mm} &
    \mbox{\epsfig{figure=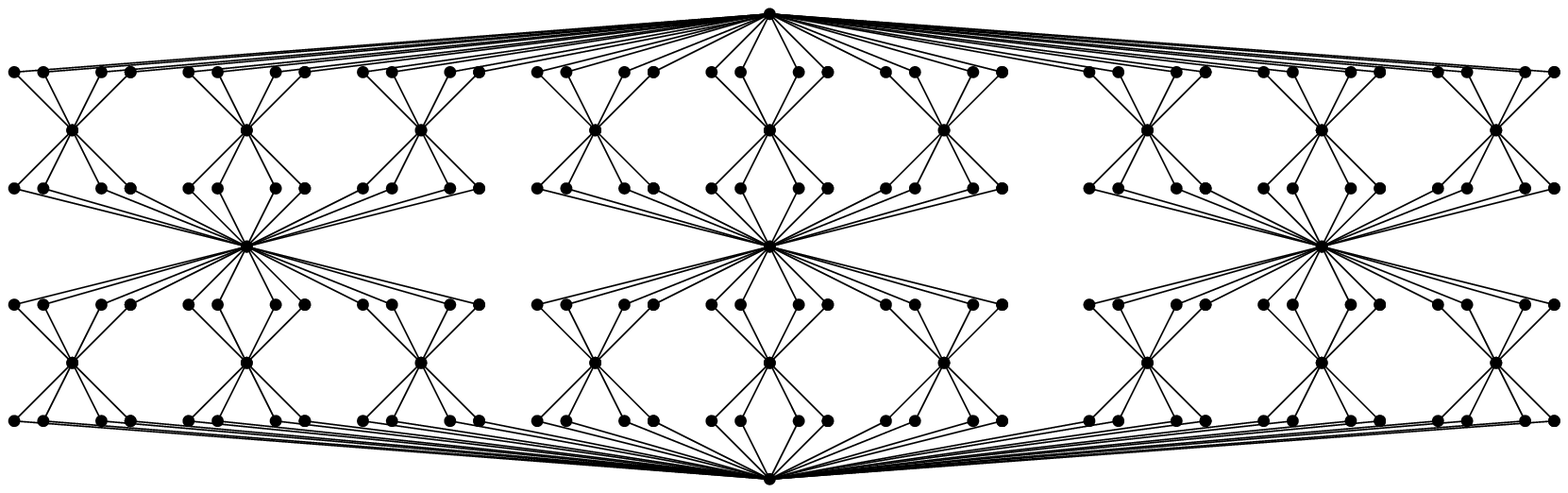,scale =0.5,clip=}} \\
        (a) \hspace{3mm} & (b) \hspace{3mm} & (c)
\end{tabular}}
	\caption{The inductive construction of series-parallel graphs requiring $\Omega(2^{\sqrt{\log n}})$ height and width in any straight-line or poly-line grid drawing.}
  \label{fig:straight-line-sp-improvedgraphs}
\end{figure}

The results on the area requirements of $K_{2,n}$ are then used to construct series-parallel graphs (shown in Fig.~\ref{fig:straight-line-sp-improvedgraphs}) out of several copies of $K_{2,2^{\sqrt{\log n}}}$ and to prove that such a graph requires $\Omega(2^{\sqrt{\log n}})$ height and width in any straight-line or poly-line grid drawing.

As no sub-quadratic area upper bound is known for straight-line planar drawings of series-parallel graphs the following is open.

\begin{problem}
Close the gap between the $O(n^2)$ upper bound and the $\Omega(n 2^{\sqrt{\log n}})$ lower bound for the area requirements of straight-line drawings of series-parallel graphs.
\end{problem}

Related to the above problem, Wood~\cite{w-08} conjectures the following: Let $p_1,\dots,p_k$ be positive integers. Let $G(p_1)$ be the graph obtained from $K_3$ by adding $p_1$ new vertices adjacent to $v$ and $w$ for each edge $(v,w)$ of $K_3$. For $k \geq 2$, let $G(p_1,p_2,\dots,p_k)$ be the graph obtained from $G(p_1,p_2,\dots,p_{k-1})$ by adding $p_k$ new vertices adjacent to $v$ and $w$ for each edge $(v,w)$ of $G(p_1,p_2,\dots,p_{k-1})$. Observe that $G(p_1,p_2,\dots,p_k)$ is a series-parallel graph.

\begin{conjecture} (D. R. Wood) Every straight-line grid drawing of $G(p_1,p_2,\dots,p_k)$ requires $\Omega(n^2)$ area for some choice of $k$ and $p_1,p_2,\dots,p_k$.
\end{conjecture}

\subsection{Outerplanar Graphs} \label{se:straight-outerplanar}

In this section, we discuss algorithms and bounds for constructing small-area planar straight-line drawings of outerplanar graphs.

The first non-trivial bound appeared in~\cite{gr-aepsdog-07}, where Garg and Rusu proved that every outerplanar graph with maximum degree $d$ has a straight-line drawing with $O(dn^{1.48})$ area. Such a result is achieved by means of an algorithm that works by induction on the dual tree $T$ of the outerplanar graph $G$. Namely, the algorithm finds a path $P$ in $T$, it removes from $G$ the subgraph $G_P$ that has $P$ as a dual tree, it inductively draws the outerplanar graphs that are disconnected by such a removal, and it puts all the drawings of such outerplanar graphs together with a drawing of $G_P$, obtaining a drawing of the whole outerplanar graph.

\begin{figure}[htb]
\centering{
	\mbox{\epsfig{figure=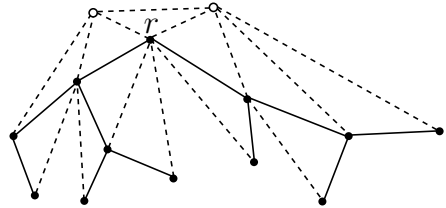,scale =0.5,clip=}}}
	\caption{A star-shaped drawing $\Gamma$ of a binary tree $T$ (with thick edges and black vertices). The dashed edges and white vertices augment $\Gamma$ into a straight-line drawing of the outerplanar graph $T$ is dual to.}
  \label{fig:straight-starshaped}
\end{figure}

\begin{figure}[htb]
\centering{
\begin{tabular} {c c}
    \mbox{\epsfig{figure=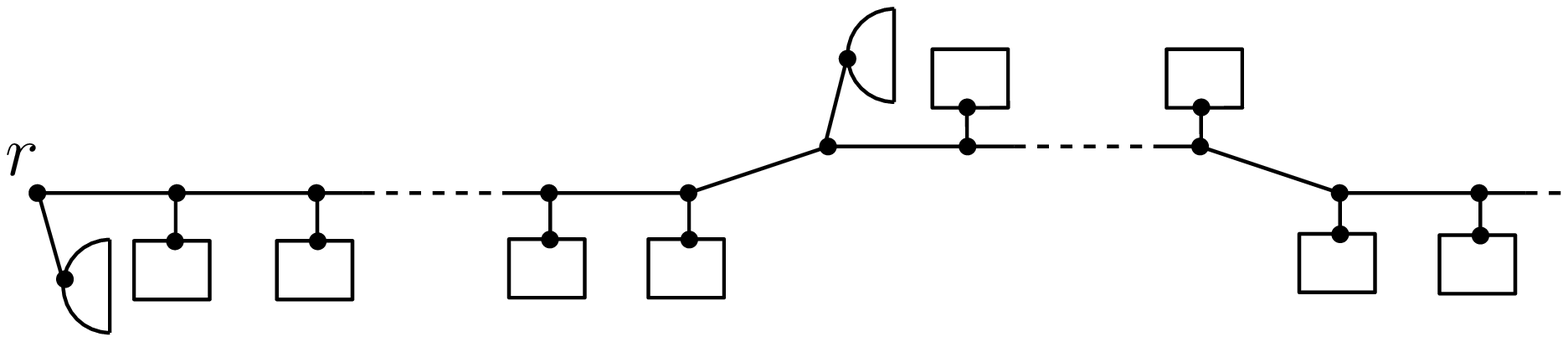,scale =0.4,clip=}} \hspace{4mm} &
    \mbox{\epsfig{figure=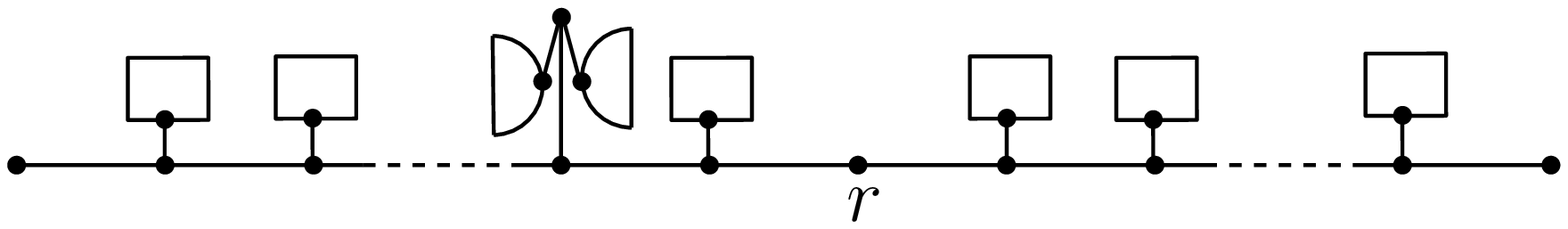,scale =0.4,clip=}}
\end{tabular}}
	\caption{Two inductive cases of the algorithm to construct star-shaped drawing of binary trees yielding an $O(n^{1.48})$ upper bound for straight-line drawings of outerplanar graphs. The rectangles and the half-circles represent subtrees recursively drawn by the construction on the right and on the left part of the figure, respectively.}
  \label{fig:straight-line-outerplanar}
\end{figure}

The first sub-quadratic area upper bound for straight-line drawings of outerplanar graphs has been proved by Di Battista and Frati in~\cite{df-sadog-j09}. The result in~\cite{df-sadog-j09} uses the following ingredients. First, it is shown that the dual binary tree $T$ of a maximal outerplanar graph $G$ is a subgraph of $G$ itself. Second, a restricted class of straight-line drawings of binary trees, called \emph{star-shaped drawings}, is defined. Star-shaped drawings are straight-line drawings in which special visibility properties among the nodes of the tree are satisfied (see Fig.~\ref{fig:straight-starshaped}).
Namely, if a tree $T$ admits a star-shaped drawing $\Gamma$, then the edges that augment $T$ into $G$ can be drawn in $\Gamma$ without creating crossings, thus resulting in a straight-line planar drawing of $G$. Third, an algorithm is shown to construct a star-shaped drawing of any binary tree $T$ in $O(n^{1.48})$ area. Such an algorithm works by induction on the number of nodes of $T$ (Fig.~\ref{fig:straight-line-outerplanar} depicts two inductive cases of such a construction), making use of a strong combinatorial decomposition of ordered binary trees introduced by Chan in~\cite{c-nlabdbt-02} (discussed in Sect.~\ref{se:straight-trees}).

\begin{figure}[htb]
\centering{
	\mbox{\epsfig{figure=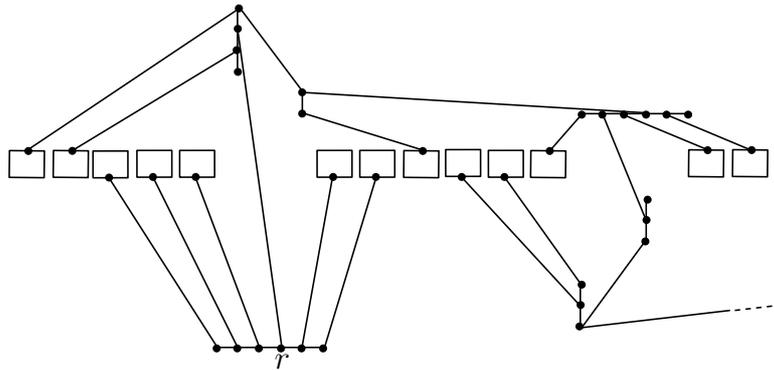,scale =0.5,clip=}}}
	\caption{The inductive construction of star-shaped drawings of binary trees yielding an $O(dn\log n)$ upper bound for straight-line drawings of outerplanar graphs with degree $d$. The rectangles represent recursively constructed star-shaped drawings of subtrees.}
  \label{fig:straight-outerplanar-degreed}
\end{figure}

Frati used in~\cite{f-sdogdnlogna-07} the same approach of~\cite{df-sadog-j09}, together with a different geometric construction (shown in Fig.~\ref{fig:straight-outerplanar-degreed}), to prove that every outerplanar graph with degree $d$ has a straight-line drawing with $O(dn\log n)$ area.

As far as we know, no super-linear area lower bound is known for straight-line drawings of outerplanar graphs. In~\cite{b-dopgia-02} Biedl defined a class of outerplanar graphs, called \emph{snowflake graphs}, and conjectured that such graphs require $\Omega(n \log n)$ area in any straight-line or poly-line drawing. However, Frati disproved such a conjecture in~\cite{f-sdogdnlogna-07} by exhibiting $O(n)$ area straight-line drawings of snowflake graphs. In the same paper, he conjectured that an $O(n\log n)$ area upper bound for straight-line drawings of outerplanar graphs can not be achieved by squeezing the drawing along one coordinate direction, as stated in the following.

\begin{conjecture} (F. Frati) There exist $n$-vertex outerplanar graphs for which, for any straight-line drawing in which the longest side of the bounding-box is $O(n)$, the smallest side of the bounding-box is $\omega(\log n)$.
\end{conjecture}

The following problem remains wide open.

\begin{problem}
Close the gap between the $O(n^{1.48})$ upper bound and the $\Omega(n)$ lower bound for the area requirements of straight-line drawings of outerplanar graphs.
\end{problem}

\subsection{Trees} \label{se:straight-trees}

In this section, we present algorithms and bounds for constructing planar straight-line drawings of trees.

\begin{figure}[htb]
\centering{
	\mbox{\epsfig{figure=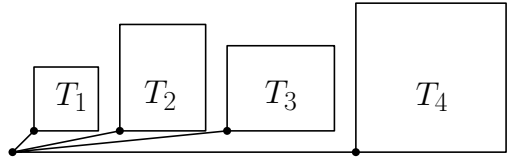,scale =0.5,clip=}}}
	\caption{Inductive construction of a straight-line drawing of a tree in $O(n\log n)$ area.}
  \label{fi:trees-hvdrawing}
\end{figure}

The best bound for constructing general trees is, as far as we know, the $O(n \log n)$ area upper bound provided by a simple modification of the $hv$-drawing algorithm of Crescenzi, Di Battista, and Piperno~\cite{cdp-noad-92}. Such an algorithm proves that a straight-line drawing of any tree $T$ in $O(n) \times O(\log n)$ area can be constructed with the further constraint that the root of $T$ is placed at the bottom-left corner of the bounding box of the drawing. If $T$ has one node, such a drawing is trivially constructed. If $T$ has more than one node, then let $T_1, \dots, T_k$ be the subtrees of $T$, where we assume, w.l.o.g., that $T_k$ is the subtree of $T$ with the greatest number of nodes. Then, the root of $T$ is placed at $(0,0)$, the subtrees $T_1, \dots, T_{k-1}$ are placed one besides the other, with the bottom side of their bounding boxes on the line $y=1$, and $T_{k}$ is placed besides the other subtrees, with the bottom side of its bounding box on the line $y=0$. The width of the drawing is clearly $O(n)$, while its height is $h(n)=\max\{h(n-1),1+h(n/2)\}=O(\log n)$, where $h(n)$ denotes the maximum height of a drawing of an $n$-node tree constructed by the algorithm. See Fig.~\ref{fi:trees-hvdrawing} for an illustration of such an algorithm. Interestingly, no super-linear area lower bound is known for the area requirements of straight-line drawings of trees.

For the special case of bounded-degree trees linear area bounds have been achieved. In fact, Garg and Rusu presented an algorithm to construct straight-line drawings of binary trees in $O(n)$ area~\cite{gr-sdbtlaaar-04} and an algorithm to construct straight-line drawings of trees with degree $O(\sqrt n)$ in $O(n)$ area~\cite{iccsa/GargR03}. Both algorithms rely on the existence of simple \emph{separators} for bounded degree trees. Namely, every binary tree $T$ has a \emph{separator edge}, that is an edge whose removal disconnects $T$ into two trees both having at most $2n/3$ vertices~\cite{Val81} and every degree-$d$ tree $T$ has a vertex whose removal disconnects $T$ into at most $d$ trees, each having at most $n/2$ nodes~\cite{iccsa/GargR03}. Such separators are exploited by Garg and Rusu to design quite complex inductive algorithms that achieve linear area bounds and optimal aspect ratio.

The following problem remains open:

\begin{problem}
Close the gap between the $O(n \log n)$ upper bound and the $\Omega(n)$ lower bound for the area requirements of straight-line drawings of trees.
\end{problem}

A lot of attention has been devoted to studying the area requirements of straight-line drawings of trees satisfying additional constraints. Table~\ref{ta:trees-straight-line} summarizes the best known area bounds for various kinds of straight-line drawings of trees.

\begin{table}[!htb]\centering\footnotesize
  \linespread{1.2}
  \selectfont
  \begin{tabular}{|c|c|c|c|c|c|c|c|c|}
   \cline{2-9}
	 \multicolumn{1}{c|}{} & \emph{Ord. Pres.} & \emph{Upw.} & \emph{Str. Upw.} & \emph{Orth.} & \emph{Upper Bound} & \emph{Refs.} & \emph{Lower Bound} & \emph{Refs.} \\ 
    \cline{2-9}
    \hline
    \emph{Binary} & & & & & $O(n)$ & \cite{gr-sdbtlaaar-04} & $\Omega(n)$ & \emph{trivial} \\
    \hline
    \emph{Binary} & \checkmark & & & & $O(n\log \log n)$ & \cite{gr-aeoppsdot-03} & $\Omega(n)$ & \emph{trivial} \\
    \hline
    \emph{Binary} & & \checkmark & & & $O(n\log \log n)$ & \cite{skc-aeastd-00} & $\Omega(n)$ & \emph{trivial} \\
    \hline
    \emph{Binary} & & & \checkmark & & $O(n\log n)$ & \cite{cdp-noad-92} & $\Omega(n \log n)$ & \cite{cdp-noad-92} \\
    \hline
    \emph{Binary} & \checkmark & & \checkmark &  & $O(n \log n)$ & \cite{gr-aeoppsdot-03} & $\Omega(n\log n)$ & \cite{cdp-noad-92}\\
    \hline
    \emph{Binary} & & & & \checkmark & $O(n \log \log n)$ & \cite{cgkt-oaarsod-02,skc-aeastd-00} & $\Omega(n)$ & \emph{trivial}\\
    \hline
    \emph{Binary} & & \checkmark & & \checkmark & $O(n \log n)$ & \cite{cdp-noad-92,cgkt-oaarsod-02} & $\Omega(n\log n)$ & \cite{cgkt-oaarsod-02}\\
    \hline
    \emph{Binary} & \checkmark & & & \checkmark & $O(n^{1.5})$ & \cite{f-sodbtt-06} & $\Omega(n)$ & \emph{trivial}\\
    \hline
    \emph{Ternary} & & & & \checkmark & $O(n^{1.631})$ & \cite{f-sodbtt-06} & $\Omega(n)$ & \emph{trivial}\\
    \hline
    \emph{Ternary} & \checkmark & & & \checkmark & $O(n^2)$ & \cite{f-sodbtt-06} & $\Omega(n^2)$ & \cite{f-sodbtt-06}\\
    \hline
    \emph{General} & & & & & $O(n \log n)$ & \cite{cdp-noad-92} & $\Omega(n)$ & \emph{trivial} \\
    \hline
    \emph{General} & \checkmark & & & & $O(n\log n)$ & \cite{gr-aeoppsdot-03} & $\Omega(n)$ & \emph{trivial} \\
    \hline
    \emph{General} & & \checkmark & & & $O(n\log n)$ & \cite{cdp-noad-92} & $\Omega(n)$ & \emph{trivial} \\
    \hline
    \emph{General} & & & \checkmark & & $O(n\log n)$ & \cite{cdp-noad-92} & $\Omega(n \log n)$ & \cite{cdp-noad-92} \\
    \hline
    \emph{General} & \checkmark & & \checkmark & & $O(n 4^{\sqrt{2\log n}})$ & \cite{c-nlabdbt-02} & $\Omega(n\log n)$ & \cite{cdp-noad-92}\\
    \hline
 \end{tabular}
 \vspace{2mm}
  \caption{\small Summary of the best known area bounds for straight-line drawings of trees. ``Ord. Pres.'', ``Upw.'', ``Str. Upw.'', and ``Orth.'' stand for order-preserving, upward, strictly-upward, and orthogonal, respectively.}
  \label{ta:trees-straight-line}
\end{table}

Concerning \emph{straight-line upward drawings}, the illustrated algorithm of Crescenzi {\em et al.}~\cite{cdp-noad-92} achieves the best known upper bound of $O(n\log n)$. For trees with constant degree, Shin \emph{et al.}~prove in~\cite{skc-aeastd-00} that upward straight-line drawings in $O(n\log \log n)$ area can be constructed. Their algorithm is based on nice inductive geometric constructions and suitable tree decompositions. No super-linear area lower bound is known, neither for binary nor for general trees, hence the following are open:

\begin{problem}
Close the gap between the $O(n \log n)$ upper bound and the $\Omega(n)$ lower bound for the area requirements of upward straight-line drawings of trees.
\end{problem}

\begin{problem}
Close the gap between the $O(n \log \log n)$ upper bound and the $\Omega(n)$ lower bound for the area requirements of upward straight-line drawings of binary trees.
\end{problem}

\begin{figure}[htb]
\centering{
	\mbox{\epsfig{figure=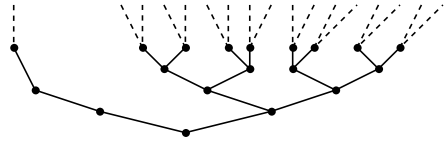,scale =0.5,clip=}}}
	\caption{A binary tree $T^*$ requiring $\Omega(n \log n)$ area in any strictly-upward drawing.}
  \label{fig:trees-upward-LB}
\end{figure}

Concerning \emph{straight-line strictly-upward drawings}, tight bounds are known. In fact, the algorithm of Crescenzi {\em et al.}~\cite{cdp-noad-92} can be suitably modified in order to obtain strictly-upward drawings (instead of aligning the subtrees of the root with their bottom sides on the same horizontal line, it is sufficient to align them with their left sides on the same vertical line). The same authors also showed a binary tree $T^*$ requiring $\Omega(n \log n)$ area in any strictly-upward drawing, hence their bound is tight. The tree $T^*$, that is shown in Fig.~\ref{fig:trees-upward-LB}, is composed of a path with $\Omega(n)$ nodes (forcing the height of the drawing to be $\Omega(n)$) and of a complete binary tree with $\Omega(n)$ nodes (forcing the width of the tree to be $\Omega(\log n)$).

Concerning \emph{straight-line order-preserving drawings}, Garg and Rusu have shown in~\cite{gr-aeoppsdot-03} how to obtain an $O(n\log n)$ area upper bound for general trees. The algorithm of Garg and Rusu inductively assumes that an \emph{$\alpha$-drawing} of a tree $T$ can be constructed, that is, a straight-line order-preserving drawing of $T$ can be constructed with the further constraints that the root $r$ of $T$ is on the upper left corner of the bounding-box of the drawing, that the children of $r$ are placed on the vertical line one unit to the right of $r$, and that the vertical distance between $r$ and any other node of $T$ is at least $\alpha$. Refer to Fig.~\ref{fig:gargrusu-ordered}(a). To construct a drawing of $T$, the algorithm considers inductively constructed drawings of all the subtrees rooted at the children of $r$, except for the node $u$ that is the root of the subtree of $r$ with the greatest number of nodes, and place such drawings one unit to the right of $r$, with their left side aligned. Further, the algorithm considers inductively constructed drawings of all the subtrees rooted at the children of $u$, except for the node $v$ that is the root of the subtree of $u$ with the greatest number of nodes, and place such drawings two units to the right of $r$, with their left side aligned. Finally, the subtree rooted at $v$ is inductively drawn, the drawing is reflected and placed with its left side on the same vertical line as $r$. Thus, the height of the drawing is clearly $O(n)$, while its width is $w(n)=\max\{w(n-1),3+w(n/2)\}=O(\log n)$, where $w(n)$ denotes the maximum width of a drawing of an $n$-node tree constructed by the algorithm. Garg and Rusu also show how to combine their described result with a decomposition scheme of binary trees due to Chan \emph{et al.}~\cite{cgkt-oaarsod-02} to obtain $O(n\log \log n)$ area straight-line order-preserving drawings of binary trees. As no super-linear lower bound is known for the area requirements of straight-line order-preserving drawings of trees, the following problems remain open:

\begin{figure}[htb]
\centering{
\begin{tabular} {c c c c}
    \mbox{\epsfig{figure=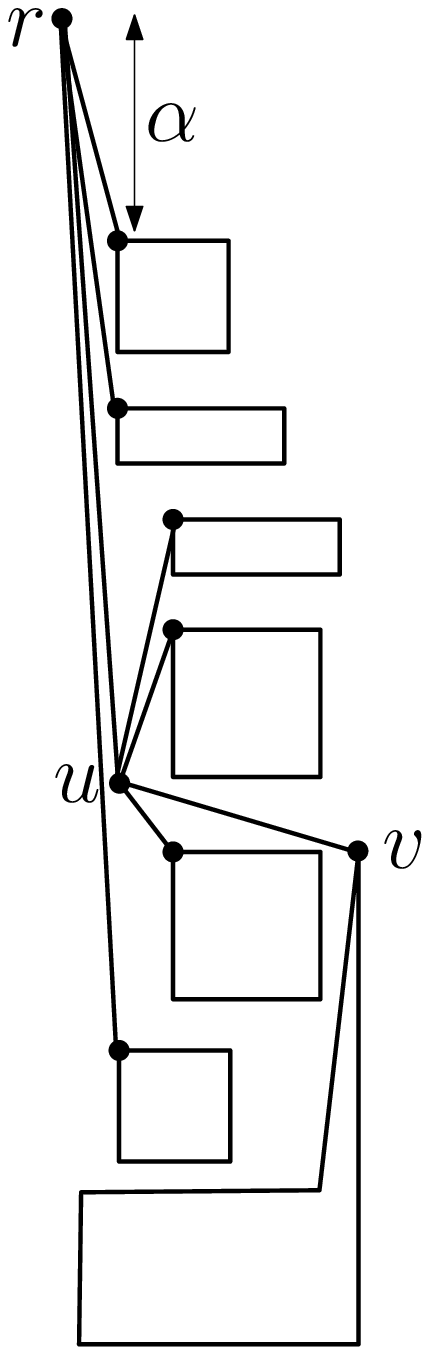,scale =0.45,clip=}} \hspace{4mm} &
    \mbox{\epsfig{figure=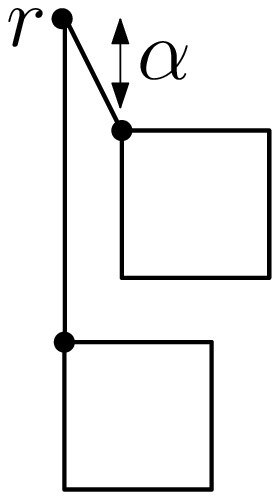,scale =0.45,clip=}} \hspace{4mm} &
    \mbox{\epsfig{figure=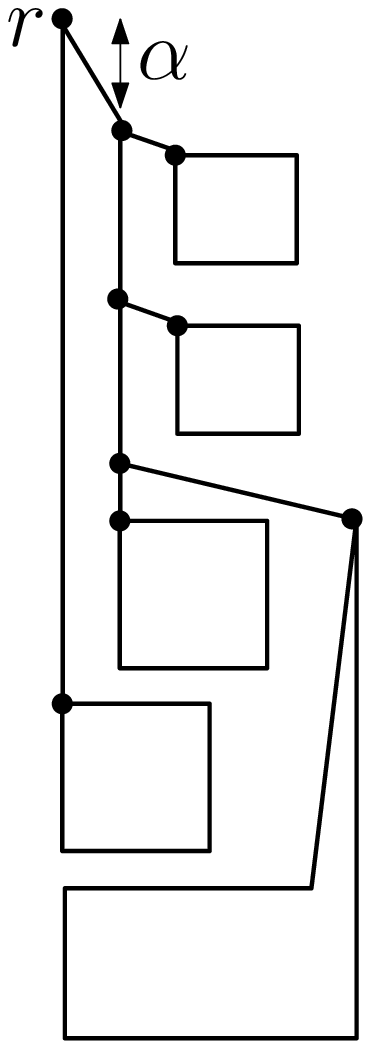,scale =0.45,clip=}} \hspace{4mm} &
    \mbox{\epsfig{figure=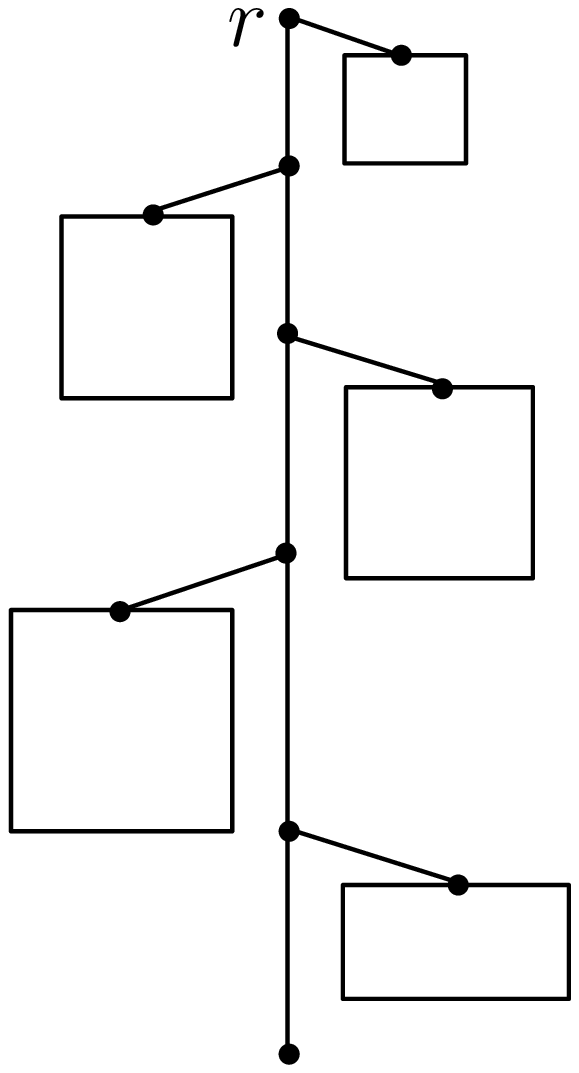,scale =0.45,clip=}} \\
    (a) \hspace{4mm} & (b) \hspace{4mm} & (c) \hspace{4mm} & (d)
\end{tabular}}
	\caption{(a) The inductive construction of a straight-line order-preserving drawing of a tree in $O(n\log n)$ area. (b)--(c) The inductive construction of a straight-line strictly-upward order-preserving drawing of a binary tree in $O(n\log n)$ area. The construction in (b) (resp. in (c)) refers to the case  in which the left (resp. the right) subtree of $r$ contains more nodes than the right (resp. the left) subtree of $r$. (d) The geometric construction of the algorithm of Chan.}
  \label{fig:gargrusu-ordered}
\end{figure}

\begin{problem}
Close the gap between the $O(n \log n)$ upper bound and the $\Omega(n)$ lower bound for the area requirements of straight-line order-preserving drawings of trees.
\end{problem}

\begin{problem}
Close the gap between the $O(n \log \log n)$ upper bound and the $\Omega(n)$ lower bound for the area requirements of straight-line order-preserving drawings of binary trees.
\end{problem}

Concerning \emph{straight-line strictly-upward order-preserving drawings}, Garg and Rusu have shown in~\cite{gr-aeoppsdot-03} how to obtain an $O(n\log n)$ area upper bound for binary trees. Observe that such an upper bound is still matched by the described $\Omega(n\log n)$ lower bound of Crescenzi {\em et al.}~\cite{cdp-noad-92}. The algorithm of Garg and Rusu, shown in Figs.~\ref{fig:gargrusu-ordered}(b)--(c), is similar to their described algorithm for constructing straight-line order-preserving drawings of trees. The results of Garg and Rusu improved upon previous results by Chan in~\cite{c-nlabdbt-02}. In~\cite{c-nlabdbt-02}, the author proved that every binary tree admits a straight-line strictly-upward order-preserving drawing in $O(n^{1+\epsilon})$ area, for any constant $\epsilon>0$. In the same paper, the author proved the best known upper bound for the area requirements of straight-line strictly-upward order-preserving drawings of trees, namely $O(n 4^{\sqrt{2\log n}})$. The approach of Chan consists of using very simple geometric constructions together with non-trivial tree decompositions. The simplest geometric construction discussed by Chan consists of selecting a path $P$ in the input tree $T$, of drawing $P$ on a vertical line $l$, and of inductively constructing drawings of the subtrees of $P$ to be placed to the left and right of $l$ (see Fig.~\ref{fig:gargrusu-ordered}(d)). Thus, denoting by $w(n)$ the maximum width of a drawing constructed by the algorithm, it holds $w(n)=1+w(n_1)+w(n_2)$, where $n_1$ and $n_2$ are the maximum number of nodes in a left subtree of $P$ and in a right subtree of $P$, respectively (assuming that $w(n)$ is monotone with $n$). Thus, depending on the way in which $P$ is chosen, different upper bounds on the asymptotic behavior of $w(n)$ can be achieved. Chan proves that $P$ can be chosen so that $w(n)=O(n^{0.48})$. Such a bound is at the base of the best upper bound for constructing straight-line drawings of outerplanar graphs (see~\cite{df-sadog-j09} and Sect.~\ref{se:straight-outerplanar}). An improvement on the following problem would be likely to improve the area upper bound on straight-line drawings of outerplanar graphs:

\begin{problem}
Let $w(n)$ be the function inductively defined as follows: $w(0)=0$, $w(1)=1$, and, for any $n>1$, let $w(n)=\max_T \{\min_P \{1 + w(n_1) + w(n_2)\}\}$, where the maximum is among all ordered rooted trees $T$ with $n$ vertices, the minimum is among all the root-to-leaf paths $P$ in $T$, where $n_1$ denotes the largest number of nodes in a left subtree of $P$, and where $n_2$ denotes the largest number of nodes in a right subtree of $P$. What is the asymptotic behavior of $w(n)$?
\end{problem}

It is easy to observe an $\Omega(\log n)$ lower bound for $w(n)$. We believe that in fact $w(n)=\Omega(2^{\sqrt{\log n}})$, but it is not clear to us whether the same bound can be achieved from above.

Turning the attention back to straight-line strictly-upward order-preserving drawings, the following problem remains open:

\begin{problem}
Close the gap between the $O(n 4^{\sqrt{2\log n}})$ upper bound and the $\Omega(n \log n)$ lower bound for the area requirements of straight-line strictly-upward order-preserving drawings of trees.
\end{problem}

Concerning \emph{straight-line orthogonal drawings}, Chan \emph{et al.}~in~\cite{cgkt-oaarsod-02} and Shin \emph{et al.}~in~\cite{skc-aeastd-00} have independently shown that $O(n \log \log n)$ area suffices for binary trees. Both algorithms are based on nice inductive geometric constructions and on non-trivial tree decompositions. Frati proved in~\cite{f-sodbtt-06} that every ternary tree admits a straight-line orthogonal drawing in $O(n^{1.631})$ area. The following problems are still open:

\begin{problem}
Close the gap between the $O(n \log \log n)$ upper bound and the $\Omega(n)$ lower bound for the area requirements of straight-line orthogonal drawings of binary trees.
\end{problem}

\begin{problem}
Close the gap between the $O(n^{1.631})$ upper bound and the $\Omega(n)$ lower bound for the area requirements of straight-line orthogonal drawings of ternary trees.
\end{problem}

\begin{figure}[htb]
\centering{
	\mbox{\epsfig{figure=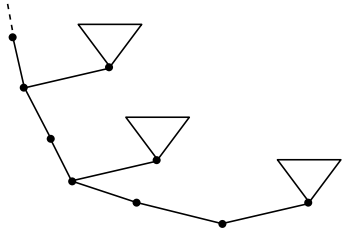,scale =0.5,clip=}}}
	\caption{A binary tree requiring $\Omega(n \log n)$ area in any straight-line upward orthogonal drawing. The tree is composed of a path $P$ and of complete binary trees with size $n^{\alpha/2}$, where $\alpha$ is some constant greater than $0$, attached to the $i$-th node of $P$, for each $i$ multiple of $n^{\alpha/2}$.}
  \label{fig:trees-orthogonal-LB}
\end{figure}

Concerning \emph{straight-line upward orthogonal drawings}, Crescenzi \emph{et al.}~\cite{cdp-noad-92} and  Chan \emph{et al.} in~\cite{cgkt-oaarsod-02} have shown that $O(n \log n)$ area suffices for binary trees. Such an area bound is worst-case optimal, as proved in~\cite{cgkt-oaarsod-02}. The tree providing the lower bound, shown in Fig.~\ref{fig:trees-orthogonal-LB}, consists of a path to which some complete binary trees are attached.

Concerning \emph{straight-line order-preserving orthogonal drawings}, $O(n^{1.5})$ and $O(n^2)$ area upper bounds are known~\cite{f-sodbtt-06} for binary and ternary trees, respectively. Once again such algorithms are based on simple inductive geometric constructions. While the bound for ternary trees is tight, no super-linear lower bound is known for straight-line order-preserving orthogonal drawings of binary trees, hence the following is open:

\begin{problem}
Close the gap between the $O(n^{1.5})$ upper bound and the $\Omega(n)$ lower bound for the area requirements of straight-line order-preserving orthogonal drawings of binary trees.
\end{problem}

\section{Poly-line Drawings}\label{se:poly-line}

In this section, we discuss algorithms and bounds for constructing small-area planar poly-line drawings of planar graphs and their subclasses.
In Sect.~\ref{se:poly-planar} we deal with general planar graphs, in Sect.~\ref{se:poly-sp} we deal with series-parallel and outerplanar graphs, and in Sect.~\ref{se:poly-trees} we deal with trees. Table~\ref{ta:poly-line} summarizes the best known area bounds for poly-line planar drawings of planar graphs and their subclasses. Observe that the lower bound of the table referring to general planar graphs hold true for {\em plane} graphs.

\begin{table}[!htb]\footnotesize
\centering
  \linespread{1.2}
  \selectfont
  \begin{tabular}{|c|c|c|c|c|}
    \cline{2-5}
    \multicolumn{1}{c|}{} & \emph{Upper Bound} & \emph{Refs.} & \emph{Lower Bound} & \emph{Refs.} \\
    \hline
    {\em General Planar Graphs} & $\frac{4(n-1)^2}{9}$ & \cite{bsm-wtr-02} & $\frac{4(n-1)^2}{9}$ & \cite{fpp-hdpgg-90}\\
    \hline
    {\em Series-Parallel Graphs} & $O(n^{1.5})$ & \cite{b-sdogspgopg-11} & $\Omega(n 2^{\sqrt{\log n}})$ & \cite{f-lbarspg-j10}\\
    \hline
    {\em Outerplanar Graphs} & $O(n \log n)$ & \cite{b-dopgia-02,b-sdogspgopg-11} & $\Omega(n)$ & \emph{trivial}\\
    \hline
    {\em Trees} & $O(n \log n)$ & \cite{cdp-noad-92} & $\Omega(n)$ & \emph{trivial}\\
    \hline
 \end{tabular}
 \vspace{2mm}
 \caption{\small A table summarizing the area requirements for poly-line planar drawings of several classes of planar graphs.}
  \label{ta:poly-line}
\end{table}

\subsection{General Planar Graphs} \label{se:poly-planar}

Every $n$-vertex plane graph admits a planar poly-line drawing on a grid with $O(n^2)$ area. In fact, this has been known since the beginning of the 80's~\cite{w-dpgt-82}. Tamassia and Tollis introduced in~\cite{tt-uavrpg-86} a technique that has later become pretty much a standard for constructing planar poly-line drawings. Namely, the authors showed that a poly-line drawing $\Gamma$ of a plane graph $G$ can be easily obtained from a visibility representation $R$ of $G$; moreover, $\Gamma$ and $R$ have asymptotically the same area. In order to obtain a visibility representation $R$ of $G$, Tamassia and Tollis design a very nice algorithm (an application is shown in Fig.~\ref{fig:tamassia-vr}). The algorithm assumes that $G$ is biconnected (if it is not, it suffices to augment $G$ to biconnected by inserting dummy edges, apply the algorithm, and then remove the inserted dummy edges to obtain a visibility representation of $G$). The algorithm consists of the following steps:
\begin{figure}[htb]
\centering{
\begin{tabular} {c c}
    \mbox{\epsfig{figure=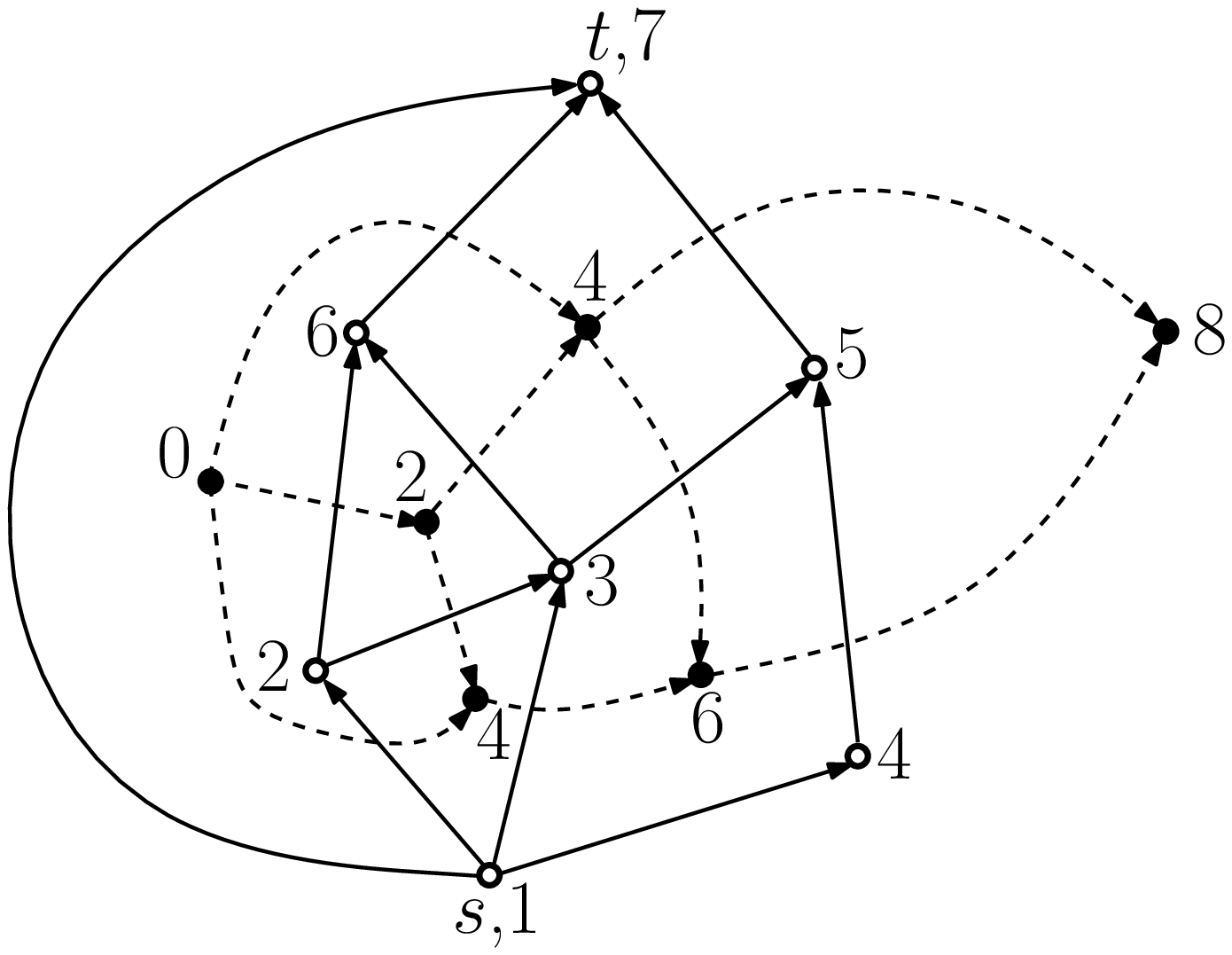,scale =0.4,clip=}} \hspace{7mm} &
    \mbox{\epsfig{figure=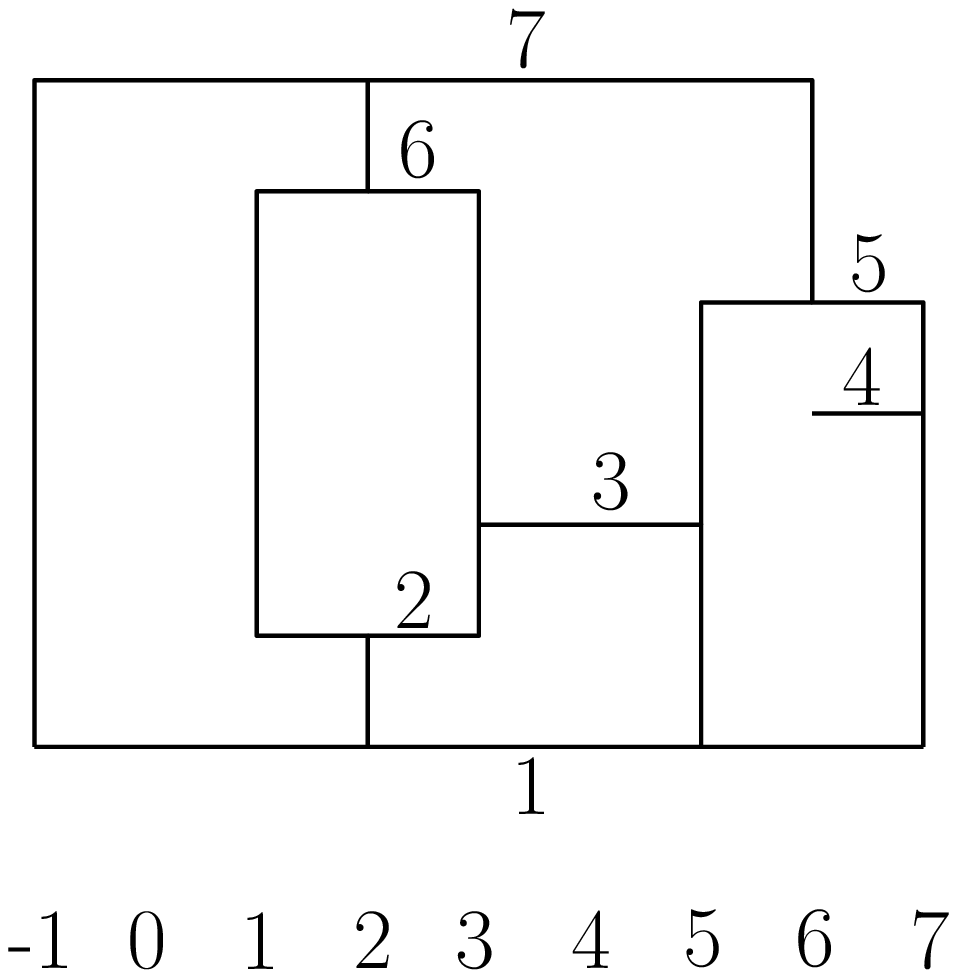,scale =0.4,clip=}} \\
    (a) \hspace{7mm} & (b)
\end{tabular}
}
\caption{An illustration for the algorithm of Tamassia and Tollis~\cite{tt-uavrpg-86}. (a) White circles and solid edges represent $G$. Black circles and dashed edges represent $G^*$. An st-numbering of $G$ (and the corresponding orientation) is shown. An orientation of $G^*$ and the number $2\psi(f)$ for each each face $f$ of $G$ is shown. (b) A visibility representation of $G$.}
  \label{fig:tamassia-vr}
\end{figure}
(1) Consider an orientation of $G$ induced by an \emph{st-numbering} of $G$, that is a bijective mapping $\phi : V(G)\rightarrow \{1,\dots,n\}$ such that, for a given edge $(s,t)$ incident to the outer face of $G$, $\phi(s)=1$, $\phi(t)=n$, and for each $u\in V(G)$ with $u \neq s,t$, there exist two neighbors of $u$, say $v$ and $w$, such that $\phi(v)<\phi(u)<\phi(w)$; (2) consider the orientation of the dual graph $G^*$ of $G$ induced by the orientation of $G$; (3) the $y$-coordinate of each vertex-segment $u$ is given by $\phi(u)$; (4) the $y$-coordinates of the endpoints of each edge-segment $(u,v)$ are given by $\phi(u)$ and $\phi(v)$; (5) the $x$-coordinate of edge-segment $(s,t)$ is set equal to $-1$; (6) the $x$-coordinate of each edge-segment $(u,v)$ is chosen to be any number strictly between $2\psi(f)$ and $2\psi(g)$, where $f$ and $g$ are the faces adjacent to $(u,v)$ in $G$ and $\psi(f)$ denotes the length of the longest path from the source to $f$ in $G^*$; (7) finally, the $x$-coordinates of the endpoints of each vertex-segment $u$ is set equal to the smallest and largest $x$-coordinates of its incident edges.

After the algorithm of Tamassia and Tollis, a large number of algorithms have been proposed to construct poly-line drawings of planar graphs (see, e.g.,~\cite{gm-ppdgar-98,gw-fdpgcp-00,cdgk-dpgca-01,zs-ppd-08,z-ppdgt-10}), proposing several tradeoffs between area requirements, number of bends, and angular resolution. Here we briefly discuss an algorithm proposed by Bonichon \emph{et al.}~in~\cite{bsm-wtr-02}, the first one to achieve optimal area, namely $\frac{4(n-1)^2}{9}$. The algorithm consists of two steps. In the first one, a deep study of Schnyder realizers (see~\cite{s-epgg-90} and Sect.~\ref{se:straight-planar} for the definition of Schnyder realizers) leads to the definition of a \emph{weak-stratification} of a realizer. Namely, given a realizer $(T_0,T_1,T_2)$ of a triangulation $G$, a weak-stratification is a layering $L$ of the vertices of $G$ such that $T_0$ (which is rooted at the vertex incident to the outer face of $G$) is upward, while $T_1$ and $T_2$ (which are rooted at the vertices incident to the outer face of $G$) are downward and some further conditions are satisfied. Each vertex will get a $y$-coordinate which is equal to its layer in the weak stratification. In the second step $x$-coordinates for vertices and bends are computed. The conditions of the weak stratification ensure that a planar drawing can in fact be obtained.

\subsection{Series-Parallel and Outerplanar Graphs} \label{se:poly-sp}

Biedl proved in~\cite{b-sdogspgopg-11} that every series-parallel graph admits a poly-line drawing with $O(n^{1.5})$ area and a poly-line drawing with $O(fn \log n)$ area, where $f$ is the fan-out of the series-parallel graph. In particular, since outerplanar graphs are series-parallel graphs with fan-out two, the last result implies that outerplanar graphs admit poly-line drawings with $O(n \log n)$ area. Biedl's algorithm constructs a visibility representation $R$ of the input graph $G$ with $O(n^{1.5})$ area; a poly-line drawing $\Gamma$ with asymptotically the same area of $R$ can then be easily obtained from $R$.
\begin{figure}[htb]
\centering{
\begin{tabular} {c c c c}
    \mbox{\epsfig{figure=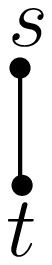,scale =0.45,clip=}} \hspace{4mm} &
    \mbox{\epsfig{figure=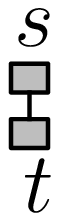,scale =0.45,clip=}} \hspace{4mm} &
    \mbox{\epsfig{figure=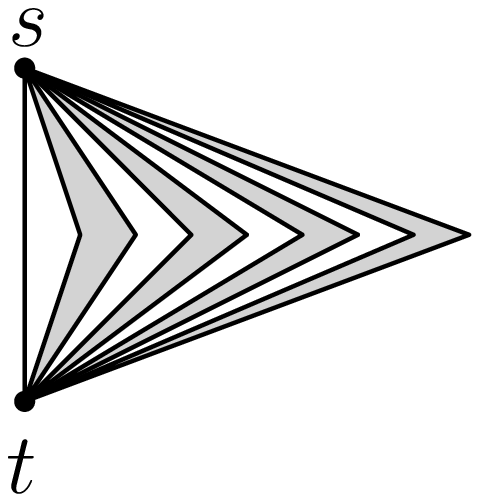,scale =0.33,clip=}} \hspace{4mm} &
    \mbox{\epsfig{figure=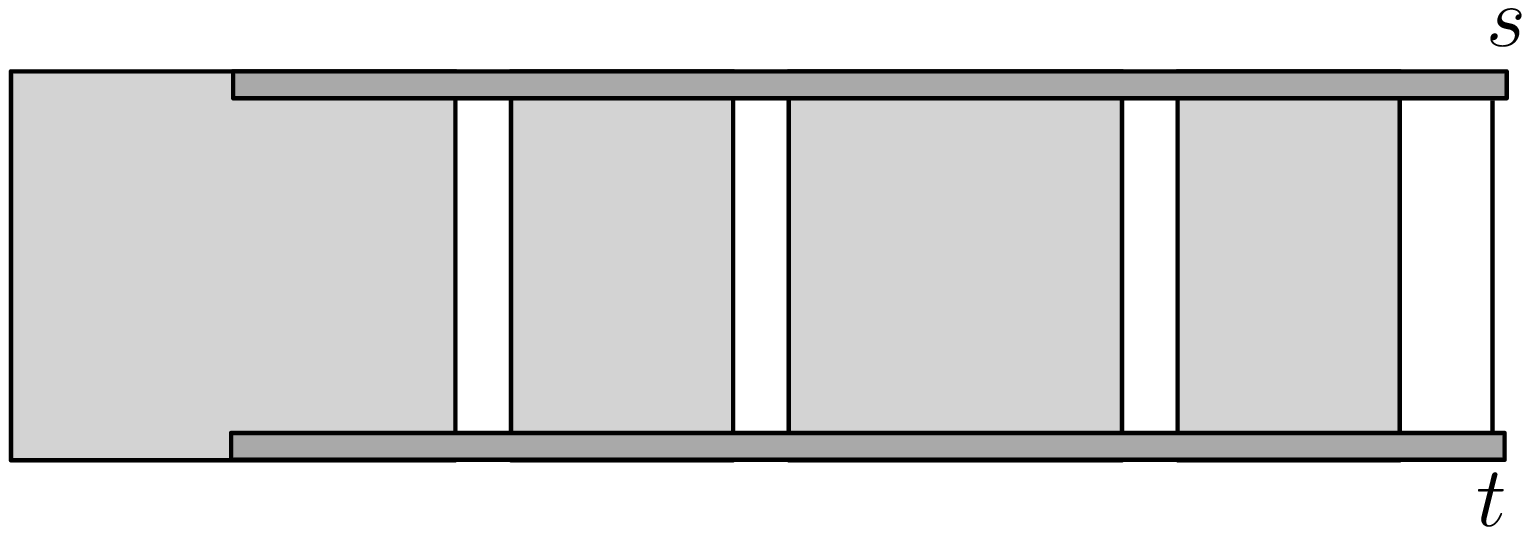,scale =0.33,clip=}} \\
    (a) \hspace{4mm} & (b) \hspace{4mm} & (c) \hspace{4mm} & (d) \vspace{3mm}\\
    \mbox{\epsfig{figure=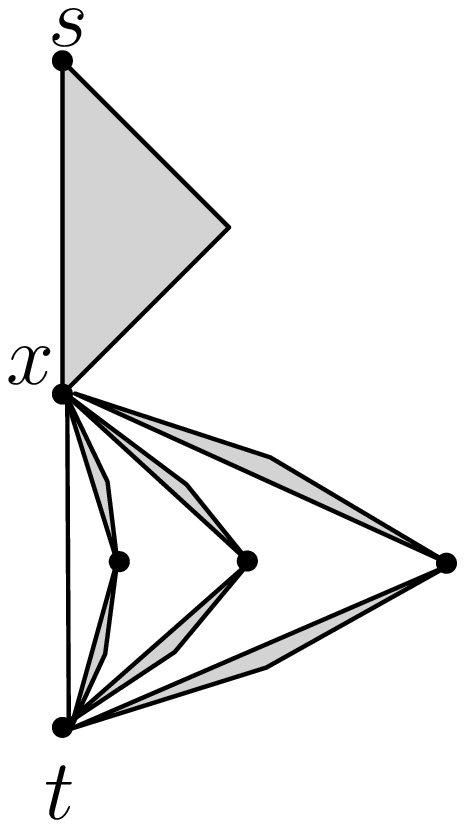,scale =0.27,clip=}} \hspace{3mm} &
    \mbox{\epsfig{figure=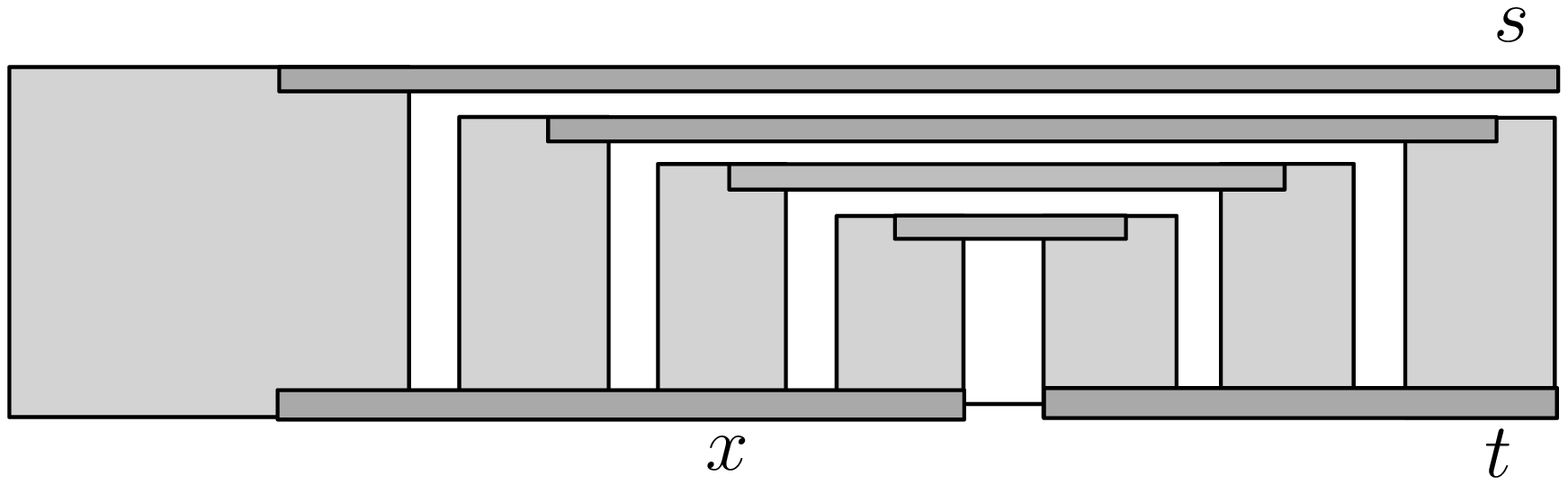,scale =0.27,clip=}} \hspace{3mm} &
    \mbox{\epsfig{figure=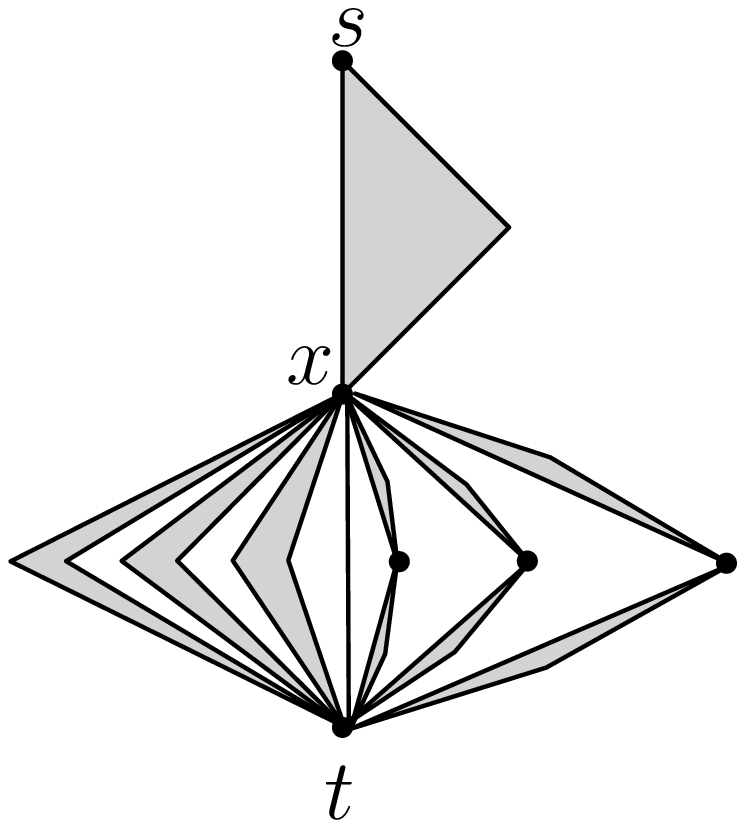,scale =0.27,clip=}} \hspace{3mm} &
    \mbox{\epsfig{figure=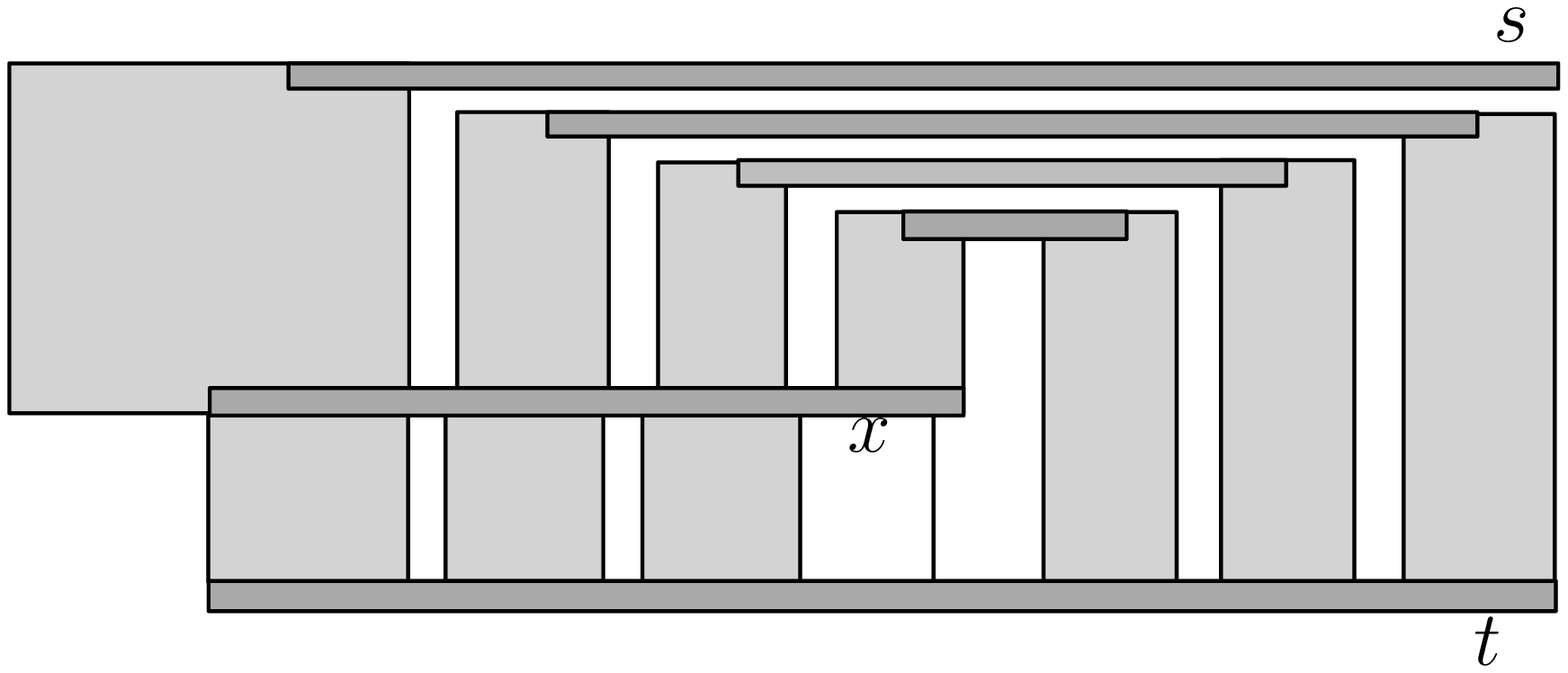,scale =0.27,clip=}} \\
    (e) \hspace{3mm} & (f) \hspace{3mm} & (g) \hspace{3mm} & (h)
\end{tabular}
}
\caption{Biedl's algorithm for constructing visibility representations of series-parallel graphs. (a)--(b) The base case. (c)--(d) The parallel case. (e)--(h) The series case.}
  \label{fig:biedl-sp}
\end{figure}
In order to construct a visibility representation $R$ of the input graph $G$, Biedl relies on a strong inductive hypothesis, namely that a small area visibility representation $R$ of $G$ can be constructed with the further constraint that the poles $s$ and $t$ of $G$ are placed at the top right corner and at the bottom right corner of the representation, respectively. Figs.~\ref{fig:biedl-sp}(a)--(b) show how this is accomplished in the base case. The parallel case is also pretty simple, as the visibility representations of the components of $G$ are just placed one besides the other (as in Figs.~\ref{fig:biedl-sp}(c)--(d)). The series case is much more involved. Namely, assuming w.l.o.g. that $G$ is the series of two components $H_1$ and $H_2$, where $H_1$ has poles $s$ and $x$ and $H_2$ has poles $x$ and $t$, and assuming w.l.o.g. that $H_2$ has more vertices than $H_1$, then if $H_2$ is the parallel composition of a ``small'' number of components, the composition shown in Figs.~\ref{fig:biedl-sp}(e)--(f) is applied, while if $H_2$ is the parallel composition of a ``large'' number of components, the composition shown in Figs.~\ref{fig:biedl-sp}(g)--(h) is applied. The rough idea behind these constructions is that if $H_2$ is the parallel composition of a small number of components, then a vertical unit can be spent for each of them without increasing much the height of the drawing; on the other hand, if $H_2$ is the parallel composition of a large number of components, then lots of such components have few vertices, hence two of them can be placed one above the other without increasing much the height of the drawing.

The following problems remain open:

\begin{problem}
Close the gap between the $O(n^{1.5})$ upper bound and the $\Omega(n 2^{\sqrt{\log n}})$ lower bound for the area requirements of poly-line drawings of series-parallel graphs.
\end{problem}

\begin{problem}
Close the gap between the $O(n \log n)$ upper bound and the $\Omega(n)$ lower bound for the area requirements of poly-line drawings of outerplanar graphs.
\end{problem}

\subsection{Trees} \label{se:poly-trees}

No algorithms are known exploiting the possibility of bending the edges of a tree to get area bounds better than the corresponding ones shown for straight-line drawings.

\begin{problem}
Close the gap between the $O(n \log n)$ upper bound and the $\Omega(n)$ lower bound for the area requirements of poly-line drawings of trees.
\end{problem}

However, better bounds can be achieved for poly-line drawings satisfying further constraints. Table~\ref{ta:trees-straight-line} summarizes the best known area bounds for various kinds of poly-line drawings of trees.

\begin{table}[!htb]\centering\footnotesize
  \linespread{1.2}
  \selectfont
  \begin{tabular}{|c|c|c|c|c|c|c|c|c|}
   \cline{2-9}
	 \multicolumn{1}{c|}{} & \emph{Ord. Pres.} & \emph{Upw.} & \emph{Str. Upw.} & \emph{Orth.} & \emph{Upper Bound} & \emph{Refs.} & \emph{Lower Bound} & \emph{Refs.} \\ 
    \cline{2-9}
    \hline
    \emph{Binary} & & & & & $O(n)$ & \cite{ggt-putdoa-96} & $\Omega(n)$ & \emph{trivial} \\
    \hline
    \emph{Binary} & \checkmark & & & & $O(n\log \log n)$ & \cite{gr-aeoppsdot-03} & $\Omega(n)$ & \emph{trivial} \\
    \hline
    \emph{Binary} & & \checkmark & & & $O(n)$ & \cite{ggt-putdoa-96} & $\Omega(n)$ & \emph{trivial} \\
    \hline
    \emph{Binary} & & & \checkmark & & $O(n\log n)$ & \cite{cdp-noad-92} & $\Omega(n \log n)$ & \cite{cdp-noad-92} \\
    \hline
    \emph{Binary} & \checkmark & & \checkmark &  & $O(n \log n)$ & \cite{ggt-putdoa-96} & $\Omega(n\log n)$ & \cite{cdp-noad-92}\\
    \hline
    \emph{Binary} & & & & \checkmark & $O(n)$ & \cite{Val81} & $\Omega(n)$ & \emph{trivial}\\
    \hline
    \emph{Binary} & & \checkmark & & \checkmark & $O(n \log \log n)$ & \cite{ggt-putdoa-96} & $\Omega(n \log \log n)$ & \cite{ggt-putdoa-96}\\
    \hline
    \emph{Binary} & \checkmark & & & \checkmark & $O(n)$ & \cite{dt-laepg-81} & $\Omega(n)$ & \emph{trivial}\\
    \hline
    \emph{Binary} & \checkmark & \checkmark & & \checkmark & $O(n \log n)$ & \cite{k-saoudbtt-95} & $\Omega(n \log n)$ & \cite{ggt-putdoa-96}\\
    \hline
    \emph{Ternary} & & & & \checkmark & $O(n)$ & \cite{Val81} & $\Omega(n)$ & \emph{trivial}\\
    \hline
    \emph{Ternary} & & \checkmark & & \checkmark & $O(n \log n)$ & \cite{k-saoudbtt-95} & $\Omega(n \log n)$ & \cite{k-saoudbtt-95}\\
    \hline
    \emph{Ternary} & \checkmark & & & \checkmark & $O(n)$ & \cite{dt-laepg-81} & $\Omega(n)$ & \emph{trivial}\\
    \hline
    \emph{Ternary} & \checkmark & \checkmark & & \checkmark & $O(n \log n)$ & \cite{k-saoudbtt-95} & $\Omega(n \log n)$ & \cite{ggt-putdoa-96}\\
    \hline
    \emph{General} & & & & & $O(n \log n)$ & \cite{cdp-noad-92} & $\Omega(n)$ & \emph{trivial} \\
    \hline
    \emph{General} & \checkmark & & & & $O(n\log n)$ & \cite{gr-aeoppsdot-03} & $\Omega(n)$ & \emph{trivial} \\
    \hline
    \emph{General} & & \checkmark & & & $O(n\log n)$ & \cite{cdp-noad-92} & $\Omega(n)$ & \emph{trivial} \\
    \hline
    \emph{General} & & & \checkmark & & $O(n\log n)$ & \cite{cdp-noad-92} & $\Omega(n \log n)$ & \cite{cdp-noad-92} \\
    \hline
    \emph{General} & \checkmark & & \checkmark & & $O(n 4^{\sqrt{2\log n}})$ & \cite{c-nlabdbt-02} & $\Omega(n\log n)$ & \cite{cdp-noad-92}\\
    \hline
 \end{tabular}
 \vspace{2mm}
  \caption{\small Summary of the best known area bounds for poly-line drawings of trees. ``Ord. Pres.'', ``Upw.'', ``Str. Upw.'', and ``Orth.'' stand for order-preserving, upward, strictly-upward, and orthogonal, respectively.}
  \label{ta:trees-straight-line}
\end{table}

Concerning \emph{poly-line upward drawings}, a linear area bound is known, due to Garg {\em et al.}~\cite{ggt-putdoa-96}, for all trees whose degree is $O(n^{\delta})$, where $\delta$ is \emph{any} constant less than $1$. The algorithm of Garg {\em et al.}~first constructs a layering $\gamma(T)$ of the input tree $T$; in $\gamma(T)$ each node $u$ is assigned a layer smaller than or equal to the layer of the leftmost child of $u$ and smaller than the layer of any other child of $u$; second, the authors show that $\gamma(T)$ can be converted into an upward poly-line drawing whose height is the number of layers and whose width is the maximum \emph{width of a layer}, that is the number of nodes of the layer plus the number of edges crossing the layer; third, the authors show how to construct a layering of every tree whose degree is $O(n^{\delta})$ so that the number of layers times the maximum width of a layer is $O(n)$. No upper bound better than $O(n \log n)$ (from the results on straight-line drawings, see~\cite{cdp-noad-92} and Sect.~\ref{se:straight-trees}) and no super-linear lower bound is known for trees with unbounded degree.

\begin{problem}
Close the gap between the $O(n \log n)$ upper bound and the $\Omega(n)$ lower bound for the area requirements of poly-line upward drawings of trees.
\end{problem}

\begin{figure}[htb]
\centering{
\begin{tabular} {c c}
    \mbox{\epsfig{figure=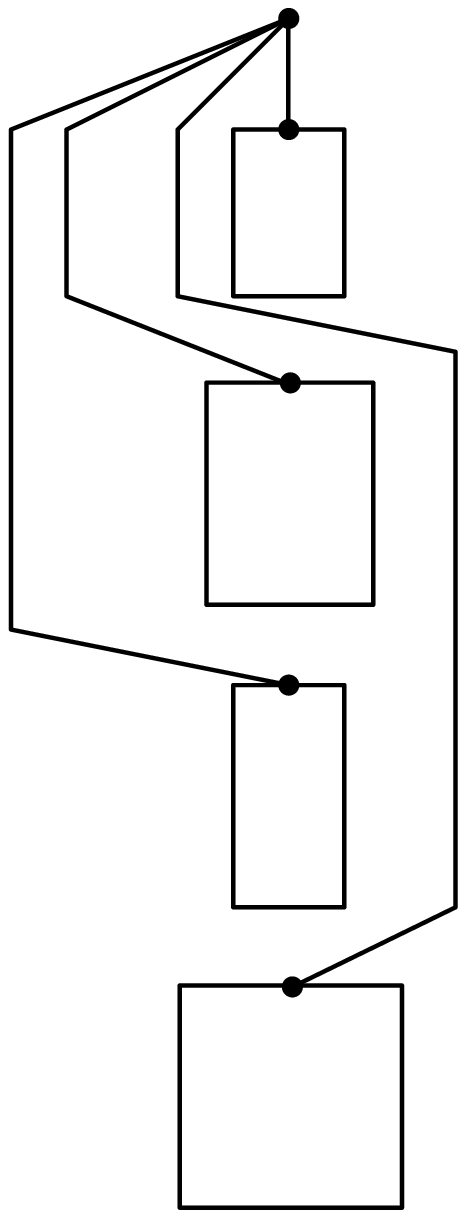,scale =0.45,clip=}} \hspace{7mm} &
    \mbox{\epsfig{figure=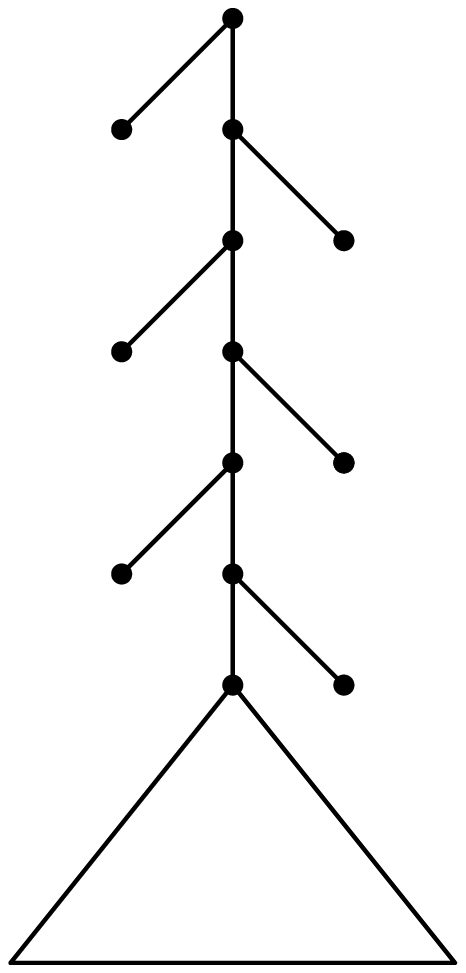,scale =0.45,clip=}} \\
    (a) \hspace{7mm} & (b)
\end{tabular}
}
\caption{(a) The construction of Garg {\em et al.}~\cite{ggt-putdoa-96} to obtain $O(n\log n)$ area poly-line order-preserving strictly-upward drawings of bounded-degree trees. (b) A tree requiring $\Omega(n \log n)$ area in any upward order-preserving drawing. The triangle represents a complete binary tree with $n/3$ nodes.}
  \label{fig:changood}
\end{figure}

Concerning \emph{poly-line order-preserving strictly-upward drawings}, Garg {\em et al.}~\cite{ggt-putdoa-96} show a simple algorithm to achieve $O(n\log n)$ area for bounded-degree trees. The algorithm, whose construction is shown in Fig.~\ref{fig:changood}(a), consists of stacking inductively constructed drawings of the subtrees of the root of the input tree $T$, in such a way that the tree with the greatest number of nodes is the bottommost in the drawing. The edges connecting the root to its subtrees are then routed besides the subtrees. The $O(n\log n)$ area upper bound is tight. Namely, there exist binary trees requiring $\Omega(n \log n)$ area in any strictly-upward order-preserving drawing~\cite{cdp-noad-92} and binary trees requiring $\Omega(n \log n)$ area in any (even non-strictly) upward order-preserving drawing~\cite{ggt-putdoa-96}. The lower bound tree of Garg {\em et al.}~is shown in Fig.~\ref{fig:changood}(b). As far as we know, no area bounds better than the ones for straight-line drawings have been proved for general trees, hence the following are open:

\begin{problem}
Close the gap between the $O(n \log n)$ upper bound and the $\Omega(n)$ lower bound for the area requirements of poly-line order-preserving drawings of trees.
\end{problem}

\begin{problem}
Close the gap between the $O(n 4^{\sqrt{\log n}})$ upper bound and the $\Omega(n \log n)$ lower bound for the area requirements of poly-line order-preserving strictly-upward drawings of trees.
\end{problem}

Concerning \emph{orthogonal drawings}, Valiant proved in~\cite{Val81} that every $n$-node ternary tree (and every $n$-node binary tree) admits a $\Theta(n)$ area orthogonal drawing. Such a result was strengthened by Dolev and Trickey in ~\cite{dt-laepg-81}, who proved that ternary trees (and binary trees) admit $\Theta(n)$ area order-preserving orthogonal drawings. The technique of Valiant is based on the use of separator edges (see~\cite{Val81} and Sect.~\ref{se:straight-trees}). The result of Dolev and Trickey is a consequence of a more general result on the construction of linear area embeddings of degree-$4$ outerplanar graphs.

\begin{figure}[htb]
\centering{
\begin{tabular} {c c c}
    \mbox{\epsfig{figure=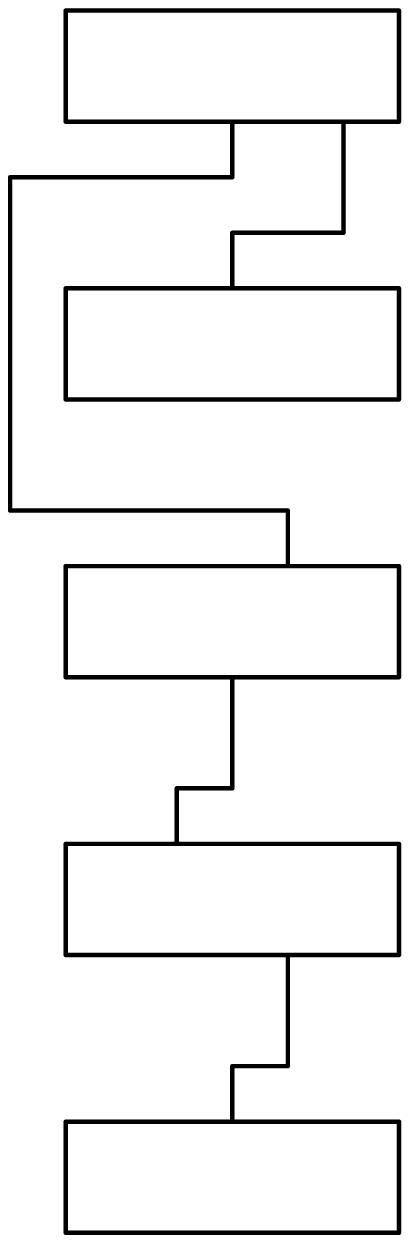,scale =0.45,clip=}} \hspace{7mm} &
    \mbox{\epsfig{figure=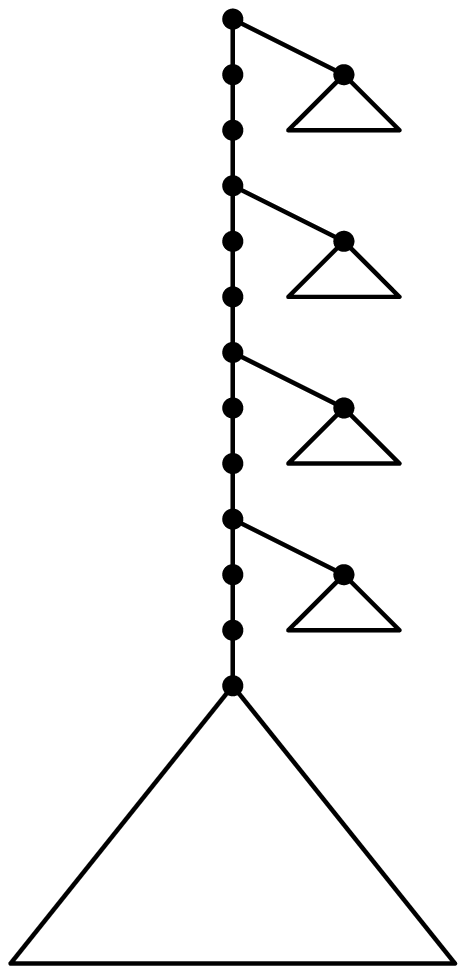,scale =0.45,clip=}} \hspace{7mm} &
    \mbox{\epsfig{figure=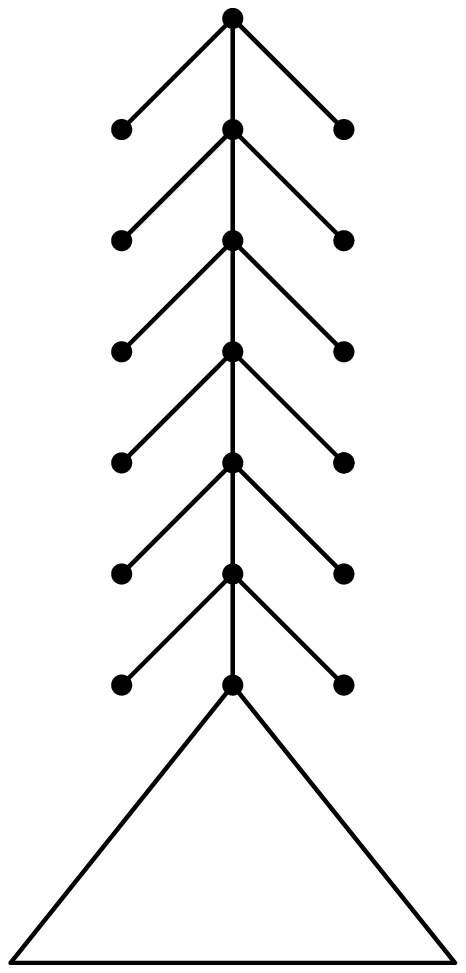,scale =0.45,clip=}} \\
    (a) \hspace{7mm} & (b) \hspace{7mm} & (c)
\end{tabular}
}
\caption{(a) The construction of Garg {\em et al.}~\cite{ggt-putdoa-96} to obtain $O(n\log \log n)$ area orthogonal upward drawings of binary trees. Rectangles represent drawings of small subtrees constructed via an $O(n \log n)$ area algorithm. (b) A binary tree requiring $\Omega(n \log \log n)$ area in any upward orthogonal drawing. The tree is composed of a chain with $n/3$ nodes, a complete binary tree with $n/3$ nodes (the large triangle in the figure), and $\frac{n}{3\sqrt{\log n}}$ subtrees (the small triangles in the figure) with $\sqrt{\log n}$ nodes rooted at the child of each $\sqrt{\log n}$-th node of the chain. (c) A ternary tree requiring $\Omega(n \log n)$ area in any upward orthogonal drawing. The tree is composed of a chain with $n/4$ nodes, two other children for each node of the chain, and a complete binary tree with $n/4$ nodes (the large triangle in the figure)}
  \label{fig:changoodorthogonal}
\end{figure}

Concerning \emph{orthogonal upward drawings}, an $O(n \log \log n)$ area bound for binary trees was proved by Garg \emph{et al.} in~\cite{ggt-putdoa-96}. The algorithm has several ingredients. (1) A simple algorithm is shown to construct orthogonal upward drawings in $O(n \log n)$ area; such drawings exhibit the further property that no vertical line through a node of degree at most two intersects the drawing below such a node. (2) The \emph{separator tree} $S$ of the input tree $T$ is constructed; such a tree represents the recursive decomposition of a tree via separator edges; namely, $S$ is a binary tree that is recursively constructed as follows: The root $r$ of $S$ is associated  with tree $T$ and with a separator edge of $T$, that splits $T$ into subtrees $T_1$ and $T_2$; the subtrees of $r$ are the separator trees associated with $T_1$ and $T_2$; observe that the leaves of $S$ are the nodes of $T$. (3) A \emph{truncated separator tree} $S'$ is obtained from $S$ by removing all the nodes of $S$ associated with subtrees of $T$ with less than $\log n$ nodes. (4) Drawings of the subtrees of $T$ associated with the leaves of $S'$ are constructed via the $O(n \log n)$ area algorithm. (5) Such drawings are stacked one on top of the other and the separator edges connecting them are routed (see Fig.~\ref{fig:changoodorthogonal}(a)). The authors prove that the constructed drawings have $O(\frac{n \log \log n}{\log n})$ height and $O(\log n)$ width, thus obtaining the claimed upper bound. The same authors also proved that the $O(n \log \log n)$ bound is tight, by exhibiting the class of trees shown in Fig.~\ref{fig:changoodorthogonal}(b). In~\cite{k-saoudbtt-95} Kim showed that $\Theta(n \log n)$ area is an optimal bound for upward orthogonal drawings of ternary trees. The upper bound comes from a stronger result on orthogonal order-preserving upward drawings cited below, while the lower bound comes from the tree shown in Fig.~\ref{fig:changoodorthogonal}(c).

\begin{figure}[htb]
\centering{
\begin{tabular} {c c c}
    \mbox{\epsfig{figure=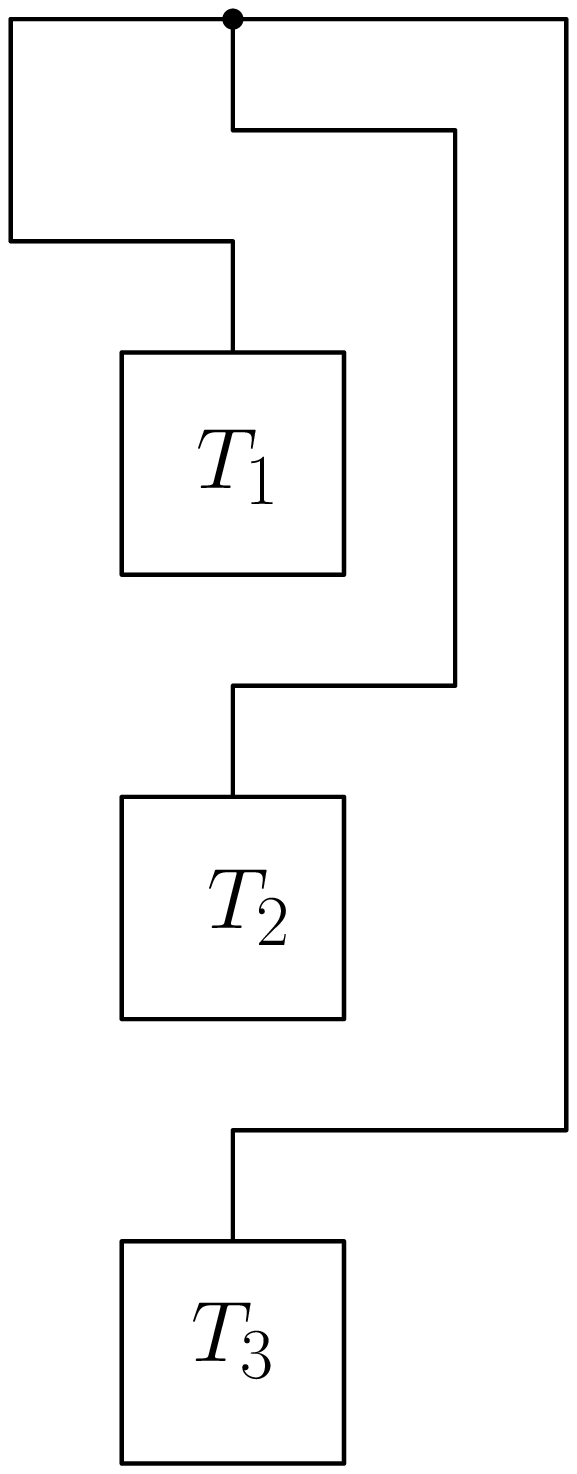,scale =0.3,clip=}} \hspace{7mm} &
    \mbox{\epsfig{figure=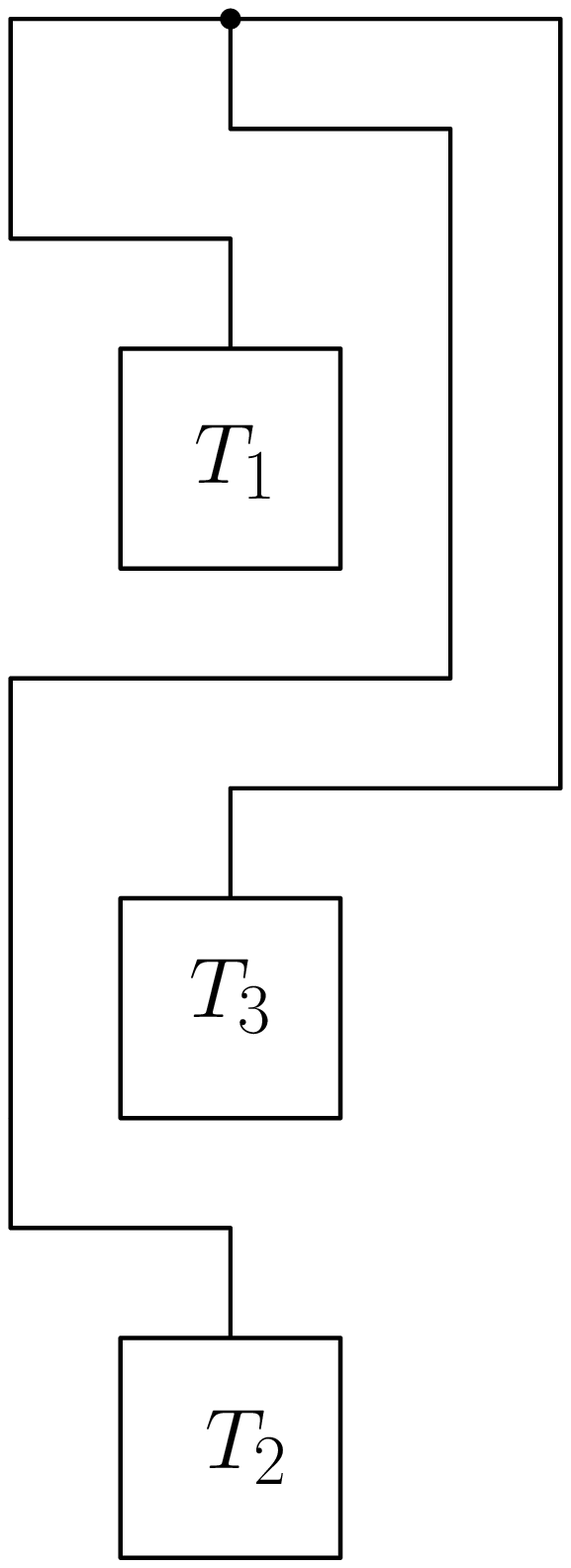,scale =0.3,clip=}} \hspace{7mm} &
    \mbox{\epsfig{figure=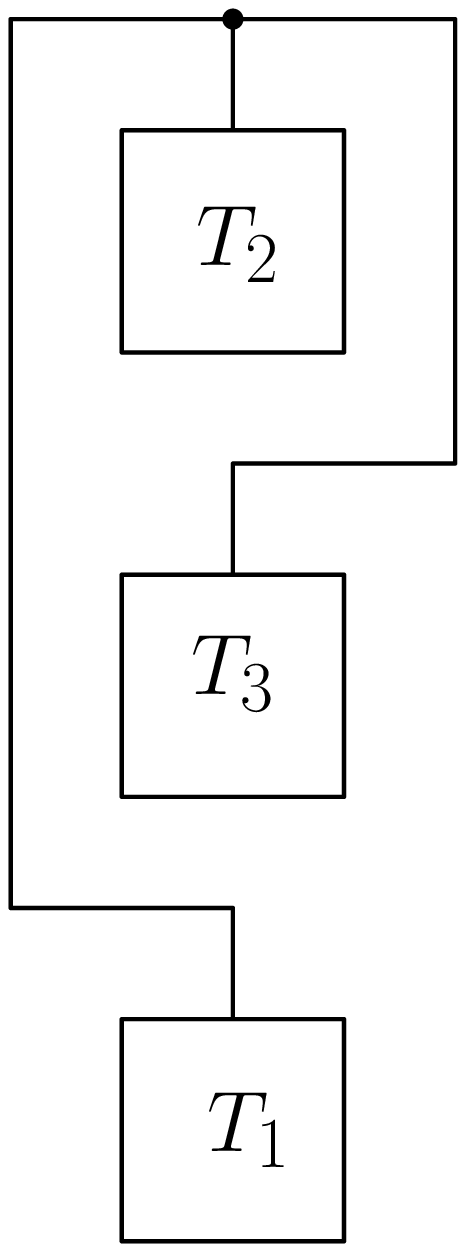,scale =0.3,clip=}} \\
    (a) \hspace{7mm} & (b) \hspace{7mm} & (c)
\end{tabular}
}
\caption{An algorithm to construct $O(n \log n)$ area orthogonal order-preserving upward drawings of ternary trees. The figures illustrate the cases in which: (a) The right subtree has the greatest number of nodes; (b) the middle subtree has the greatest number of nodes; and (b) the left subtree has the greatest number of nodes.}
  \label{fig:kim}
\end{figure}

Concerning \emph{orthogonal order-preserving upward drawings}, $\Theta(n \log n)$ is an optimal bound both for binary and ternary trees. In fact, Kim~\cite{k-saoudbtt-95} proved the upper bound for ternary trees (such a bound can be immediately extended to binary trees). The simple construction of Kim is presented in Fig.~\ref{fig:kim}. The lower bound directly comes from the results of Garg \emph{et al.}~on order-preserving upward (non-orthogonal) drawings~\cite{ggt-putdoa-96}.

\section{Upward Drawings}\label{se:upward}

In this section, we discuss algorithms and bounds for constructing small-area planar straight-line/poly-line upward drawings of upward planar directed acyclic graphs. Table~\ref{ta:upward} summarizes the best known area bounds for straight-line upward planar drawings of upward planar DAGs and their subclasses.

\begin{table}[!htb]\footnotesize
\centering
  \linespread{1.2}
  \selectfont
  \begin{tabular}{|c|c|c|c|c|}
    \cline{2-5}
    \multicolumn{1}{c|}{} & \emph{Upper Bound} & \emph{Refs.} & \emph{Lower Bound} & \emph{Refs.} \\
    \hline
    {\em General Upward Planar DAGs} & $O(c^n)$ & \cite{GargTam93} & $\Omega(b^n)$ & \cite{BattistaTT92}\\
    \hline
    {\em Fixed-Embedding Series-Parallel DAGs}  & $O(c^n)$ & \cite{GargTam93} & $\Omega(b^n)$ & \cite{BertolazziCBTT94}\\
    \hline
    {\em Series-Parallel DAGs} & $O(n^2)$ & \cite{BertolazziCBTT94} & $\Omega(n^2)$ & \emph{trivial}\\
    \hline
    {\em Bipartite DAGs} & $O(c^n)$ & \cite{GargTam93} & $\Omega(b^n)$ & \cite{f-mapudt-08}\\
    \hline
    {\em Fixed-Embedding Directed Trees} & $O(c^n)$ & \cite{GargTam93} & $\Omega(b^n)$ & \cite{f-mapudt-08}\\
    \hline
    {\em Directed Trees} & $O(n \log n)$ & \cite{f-mapudt-08} & $\Omega(n \log n)$ & \cite{f-mapudt-08}\\
    \hline
 \end{tabular}
 \vspace{2mm}
 \caption{\small A table summarizing the area requirements for straight-line upward planar drawings of upward planar DAGs; $b$ and $c$ denote constants greater than $1$.}
  \label{ta:upward}
\end{table}

It is known that testing the upward planarity of a DAG is an NP-complete problem if the DAG has a variable embedding~\cite{GargT01}, while it is polynomial-time solvable if the embedding of the DAG is fixed~\cite{BertolazziBLM94}, if the underlying graph is an outerplanar graph~\cite{Papakostas94}, if the DAG has a single source~\cite{HuttonL96}, or if it is bipartite~\cite{BattistaLR90}. Di Battista and Tamassia~\cite{BattistaT88} showed that a DAG is upward planar if and only if it is a subgraph of an st-planar DAG. Some families of DAGs are always upward planar, like the series-parallel DAGs and the directed trees.

\begin{figure}[htb]
  \centering
  \begin{tabular}{c c c c}
	\mbox{\epsfig{figure=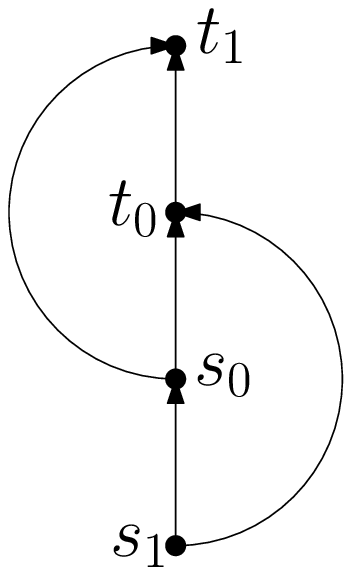,scale=0.55,clip=}} \hspace{2mm} &
	\mbox{\epsfig{figure=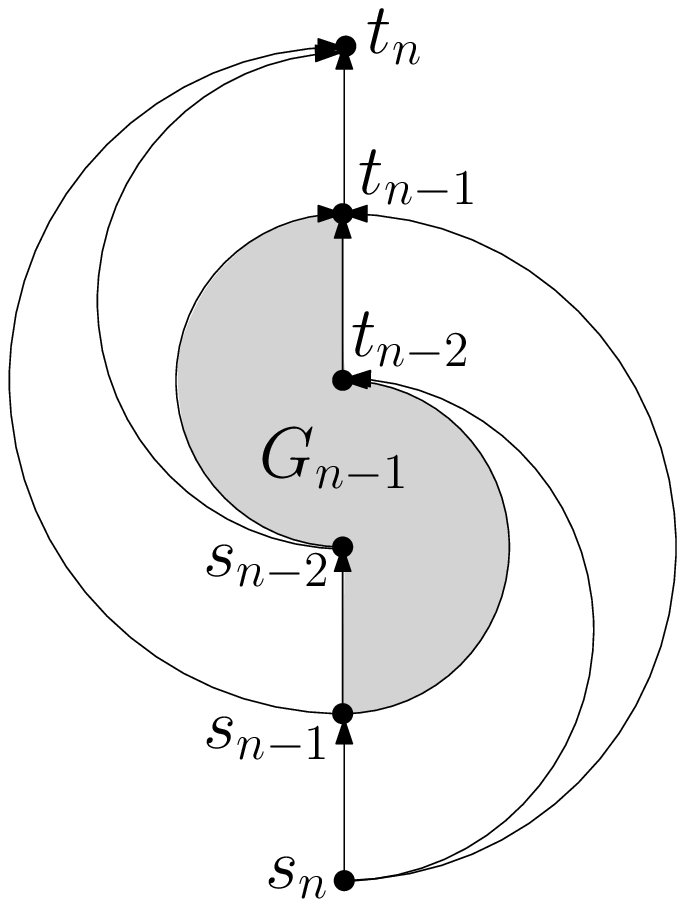,scale=0.55,clip=}} \hspace{2mm} &
	\mbox{\epsfig{figure=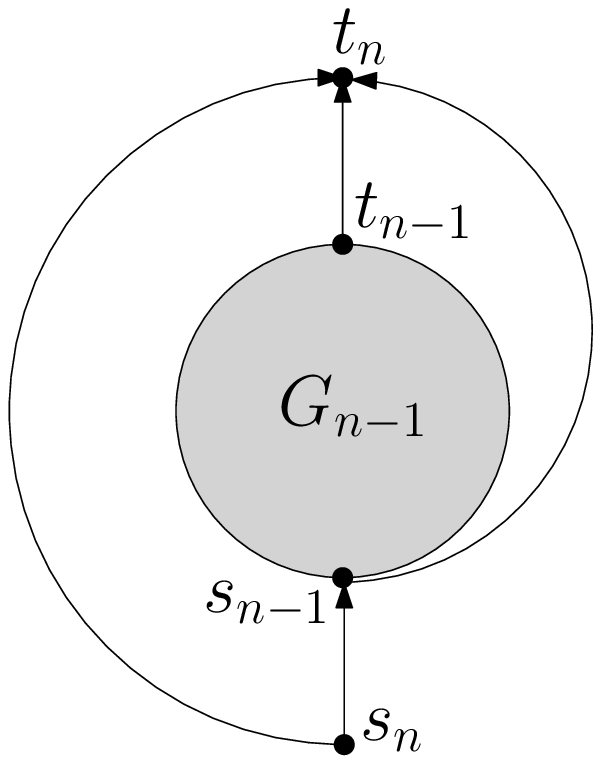,scale=0.55,clip=}} \hspace{2mm} &
    \mbox{\epsfig{figure=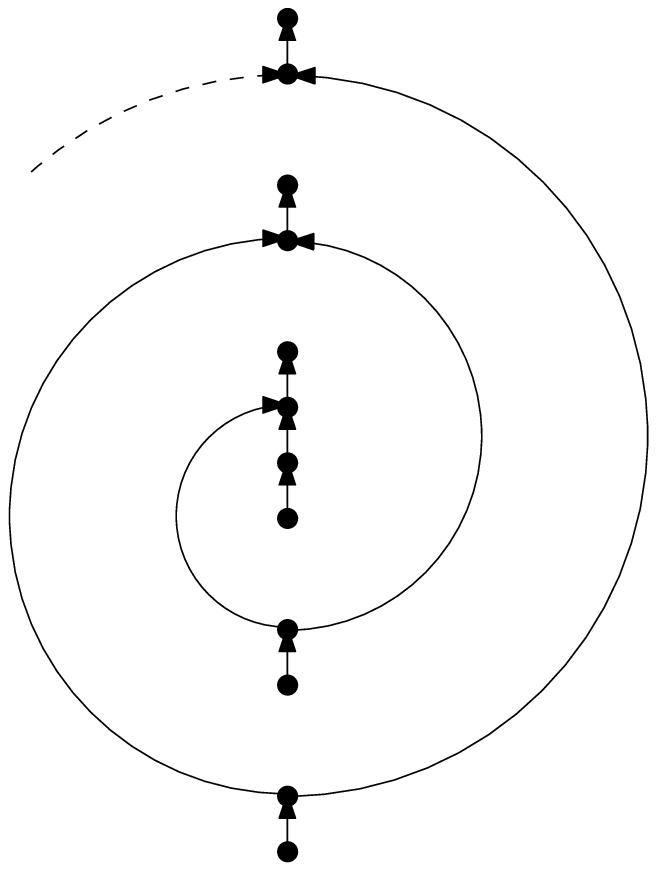,scale=0.55,clip=}}\\
        (a) \hspace{2mm} & (b) \hspace{2mm} & (c) \hspace{2mm} & (d) \\
  \end{tabular}
  \caption{(a)-(b) Inductive construction of a class $G_n$ of upward planar DAGs requiring exponential area in any planar straight-line upward drawing. (c) Inductive construction of a class of series-parallel DAGs requiring exponential area in any planar straight-line upward drawing respecting a fixed embedding. (d) A class of directed trees requiring exponential area in any planar straight-line upward drawing respecting a fixed embedding.}
  \label{fig:upward}
\end{figure}

Di Battista and Tamassia proved in~\cite{BattistaT88} that every upward planar DAG admits an upward straight-line drawing. Such a result is achieved by means of an algorithm similar to F\'ary's algorithm for constructing planar straight-line drawings of undirected planar graphs (see Sect.~\ref{se:straight-planar}). However, while planar straight-line drawings of undirected planar graphs can be constructed in polynomial area, Di Battista \emph{et al.}~proved in~\cite{BattistaTT92} that there exist upward planar DAGs that require exponential area in any planar straight-line upward drawing. Such a result is achieved by considering the class $G_n$ of DAGs whose inductive construction is shown in Fig.~\ref{fig:upward}(a)--(b) and by using some geometric considerations to prove that the area of the smallest region containing an upward planar straight-line drawing of $G_{n}$ is a constant number of times larger than the area of a region containing an upward planar straight-line drawing of $G_{n-1}$. The techniques introduced by Di Battista \emph{et al.}~in~\cite{BattistaTT92} to prove the exponential lower bound for the area requirements of upward planar straight-line drawings of upward planar DAGs have later been strengthened by Bertolazzi \emph{et al.}~in~\cite{BertolazziCBTT94} and by Frati in~\cite{f-mapudt-08} to prove, respectively, that there exist series-parallel DAGs with fixed embedding (see Fig.~\ref{fig:upward}(c)) and there exist directed trees with fixed embedding (see Fig.~\ref{fig:upward}(d)) requiring exponential area in any upward planar straight-line drawing. Similar lower bound techniques have also been used to deal with straight-line drawings of clustered graphs (see Sect.~\ref{se:clustered}).

On the positive side, area-efficient algorithms exist for constructing upward planar straight-line drawings for restricted classes of upward planar DAGs. Namely, Bertolazzi \emph{et al.}~in~\cite{BertolazziCBTT94} have shown how to construct upward planar straight-line drawings of series-parallel DAGs in optimal $\Theta(n^2)$ area, and Frati~\cite{f-mapudt-08} has shown how to construct upward planar straight-line drawings of directed trees in optimal $\Theta(n \log n)$ area. Both algorithms are based on the inductive construction of upward planar straight-line drawings satisfying some additional geometric constraints. We remark that for upward planar DAGs whose underlying graph is a series-parallel graph neither an exponential lower bound nor a polynomial upper bound is known for the area requirements of straight-line upward planar drawings. Observe that testing upward planarity for this family of graphs can be done in polynomial time~\cite{dgl-usupt-05}.

\begin{problem}
What are the area requirements of straight-line upward planar drawings of upward planar DAGs whose underlying graph is a series-parallel graph?
\end{problem}

Algorithms have been provided to construct upward planar poly-line drawings of upward planar DAGs. The first $\Theta(n^2)$ optimal area upper bound for such drawings has been established by Di Battista and Tamassia in~\cite{BattistaT88}. Their algorithm consists of first constructing an upward visibility representation of the given upward planar DAG and then of turning such a representation into an upward poly-line drawing. Such a technique has been discussed in Sect.~\ref{se:poly-line}.

\section{Convex Drawings}\label{se:convex}

In this section, we discuss algorithms and bounds for constructing small-area convex and strictly-convex drawings of planar graphs. Table~\ref{ta:convex} summarizes the best known area bounds for convex and strictly-convex drawings of planar graphs.

\begin{table}[!htb]\footnotesize
\centering
  \linespread{1.2}
  \selectfont
  \begin{tabular}{|c|c|c|c|c|}
    \cline{2-5}
    \multicolumn{1}{c|}{} & \emph{Upper Bound} & \emph{Refs.} & \emph{Lower Bound} & \emph{Refs.} \\
    \hline
    {\em Convex} & $n^2+O(n)$ & \cite{ChrobakK97,st-cdpg-92,BattistaTV99,BonichonFM07} & $\frac{4n^2}{9} - O(n)$ & \cite{Val81,fpp-hdpgg-90,FratiP07,mnra-madp3t-10}\\
    \hline
    {\em Strictly-Convex}  & $O(n^4)$ & \cite{br-scdpg-06} & $\Omega(n^3)$ & \cite{a-lbvscb-63,r-baclp-88,BaranyP92,BaranyT04} \\
    \hline
 \end{tabular}
 \vspace{2mm}
 \caption{\small A table summarizing the area requirements for convex and strictly-convex drawings of triconnected plane graphs.}
  \label{ta:convex}
\end{table}

Not every planar graph admits a convex drawing. Tutte~\cite{t-crg-60,t-hdg-63} proved that every triconnected planar graph $G$ admits a strictly-convex drawing in which its outer face is drawn as an arbitrary strictly-convex polygon $P$. His algorithm consists of first drawing the outer face of $G$ as $P$ and then placing each vertex at the barycenter of the positions of its adjacent vertices. This results in a set of linear equations that always admits a unique solution.

Characterizations of the plane graphs admitting convex drawings were given by Tutte in~\cite{t-crg-60,t-hdg-63}, by Thomassen in~\cite{Thomassen80,t-prg-84}, by Chiba, Yamanouchi, and Nishizeki in~\cite{cyn-lacdp-84}, by Nishizeki and Chiba in~\cite{nc-pgta-88}, by Di Battista, Tamassia, and Vismara in~\cite{BattistaTV01}. Roughly speaking, the plane graphs admitting convex drawings are biconnected, their separation pairs are composed of vertices both incident to the outer face, and distinct separation pairs do not ``nest''. Chiba, Yamanouchi, and Nishizeki presented in~\cite{cyn-lacdp-84} a linear-time algorithm for testing whether a graph admits a convex drawing and producing a convex drawing if the graph allows for one. The area requirements of convex and strictly-convex grid drawings have been widely studied, especially for triconnected plane graphs.

\begin{figure}[htb]
  \centering
	\mbox{\epsfig{figure=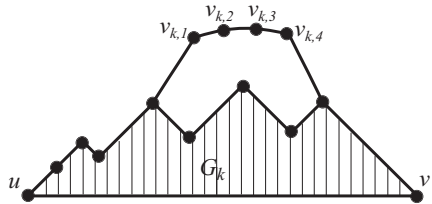,scale=0.35,clip=}}
  \caption{An illustration of the canonical ordering of a triconnected plane graph.}
  \label{fig:kant}
\end{figure}

Convex grid drawings of triconnected plane graphs can be realized on a quadratic-size grid. This was first shown by Kant in~\cite{Kant96}. In fact, Kant proved that such drawings can always be realized on a $(2n-4)\times(n-2)$ grid. The result is achieved by defining a stronger notion of canonical ordering of a plane graph (see Sect.~\ref{se:straight-planar}). Such a strengthened canonical ordering allows to construct every triconnected plane graph $G$ starting from a cycle delimiting an internal face of $G$ and repeatedly adding to the previously constructed biconnected graph $G_{k}$ a vertex or a path in the outer face of $G_{k}$ so that the newly formed graph $G_{k+1}$ is also biconnected (see Fig.~\ref{fig:kant}). Observe that this generalization of the canonical ordering allows to deal with plane graphs containing non-triangular faces. Similarly to de Fraysseix \emph{et al.}'s algorithm~\cite{fpp-hdpgg-90}, Kant's algorithm exploits a canonical ordering of $G$ to incrementally construct a convex drawing of $G$ in which the outer face of the currently considered graph $G_k$ is composed of segments whose slopes are either $-45^{\degree}$, or $0^{\degree}$, or $45^{\degree}$.

The bound of Kant was later improved down to $(n-2)\times(n-2)$ by Chrobak and Kant~\cite{ChrobakK97}, and independently by Schnyder and Trotter~\cite{st-cdpg-92}. The result of Chrobak and Kant again relies on a canonical ordering. On the other hand, the result of Schnyder and Trotter relies on a generalization of the Schnyder realizers (see Sect.~\ref{se:straight-planar}) in order to deal with triconnected plane graphs. Such an extension was independently shown by Di~Battista, Tamassia, and Vismara~\cite{BattistaTV99}, who proved that every triconnected plane graph has a convex drawing on a $(f-2)\times(f-2)$ grid, where $f$ is the number of faces of the graph. The best bound is currently, as far as we know, an $(n-2-\Delta)\times(n-2-\Delta)$ bound achieved by Bonichon, Felsner, and Mosbah in~\cite{BonichonFM07}. The bound is again achieved using Schnyder realizers. The parameter $\Delta$ is dependent of the Schnyder realizers, and can vary among $0$ and $\frac{n}{2}-2$. The following remains open:

\begin{problem}
Close the gap between the $(n-2-\Delta)\times(n-2-\Delta)$ upper bound and the $\frac{4n^2}{9}-O(n)$ lower bound for the area requirements of convex drawings of triconnected plane graphs.
\end{problem}

Strictly-convex drawings of triconnected plane graphs might require $\Omega(n^3)$ area. In fact, an $n$-vertex cycle needs $\Omega(n^3)$ area in any grid realization (see, e.g.,~\cite{a-lbvscb-63,BaranyP92,BaranyT04}). The currently best lower bound for the area requirements of a strictly-convex polygon drawn on the grid, which has been proved by Rabinowitz in~\cite{r-baclp-88}, is $\frac{n^3}{8\pi^2}$. The first polynomial upper bound for strictly-convex drawings of triconnected plane graphs has been proved by Chrobak, Goodrich, and Tamassia in~\cite{ChrobakGT96}. The authors showed that every triconnected plane graph admits a strictly-convex drawing in an $O(n^3)\times O(n^3)$ grid. Their idea consists of first constructing a (non-strictly-) convex drawing of the input graph, and of then perturbing the positions of the vertices in order to achieve strict convexity. A more elaborated technique relying on the same idea allowed Rote to achieve an $O(n^{7/3})\times O(n^{7/3})$ area upper bound in~\cite{Rote05}, which was further improved by B\'ar\'any and Rote to $O(n^2)\times O(n^2)$ and to $O(n)\times O(n^3)$ in~\cite{br-scdpg-06}. The last ones are, as far as we know, the best known upper bounds. One of the main differences between the Chrobak \emph{et al.}'s algorithm, and the B\'ar\'any and Rote's ones is that the former one constructs the intermediate non-strictly-convex drawing by making use of a canonical ordering of the graph, while the latter ones by making use of the Schnyder realizers. The following is, in our opinion, a very nice open problem:

\begin{problem}
Close the gap between the $O(n^4)$ upper bound and the $\Omega(n^3)$ lower bound for the area requirements of strictly-convex drawings of triconnected plane graphs.
\end{problem}

\section{Proximity Drawings}\label{se:proximity}

In this section, we discuss algorithms and bounds for constructing small-area proximity drawings of planar graphs.

Characterizing the graphs that admit a proximity drawing, for a certain definition of proximity, is a difficult problem. For example, despite several research efforts (see, e.g.,~\cite{d-rdt-90,ll-pdog-96,ds-gtcidr-96}), characterizing the graphs that admit a \emph{realization} (word which often substitutes \emph{drawing} in the context of proximity graphs) as Delaunay triangulations is still an intriguing open problem. Dillencourt showed that every maximal outerplanar graph can be realized as a Delaunay triangulation~\cite{d-rdt-90} and provided examples of small triangulations which can not. The decision version of several realizability problems (that is, given a graph $G$ and a definition of proximity, can $G$ be realized as a proximity graph?) is $\mathcal{NP}$-hard. For example, Eades and Whitesides proved that deciding whether a tree can be realized as a minimum spanning tree is an $\mathcal{NP}$-hard problem~\cite{EadesW96}, and that deciding whether a graph can be realized as a nearest neighbor graph is an $\mathcal{NP}$-hard problem~\cite{ew-lerpnng-96}, as well. Both proofs rely on a mechanism for providing the hardness of graph drawing problems, called \emph{logic engine}, which is interesting by itself. On the other hand, for several definitions of proximity graphs (such as Gabriel graphs and relative neighborhood graphs), the realizability problem is polynomial-time solvable for trees, as shown by Bose, Lenhart, and Liotta~\cite{bll-cpt-96}; further, Lubiw and Sleumer proved that maximal outerplanar graphs can be realized as relative neighborhood graphs and Gabriel graphs~\cite{ls-mogrng-93}, a result later extended by Lenhart and Liotta to all biconnected outerplanar graphs~\cite{ll-pdog-96}. For more results about proximity drawings, see~\cite{gll-pds-94,l-cpdg-95,gd-handbook}.

\begin{figure}[htb]
  \centering
  \begin{tabular}{c c}
	\mbox{\epsfig{figure=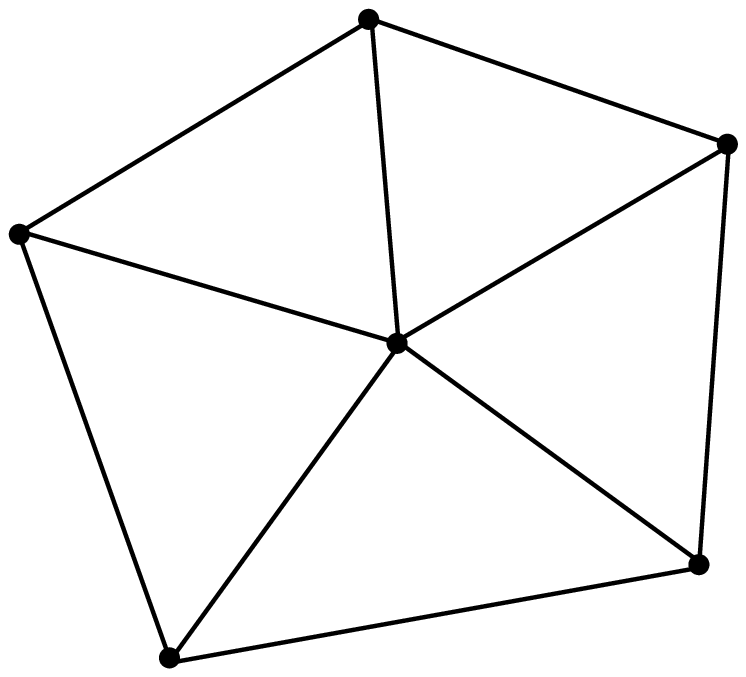,scale=0.55,clip=}} \hspace{5mm} &
	\mbox{\epsfig{figure=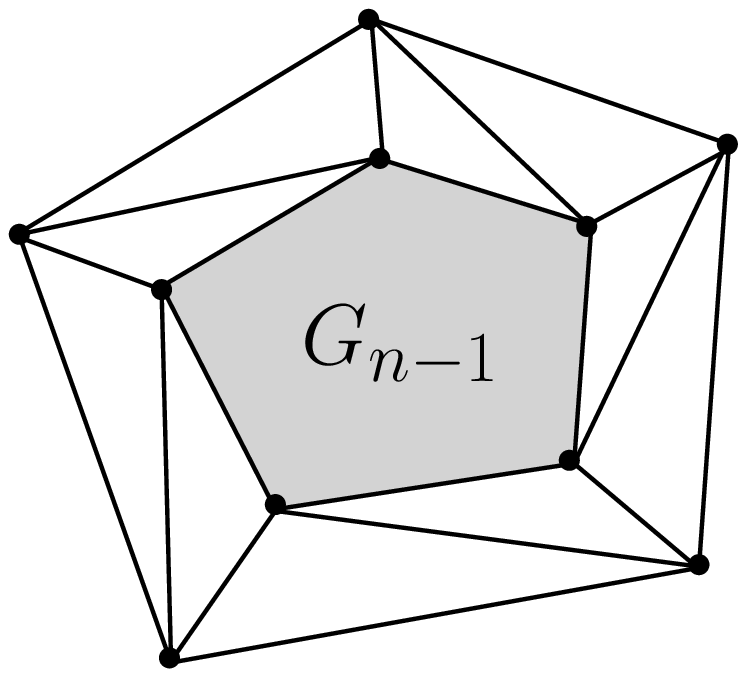,scale=0.55,clip=}}
  \end{tabular}
  \caption{Inductive construction of a class $G_n$ of graphs requiring exponential area in any Gabriel drawing, in any weak Gabriel drawing, and in any $\beta$-drawing.}
  \label{fig:gabriel}
\end{figure}

Most of the known algorithms to construct proximity drawings produce representations whose size increases exponentially with the number of vertices (see, e.g.,~\cite{ls-mogrng-93,bll-cpt-96,ll-pdog-96,dlw-swp-06}). This seems to be unavoidable for most kinds of proximity drawings, although few exponential area lower bounds are known. Liotta \emph{et al.}~\cite{ltt-argd-97} showed a class of graphs (whose inductive construction is shown in Fig.~\ref{fig:gabriel}) requiring exponential area in any Gabriel drawing, in any weak Gabriel drawing, and in any $\beta$-drawing.
Their proof is based on the observation that the circles whose diameters are the segments representing the edges incident to the outer face of $G_n$ can not contain any point in their interior. Consequently, the vertices of $G_{n-1}$ are allowed only to be placed in a region whose area is a constant number of times smaller than the area of $G_{n}$. On the other hand, Penna and Vocca~\cite{pv-pdpav-04} showed algorithms to construct polynomial-area weak Gabriel drawings and weak $\beta$-drawings of binary and ternary trees.

\begin{figure}[htb]
  \centering
	\mbox{\epsfig{figure=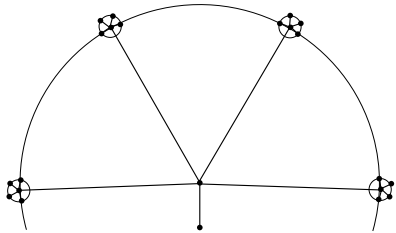,scale=0.35,clip=}}
  \caption{An illustration of the algorithm of Monma and Suri to construct realizations of degree-$5$ trees as Euclidean minimum spanning trees.}
  \label{fig:monmasuri}
\end{figure}

\begin{figure}[htb]
  \centering
	\mbox{\epsfig{figure=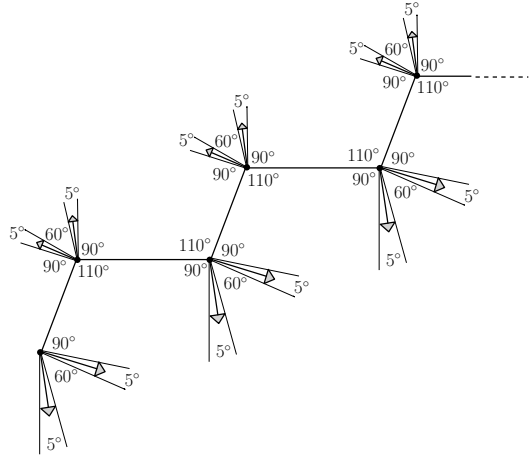,scale=0.4,clip=}}
  \caption{An illustration of the algorithm of Frati and Kaufmann to construct polynomial-area realizations of degree-$4$ trees as Euclidean minimum spanning tree realizations.}
  \label{fig:mstupper}
\end{figure}

A particular attention has been devoted to the area requirements of Euclidean minimum spanning trees. In their seminal paper on Euclidean minimum spanning trees, Monma and Suri~\cite{MonmaS92} proved that any tree of maximum degree $5$ admits a planar embedding as a Euclidean minimum spanning tree. Their algorithm, whose inductive construction is shown in Fig.~\ref{fig:monmasuri}, consists of placing the neighbors $r_i$ of the root $r$ of the tree on a circumference centered at $r$, of placing the neighbors of $r_i$ on a much smaller circumference centered at $r_i$, and so on. Monma and Suri~\cite{MonmaS92} proved that the area of the realizations constructed by their algorithm is $2^{\Omega(n^2)}$ and conjectured that exponential area is sometimes required to construct realizations of degree-$5$ trees as Euclidean minimum spanning trees.
Frati and Kaufmann~\cite{fk-pabmemstrt-08} showed how to construct polynomial area realizations of degree-$4$ trees as Euclidean minimum spanning trees. Their technique consists of using a decomposition of the input tree $T$ (similar to the ones presented in Sect.s~\ref{se:straight-outerplanar} and~\ref{se:straight-trees}) in which a path $P$ is selected such that every subtree of $P$ has at most $n/2$ nodes. Euclidean minimum spanning tree realizations of such subtrees are then inductively constructed and placed together with a drawing of $P$ to get a drawing of $T$. Suitable angles and lengths for the edges in $P$ have to be chosen to ensure that the resulting drawing is a Euclidean minimum spanning tree realization of $T$. The sketched geometric construction is shown in Fig.~\ref{fig:mstupper}.
Very recently, Angelini~\emph{et al.}~proved in~\cite{abcfks-aremst-11} that in fact there exist degree-$5$ trees requiring exponential area in any realization as a Euclidean minimum spanning tree. The tree $T^*$ exhibited by Angelini~\emph{et al.}, which is shown in Fig.~\ref{fig:treelowerbound}, consists of a degree-$5$ complete tree $T_c$ with a constant number of vertices and of a set of degree-$5$ caterpillars, each one attached to a distinct leaf of $T_c$. The complete tree $T_c$ forces the angles incident to an end-vertex of the backbone of at least one of the caterpillars to be very small, that is, between $60^{\degree}$ and $61^{\degree}$. Using this as a starting point, Angelini~\emph{et al.}~prove that each angle incident to a vertex of the caterpillar is either very small, that is, between $60^{\degree}$ and $61^{\degree}$, or is very large, that is, between $89.5^{\degree}$ and $90.5^{\degree}$. As a consequence, the lengths of the edges of the backbone of the caterpillar decrease exponentially along the caterpillar, thus obtaining the area bound. There is still some distance between the best known lower and upper bounds, hence the following is open:

 \begin{figure}[htb]
\centering{
\mbox{\epsfig{figure=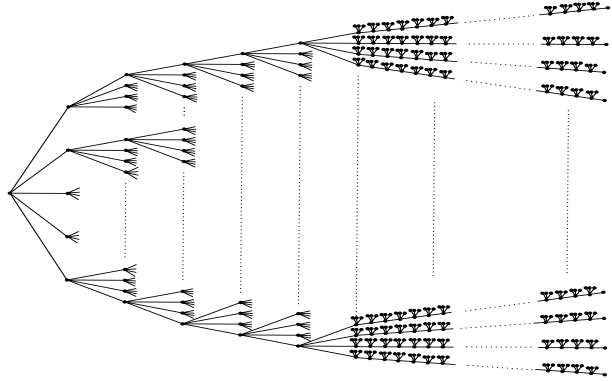,width=0.5\textwidth,height=5cm,clip=}}}
\caption{A tree $T^*$ requiring $2^{\Omega(n)}$ area in any Euclidean minimum spanning tree realization.}
\label{fig:treelowerbound}
\end{figure}

\begin{problem}
Close the gap between the $2^{O(n^2)}$ upper bound and the $2^{\Omega(n)}$ lower bound for the area requirements of Euclidean minimum spanning tree realizations.
\end{problem}

Greedy drawings are a kind of proximity drawings that recently attracted lot of attention, due to their application to network routing. Namely, consider a network in which each node $a$ that has to send a packet to some node $b$ forwards the packet to any node $c$ that is closer to $b$ than $a$ itself. If the position of any node $u$ is not its real geographic location, but rather the pair of coordinates of $u$ in a drawing $\Gamma$ of the network, it is easy to see that routing protocol never gets stuck if and only if $\Gamma$ is a greedy drawing. Greedy drawings were introduced by Rao \emph{et al.}~in~\cite{rpss-grwli-03}. A lot of attention has been devoted to a conjecture of~\cite{pr-crgr-05} stating that every triconnected planar graph has a greedy drawing. Dhandapani verified the conjecture for triangulations in~\cite{d-gdt-10}, and later Leighton and Moitra~\cite{lm-srgems-44} and independently Angelini~\emph{et al.}~\cite{afg-acgdt-10} completely settled the conjecture in the positive. The approach of Leighton and Moitra (the one of  Angelini~\emph{et al.}~is amazingly similar) consists of finding a certain subgraph of the input triconnected planar graph, called a \emph{cactus graph}, and of constructing a drawing of the cactus by induction. Greedy drawings have been proved to exist for every graph if the coordinates are chosen in the hyperbolic plane~\cite{k-gruhs-07}. Research efforts have also been devoted to construct greedy drawings in small area. More precisely, because of the routing applications, attention has been devoted to the possibility of encoding the coordinates of a greedy drawing with a small number of bits. When this is possible, the drawing is called \emph{succinct}. Eppstein and Goodrich~\cite{eg-sgghp-09} and Goodrich and Strash~\cite{gs-sggrep-09} showed how to modify the algorithm of Kleinberg~\cite{k-gruhs-07} and the algorithm of Leighton and Moitra~\cite{lm-srgems-44}, respectively, in order to construct drawings in which the vertex coordinates are represented by a logarithmic number of bits. On the other hand, Angelini~\emph{et al.}~\cite{adf-sgddae-10} proved that there exist trees requiring exponential area in any greedy drawing (or equivalently requiring a polynomial number of bits to represent their Cartesian coordinates in the Euclidean plane). The following is however open:

\begin{problem}
Is it possible to construct greedy drawings of triconnected planar graphs in the Euclidean plane in polynomial area?
\end{problem}

Partially positive results on the mentioned open problem were achieved by He and Zhang, who proved in~\cite{hz-scgd3pg-11} that succinct convex \emph{weekly greedy} drawings exist for all triconnected planar graphs, where weekly greedy means that the distance between two vertices $u$ and $v$ in the drawing is not the usual Euclidean distance $D(u,v)$ but a function $H(u,v)$ such that $D(u,v) \leq H(u,v) \leq 2 \sqrt 2 D(u,v)$. On the other hand, Cao et al.~proved in~\cite{csz-sggrep-09} that there exist triconnected planar graphs requiring exponential area in any \emph{convex} greedy drawing in the Euclidean plane.

\section{Clustered Graph Drawings}\label{se:clustered}

In this section, we discuss algorithms and bounds for constructing small-area $c$-planar drawings of clustered graphs. Table~\ref{ta:clustered} summarizes the best known area bounds for $c$-planar straight-line drawings of clustered graphs.

\begin{table}[!htb]\footnotesize
\centering
  \linespread{1.2}
  \selectfont
  \begin{tabular}{|c|c|c|c|c|}
    \cline{2-5}
    \multicolumn{1}{c|}{} & \emph{Upper Bound} & \emph{Refs.} & \emph{Lower Bound} & \emph{Refs.}\\
    \hline
    {\em Clustered Graphs} & $O(c^n)$ & \cite{EadesFLN06,afk-srdcg-11} & $\Omega(b^n)$ & \cite{cocoon/FengCE95}\\
    \hline
    {\em $c$-Connected Trees}  & $O(n^2)$ & \cite{BattistaDF07} & $\Omega(n^2)$ & \cite{BattistaDF07} \\
    \hline
    {\em Non-$c$-Connected Trees}  & $O(c^n)$ & \cite{EadesFLN06,afk-srdcg-11} & $\Omega(b^n)$ & \cite{BattistaDF07} \\
    \hline
 \end{tabular}
 \vspace{2mm}
 \caption{\small A table summarizing the area requirements for $c$-planar straight-line drawings of clustered graphs in which clusters are convex regions; $b$ and $c$ denote constants greater than $1$.}
  \label{ta:clustered}
\end{table}

Given a clustered graph, testing whether it admits a $c$-planar drawing is a problem of unknown complexity, and is perhaps the most studied problem in the Graph Drawing community during the last ten years~\cite{FengCE95,feng97algorithms,Dahlhaus98,GutwengerJLMPW02,GoodrichLS05,CorteseBPP05,cdpp-cccc-05,CornelsenW06,JelinkovaKKPSV07,df-ectefcgsf-07,cdfpp-cccg-j-08,JelinekSTV08,JelinekJKL08,afp-scgcp-09}.

Suppose that a $c$-planar clustered graph $C$ is given together with a $c$-planar embedding. How can the graph be drawn? Such a problem has been intensively studied in the literature and a number of papers have been presented for constructing $c$-planar drawings of $c$-planar clustered graphs within many drawing conventions.

\begin{figure}[htb]
  \centering
  \begin{tabular}{c c}
	\mbox{\epsfig{figure=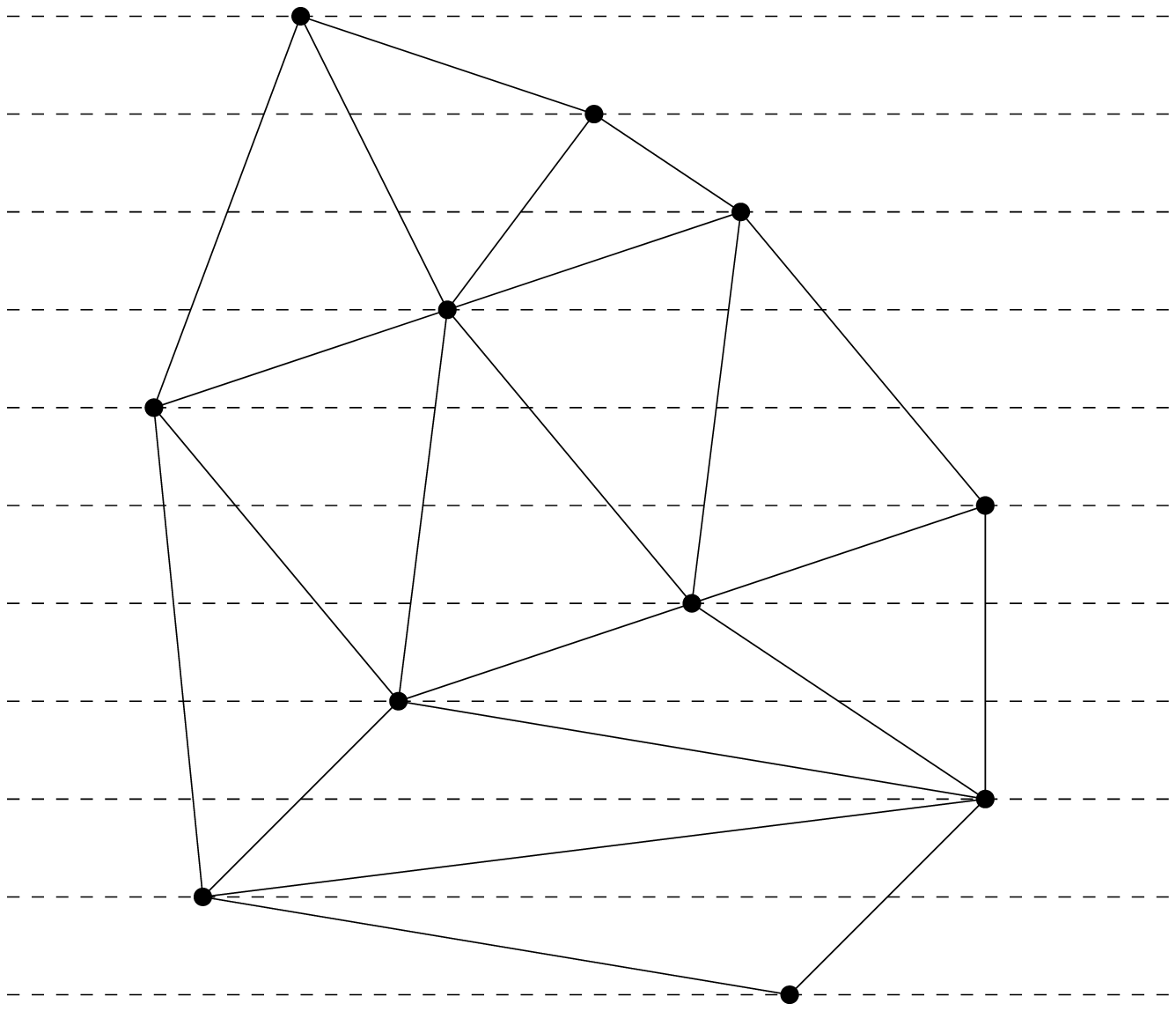,scale=0.3,clip=}} \hspace{8mm} &
    \mbox{\epsfig{figure=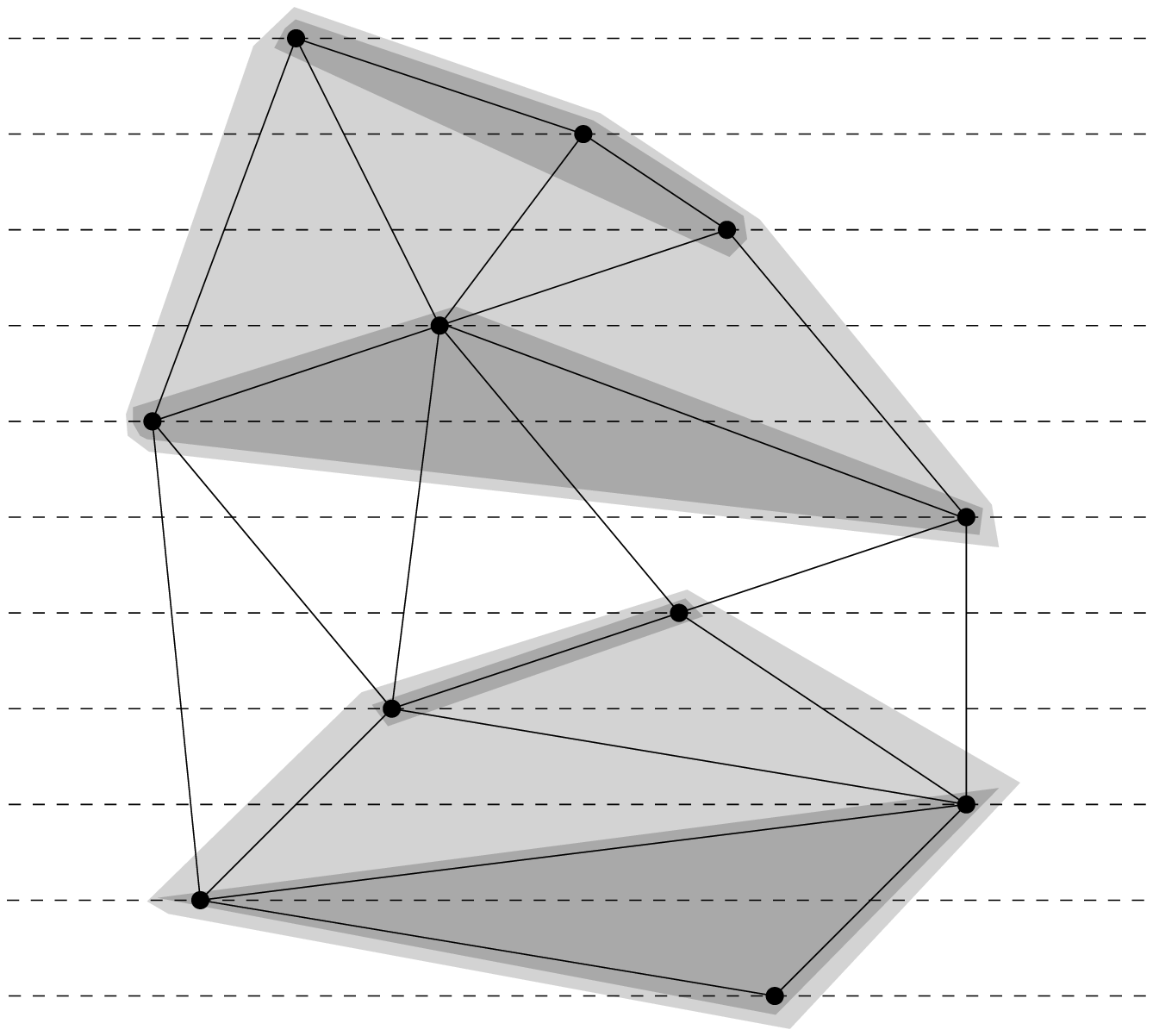,scale=0.3,clip=}}\\
        (a) \hspace{8mm} & (b)\\
  \end{tabular}
  \caption{(a) A planar straight-line drawing of a hierarchical graph $H$. Graph $H$ is obtained from a clustered graph $C$ by assigning consecutive layers to vertices of the same cluster. (b) A straight-line $c$-planar drawing of $C$.}
  \label{fig:clustered-convex}
\end{figure}

Eades \emph{et al.}~show in~\cite{EadesFLN06} an algorithm for constructing $c$-planar straight-line drawings of $c$-planar clustered graphs in which each cluster is drawn as a convex region. Such a result is achieved by first studying how to construct planar straight-line drawings of hierarchical graphs. A \emph{hierarchical graph} is a graph such that each vertex $v$ is assigned a number $y(v)$, called the \emph{layer} of $v$; a drawing of a hierarchical graph has to place each vertex $v$ on the horizontal line $y=y(v)$. Eades \emph{et al.}~show an inductive algorithm to construct a planar straight-line drawing of any hierarchical-planar graph. Second, Eades \emph{et al.}~show how to turn a $c$-planar clustered graph $C$ into a hierarchical graph $H$ such that, for each cluster $\mu$ in $C$, all the vertices in $\mu$ appear in consecutive layers of the hierarchy. This implies that, once a planar straight-line drawing of $H$ has been constructed, as in Fig.~\ref{fig:clustered-convex}(a), each cluster $\mu$ can be drawn as a region surrounding the convex hull of the vertices in $\mu$, resulting in a straight-line $c$-planar drawing of $C$ in which each cluster is drawn as a convex region, as in Fig.~\ref{fig:clustered-convex}(b).

\begin{figure}[htb]
  \centering
  \begin{tabular}{c c}
	\mbox{\epsfig{figure=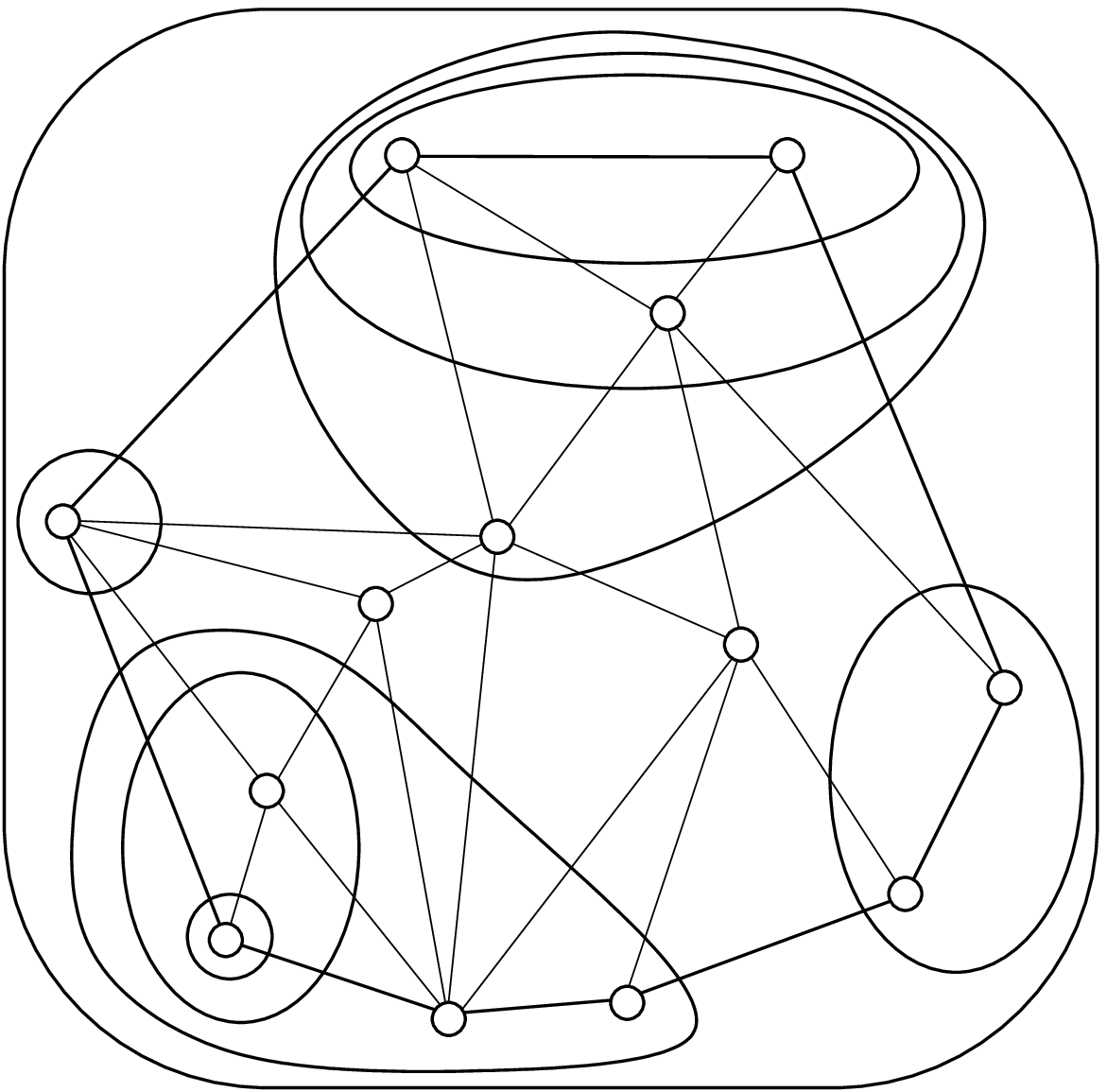,scale=0.38,clip=}} \hspace{8mm} &
    \mbox{\epsfig{figure=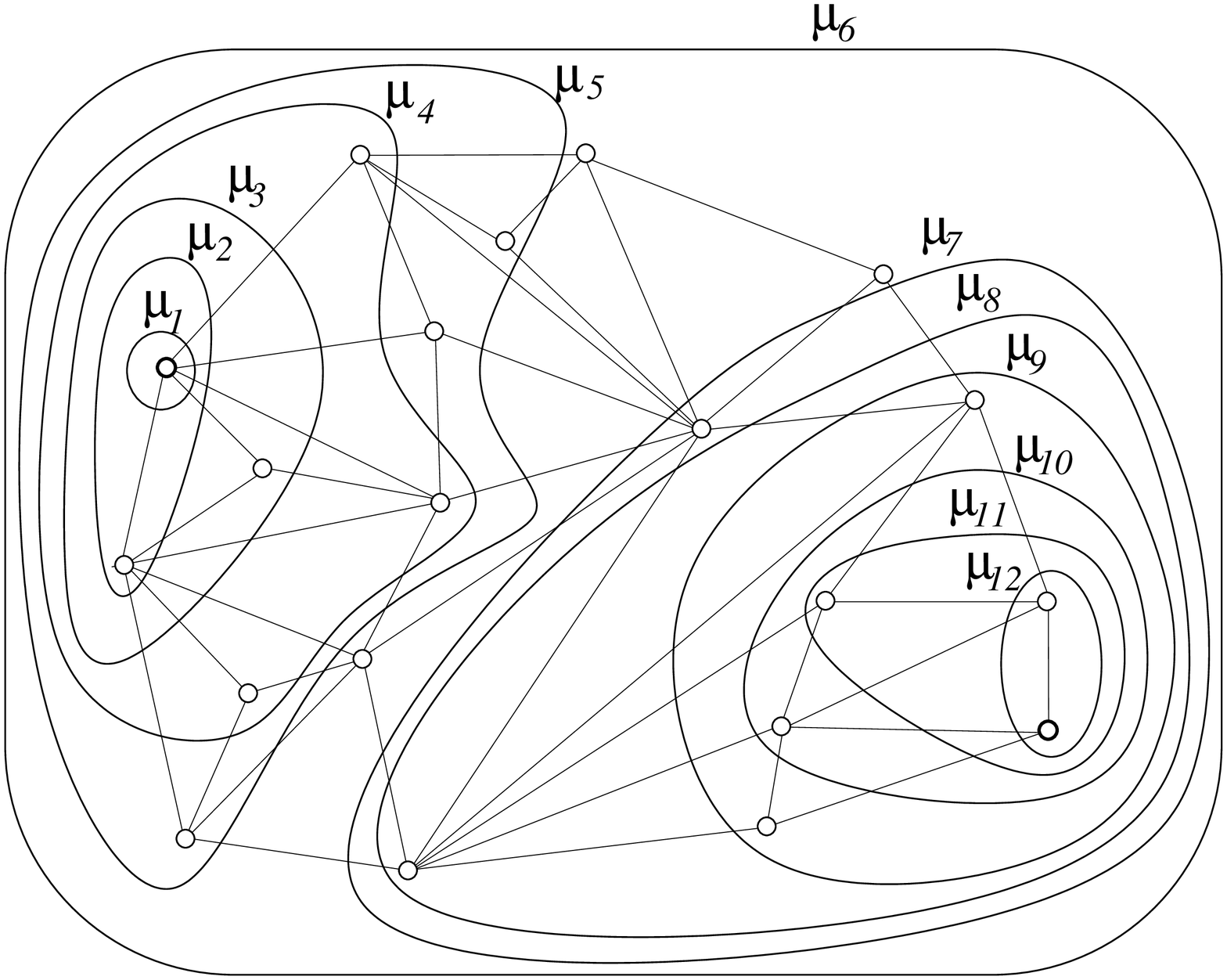,scale=0.25,clip=}}\\
        (a) \hspace{8mm} & (b)\\
  \end{tabular}
  \caption{(a) An outerclustered graph. (b) A linearly-ordered outerclustered graph. Any two consecutive clusters in the sequence $\mu_1,\dots,\mu_{12}$ are one the parent of the other.}
  \label{fig:clustered-rectangular}
\end{figure}

Angelini \emph{et al.}, improving upon the described result of Eades \emph{et al.}~in~\cite{EadesFLN06} and answering a question posed in~\cite{EadesFLN06}, show in~\cite{afk-srdcg-11} an algorithm for constructing a \emph{straight-line rectangular drawing} of any clustered graph $C$, that is, a $c$-planar straight-line drawing of $C$ in which each cluster is drawn as an axis-parallel rectangle (more in general, the algorithm of Angelini \emph{et al.}~constructs straight-line $c$-planar drawings in which each cluster is an arbitrary convex shape). The algorithm of Angelini \emph{et al.}~is reminiscent of F\'ary's algorithm (see \cite{f-srpg-48} and Sect.~\ref{se:straight-planar}). Namely, the algorithm turns a clustered graph $C$ into a smaller clustered graph $C'$ by either removing a cluster, or splitting $C$ in correspondence of a separating $3$-cycle, or contracting an edge of $C$. A straight-line rectangular drawing of $C'$ can then be inductively constructed and easily augmented to a straight-line rectangular drawing of $C$. When none of the inductive cases applies, the clustered graph is an \emph{outerclustered graph}, that is, every cluster contains a vertex incident to the outer face (see Fig.~\ref{fig:clustered-rectangular}(a)). In order to draw an outerclustered graph $C$, Angelini \emph{et al.} show how to split $C$ into three \emph{linearly-ordered outerclustered graphs}, that are outerclustered graphs such that the graph induced by the ``direct containment'' relationship among clusters is a path (see Fig.~\ref{fig:clustered-rectangular}(b)), where a cluster $\mu$ \emph{directly contains} a cluster $\nu$ if $\mu$ contains $\nu$ and $\mu$ contains no cluster $\rho$ containing $\nu$. Moreover, they show how to combine the drawings of such graphs to get a straight-line rectangular drawing of $C$. Finally, Angelini \emph{et al.}~show an inductive algorithm for constructing a straight-line rectangular drawing of any linearly-ordered outerclustered graphs $C$. Such an algorithm finds a subgraph of $C$ (a path plus an edge) that splits $G$ into smaller linearly-ordered outerclustered graphs, inductively draws such subgraphs and combines their drawings to get a straight-line rectangular drawing of $C$.

Both the algorithm of Eades \emph{et al.}~and the algorithm of Angelini \emph{et al.}~construct drawings requiring, in general, exponential area. However, Feng \emph{et al.}~proved in~\cite{cocoon/FengCE95} that there exists a clustered graph $C$ requiring exponential area in any straight-line $c$-planar drawing in which the clusters are represented by convex regions. The proof of such a lower bound is strongly based on the proof of Di Battista \emph{et al.}~that there exist directed graphs requiring exponential area in any upward straight-line drawing (see~\cite{BattistaTT92} and Sect.~\ref{se:upward}). Eades \emph{et al.}~showed in~\cite{EadesFN99} how to construct $O(n^2)$ area $c$-planar orthogonal drawings of clustered graphs with maximum degree $4$; the authors first construct a visibility representation of the given clustered graph and then turn such a representation into an orthogonal drawing. Di Battista \emph{et al.}~\cite{BattistaDF07} show algorithms for drawing clustered trees in small area. In particular, they show an inductive algorithm to construct straight-line rectangular drawings of $c$-connected clustered trees in $O(n^2)$ area; however, they prove that there exist non-$c$-connected trees requiring exponential area in any straight-line drawing in which the clusters are represented by convex regions, again using the tools designed by Di Battista \emph{et al.}~in~\cite{BattistaTT92}. The following problem has been left open by Di Battista \emph{et al.}~\cite{BattistaTT92}.

\begin{problem}
What are the area requirements of order-preserving straight-line $c$-planar drawings of clustered trees in which clusters are represented by convex regions?
\end{problem}

\bibliography{bibliography}
\bibliographystyle{plain}

\end{document}